\documentclass{aa}  
\usepackage{graphicx}
\usepackage{txfonts}
\usepackage[]{natbib}
\usepackage{longtable,lscape}
\newcommand{\spitzer}{{\it Spitzer}}
\newcommand{\micron}{{$\mu$m}}
\newcommand{\av}{{A$_V$}}
\newcommand{\hii}{H{\sc ii}}

\newcommand{\mm}{M\,33}
\newcommand{\dusty}{{\it DUSTY}}
\newcommand{\tin}{T$_{\rm in}$}
\newcommand{\yout}{Y$_{\rm out}$}
\newcommand{\rout}{R$_{\rm out}$}
\newcommand{\dr}{{$\Delta {\rm r}$}}
\newcommand{\drmax}{{\dr $_{\rm max}$}}
\begin{document}
   \title{Star formation in \mm: \spitzer\ photometry of discrete sources}


   \author{S. Verley
          \inst{1}
          \and
          L. K. Hunt
          \inst{2}
          \and
          E. Corbelli
          \inst{1}
	  \and
	  C. Giovanardi
	  \inst{1}
}


   \institute{Osservatorio Astrofisico di Arcetri, Largo E. Fermi, 5 - 50125 Firenze - Italy\\
              \email{simon, edvige, giova@arcetri.astro.it}
         \and
              INAF - Istituto di Radioastronomia-Sezione Firenze, Largo E. Fermi 5, 50125 Firenze, Italy\\
              \email{hunt@arcetri.astro.it}
             }

\date{Received; accepted}

 
  \abstract
   {}
   {Combining the relative vicinity of the Local Group spiral galaxy 
\mm\ with the \spitzer\ images, we investigate the properties of infrared (IR) 
emission sites and assess the 
reliability of the IR emission as a star formation tracer.
}
{The mid- and far-IR emission of \mm\ was obtained from  
IRAC and MIPS images from the \spitzer\ archive.  We compared the 
photometric results for several samples of three known types of discrete sources 
(\hii\ regions, supernovae remnants and planetary nebulae) with theoretical 
diagnostic diagrams, and derived 
the spectral energy distribution (from 3.6 to 24~\micron) 
of each type of object. Moreover,
we generated a catalogue of 24~\micron\ sources and inferred their nature  
from the observed and theoretical colours of the 
known type sources. We estimated the star formation rate in 
\mm\ both globally and locally, from the IR emission 
and from the H$\alpha$ emission line.
}
   {
The colours of the typical IR emissions of \hii\ regions, 
supernovae remnants and planetary nebulae are 
continuous among the different samples, 
with overlapping regions in the diagnostic diagrams. 
The comparison between the model results and the colours of
\hii\ regions indicates a dusty envelope at relatively high
temperatures $\sim$600~K, and moderate extinction \av $\la$ 10.
The 24~\micron\ sources IR colours follow 
the regions observationally defined by the three classes 
of known objects but the majority of them represent \hii\ regions.
The derived total IR luminosity function is in fact very similar to 
the \hii\ luminosity function observed in the Milky Way and in other 
late type spirals. Even though our completeness limit is 
$5\times10^{37}$~ergs s$^{-1}$, 
in low density regions we are able to 
detect sources five times fainter than this, 
corresponding to the faintest possible \hii\ region. 
The 8 and 24~\micron\ luminosities within the central 5~kpc of \mm\ are 
comparable and of order $4\times10^{28}$ ergs s$^{-1}$ Hz$^{-1}$
($\nu L_\nu(8)\,=\,1.5\times10^{42}$ and $\nu L_\nu(24)\,=\,4.4\times10^{41}$~ergs s$^{-1}$).
We estimate the total IR emission in the same region of \mm\ to be 10$^9$~L$_\odot$. 
The discrete sources account for about one third of the 24~\micron\ 
emission while the rest is diffuse. 
From the IR emission, we derive a star formation 
rate for the inner disk equal to 0.2~M$_\odot$~yr$^{-1}$, 
consistent with the star formation rate obtained from the H$\alpha$ emission.}
   {}


   \maketitle

\section{Introduction}

The interstellar medium (ISM) in galaxies provides the 
raw material from which stars form, and is the repository 
of the products of stellar evolution. 
Stars form in condensations of cool molecular gas, 
eventually destroying their birth sites through 
energetic photons, massive stellar winds, and supernova (SN) explosions. 
Dust is formed through coagulation in stellar winds 
and supernova ejecta, and over time, becomes an integral part of the ISM. 
Radiative and mechanical energy are input to the ISM 
through massive star formation and evolution, 
in the form of ionising photons, supernova explosions, 
and magnetic fields.
The ISM properties, such as magnetic field, gas, 
and dust, affect star formation processes on all scales.
At the same time, star formation drives 
the evolution of the ISM.
On the long term, local processes of star formation can 
influence large-scale structure in the ISM and 
determine how galaxies evolve. 
Ultraviolet and optical wavelengths are not the best 
wavelengths for SF studies, because dust can hide massive star 
formation, mask the effects of feedback, and redistribute
the ISM energy and radiation.
 However, first {\it IRAS} \citep{1984ApJ...278L...1N}, 
then {\it ISO} \citep{1996A&A...315L..27K}, 
and now \spitzer\ \citep{2004ApJS..154....1W} have been able 
to penetrate the dust in the ISM, and thus better study 
the interaction among its constituents.

In this series of papers, we examine star formation properties
of the Local Group late type spiral \mm. 
The proximity of Local Group members together with  \spitzer\ 
resolution makes possible the resolution of
structures on spatial scales of 5--10\,pc, 
and enable us to effectively study individual star forming sites 
and their surrounding environment in a galaxy different from our own. 
\mm, is at a distance of 840\,kpc \citep{1991ApJ...372..455F}, 
its mass and apparent size are smaller than that of M\,31 but \mm\ 
hosts the brightest \hii\ complex in the Local Group. This,
together with its blue colour assures us that star formation is
still very active throughout its disk. \mm\ bears no signs of
recent mergers and therefore gives us a unique 
opportunity for investigating the interplay of gas, dust, and 
star formation in isolated galaxy disks. This is possible thanks 
also to high resolution and sensitive observations now becoming 
available for this nearby galaxy at all wavelengths. Chemical 
abundances have been measured in stellar populations of different ages 
\citep[e.g.][]{2007arXiv0704.3187M,2007arXiv0705.3116M}, and 
surveys of atomic and molecular hydrogen point out the location of 
massive gas clumps \citep{2003ApJS..149..343E,2004ApJ...602..723H}. 
Through detailed dynamical mass modeling of the galaxy we can now trace 
the mass distribution of visible and dark matter 
\citep{2000MNRAS.311..441C,2003MNRAS.342..199C}. 

Our focus here is first on \mm 's population of discrete known
type of point sources: \hii\ regions, supernova remnants (SNRs), 
and planetary nebulae (PNe). 
In principle, these sources track different phases of the 
star formation process, from the generation of ionising 
photons and stellar winds in massive stars,  
to subsequent explosion as SNe and finally to the more evolved 
stage of PNe. 
Hence, our immediate goal is to assess the physical conditions
of the stellar environment at different stages.
The second purpose of this paper is the generation of 
an IR source catalogue from the \spitzer\ 
MIPS observations at 24~\micron. 
Here we shall use IRAC-MIPS diagnostic diagrams of known sources 
to better understand the nature of these sources.
In a subsequent paper we will relate the IR properties with
detailed observations of the surrounding ISM at other wavelengths 
and to the disk large-scale structure. 
\spitzer\ data and the vast multi-wavelength dataset available
for \mm\ will help understand how star formation proceeds in low
luminosity spiral galaxies and how this relates to other global
galaxy properties.

The present paper is structured as follows: 
in Sect.~\ref{sec:data}, we present the \spitzer\ 
IRAC and MIPS data and their reduction process. 
The large scale structure of the dust emission in \mm\ is 
investigated through colour images in Sect.~\ref{sec:largeScale}. 
We analyse the IRAC and MIPS 24 photometry of discrete 
type-known sources (\hii\ regions, PNe, SNRs) in Sect.~\ref{sec:knownSources} 
and compare their colours to theoretical diagnostic diagrams 
in Sect.~\ref{sec:diagnostic}. 
In Sect.~\ref{sec:S24}, we present a catalogue of the sources 
emitting in the 24~\micron\ MIPS band and interpret their 
nature in light of the results obtained in the previous Section. 
The reliability of the IR emission as a star formation tracer, 
through comparison with H$\alpha$, is investigated in Sect.~\ref{sec:24Ha}. 
Finally, the summary and conclusions of our study are given in Sect.~\ref{sec:disc}.

\section{Observations and data reduction} \label{sec:data}

We retrieved images from the Guaranteed Time Observations (PID 5,
PI R. Gehrz) in the
\spitzer\ Space Telescope \citep{2004ApJS..154....1W} data archive. 
The \spitzer\ Space Telescope carries two photometric cameras: 
the Infrared Array Camera (IRAC) and the Multiband Imaging Photometer (MIPS).

\subsection{IRAC data} \label{sub:IRAC}

The Infrared Array Camera \citep{2004ApJS..154...10F} 
is equipped with two detector arrays of $256 \times 256$ pixels. 
The field of view of a single array is $5\farcm2 \times 5\farcm2$. 
We analyze here eight sets of IRAC observations of \mm\
(AORs 3636224, 3636480, 3637760, 3638016, 3638784, 3639040, 3640320, 3640576) 
in all four IRAC bands. 
The Basic Calibrated Data (BCD) were created by the \spitzer\ 
Science Center (SSC) pipeline, version S14.0.0 for all AORs. 
As the first data frame of each observation sequence has a shorter 
integration time than the regular exposure time (10.4 s), 
we discarded those frames. 
A more complete description of the observations is given by
\citet{2004ApJS..154..259H}, \citet{2007ApJ...664..850M}, and
\citet{2007A&A...466..509T}.

The mosaics were assembled gathering all the BCDs for each wavelength. 
The individual calibrated frames were processed using 
Mopex \citep{2005ASPC..347...81M}, with a cosmic-ray rejection 
and a background matching applied between overlapping fields of view. 
During the reduction, 
we used a common value of the pixel size equal to $1\farcs20$. 
The final number of individual BCDs used is about 1700, in each IRAC channel. 
The final dimension of the mosaics is approximately 
$62\farcm4 \times 91\farcm7$.
Because of the overlap between adjacent BCDs, the mean 
redundancy is 12 (from 6 to 30).

The obtained IRAC mosaics were background subtracted, 
estimating 20 sky levels (the median values in 
$5 \times 5$ pixels boxes) near the edges of the mosaics, 
the farthest from the galaxy centre. 
The 5.8 and 8.0 ~\micron\ channels presented a gradient in the background, 
and we removed it using a fitted surface with the IRAF task imsurfit 
(images.imfit). 
In each channel, the final background value subtracted is 
the mean of the 20 median values: 0.059 at 3.6~\micron, 
0.146 at 4.5~\micron, 1.475 at 5.8~\micron, and 5.389 at 8.0~\micron. 
The mosaics were then aligned using the IRAF tasks geomap 
and geotran (in immatch) using 15 point sources.

\begin{figure*}
\includegraphics[width=\columnwidth]{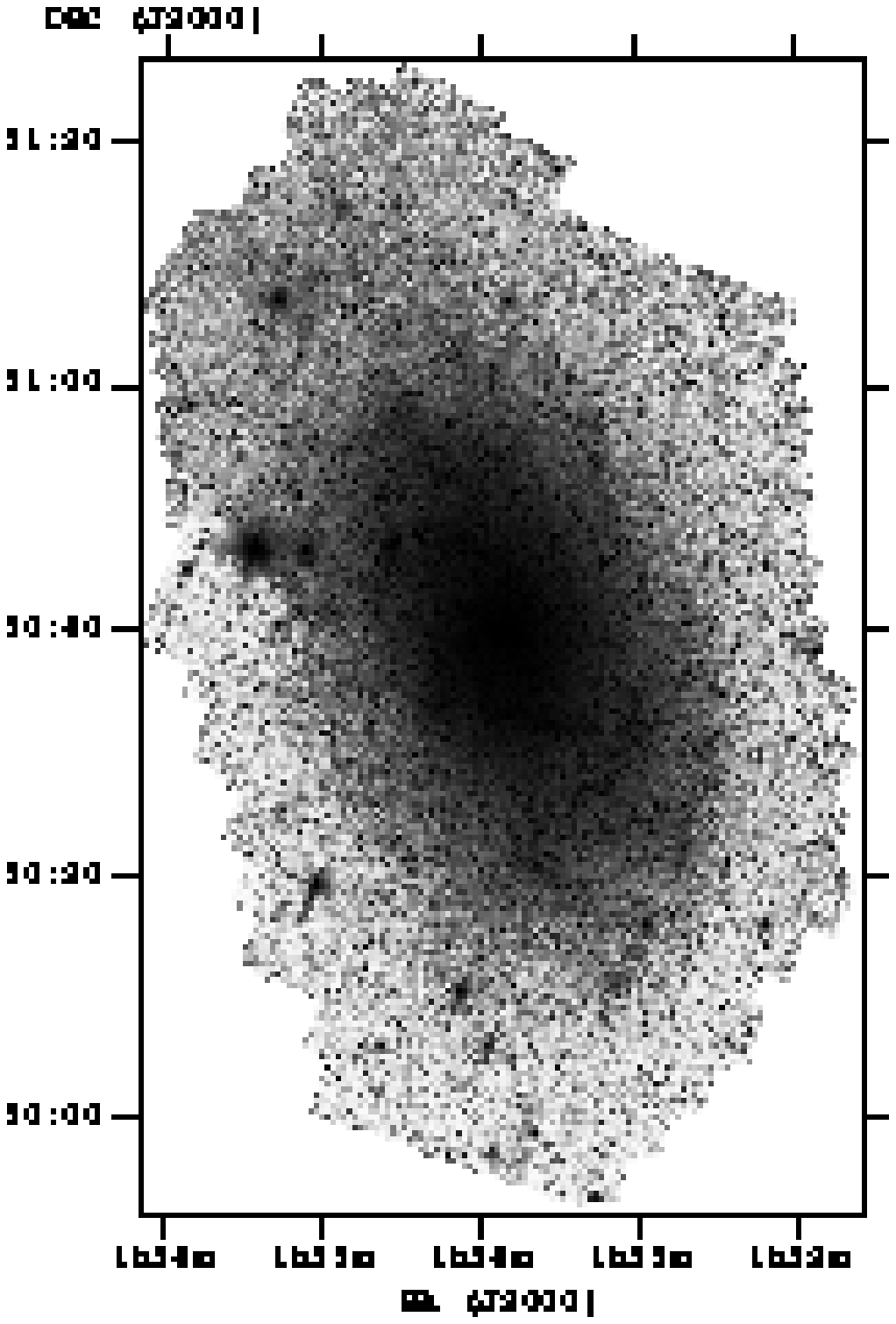}
\includegraphics[width=\columnwidth]{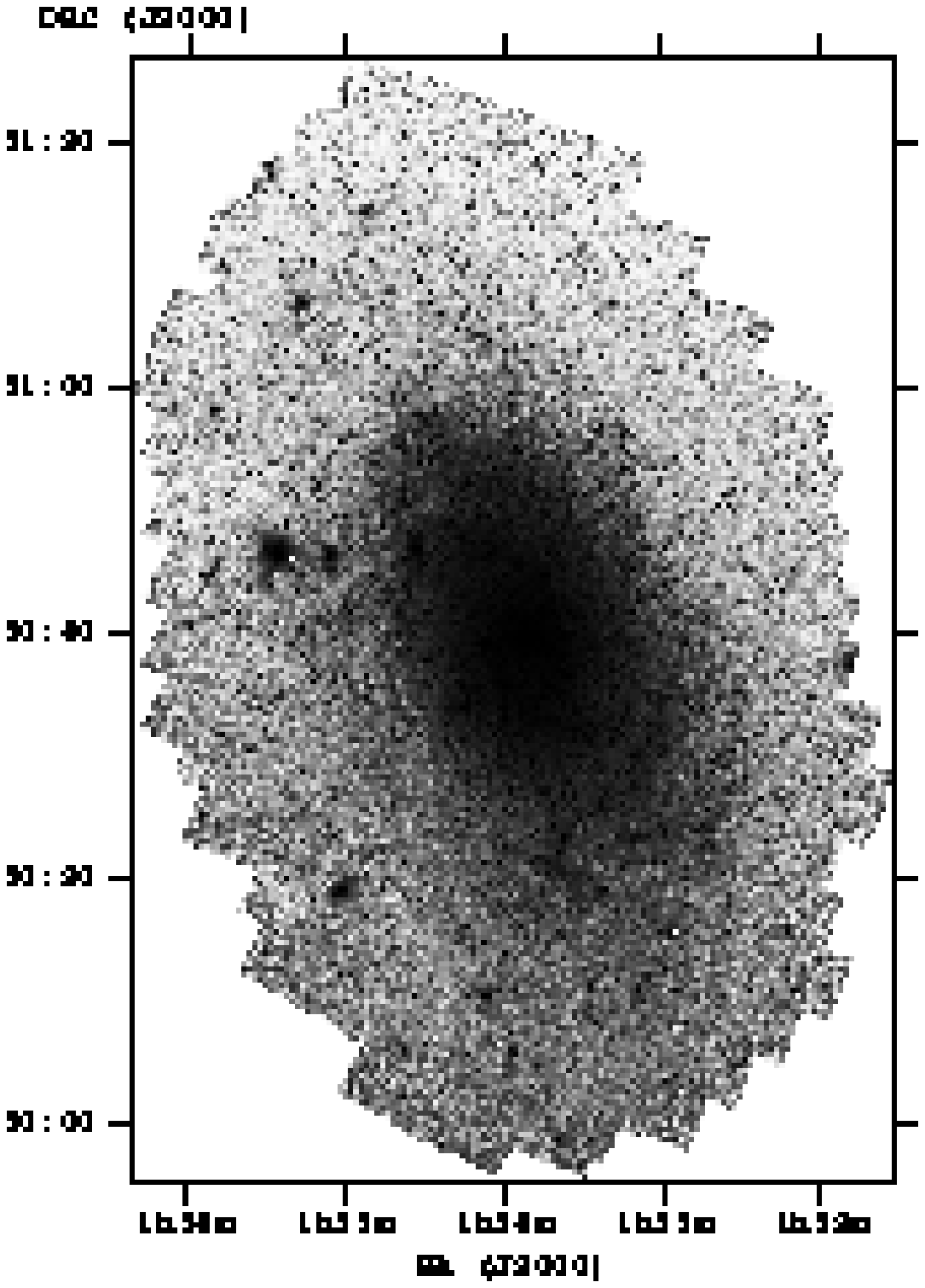}
\includegraphics[width=\columnwidth]{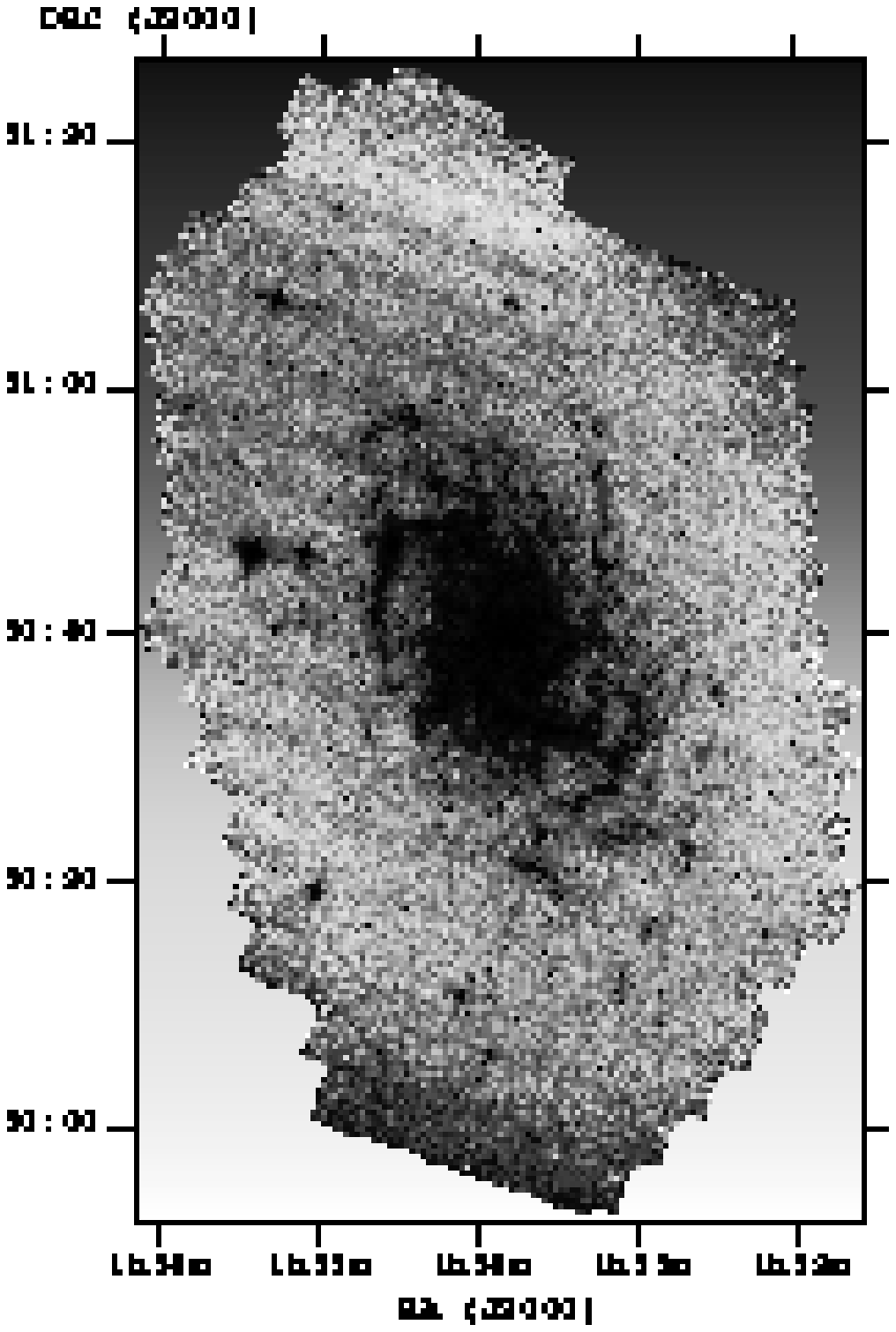}
\includegraphics[width=\columnwidth]{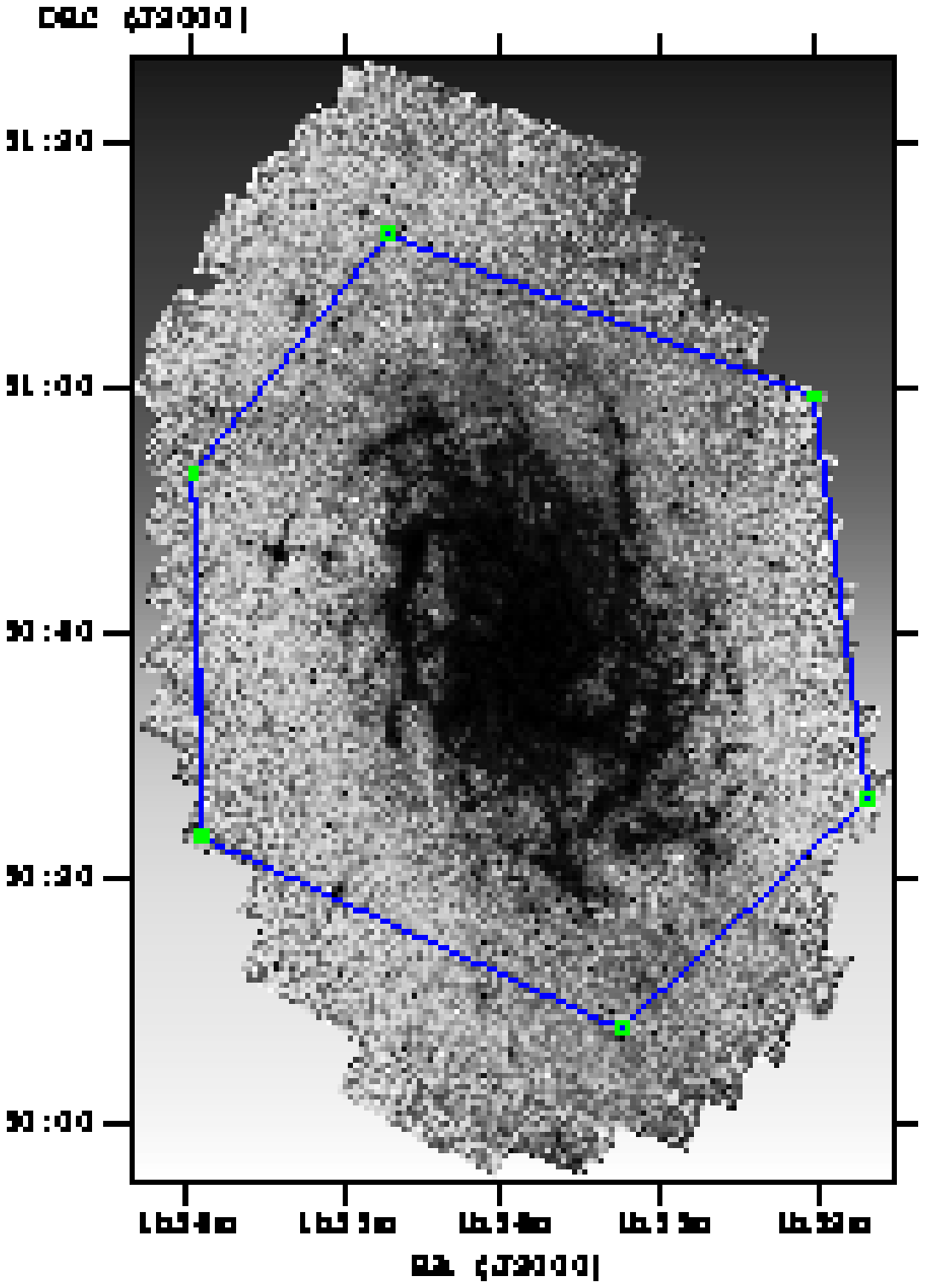}
\caption{Final IRAC mosaics of \mm. The upper left image displays 
the 3.6~\micron, the upper right 4.5~\micron, the lower 
left 5.8~\micron\ and the lower right the 8.0~\micron\ channel. 
The images are about 1\degr\ by 1.5\degr, north is up 
and east is to the left. The region enclosed by the continuous line will be further investigated in the present paper.}
\label{fig:IRAC}
\end{figure*}

Because of the diffraction limit of the telescope, 
the images have a resolution of 1\farcs7 to 2\farcs0 
(see IRAC Observing Manual) which translate into
6.7 to 8\,pc at the distance of \mm. 
The final IRAC mosaics at 3.6, 4.5, 5.8, 
and 8.0~\micron\ are shown in Fig.~\ref{fig:IRAC}.

\subsection{MIPS data} \label{sub:MIPS}

Images of \mm\ at wavelengths 24, 70, and 160~\micron\ 
(with bandwidths of 5, 19, and 35~\micron, respectively) 
were obtained using data collected by the Multiband 
Imaging Photometer \citep{2004ApJS..154...25R}. 
MIPS uses true detector arrays: 128 $\times 128$ pixels 
at 24~\micron\ (Si:As detector), $32 \times 32$ pixels 
at 70 $\mu$m (Ge:Ga device) and $2 \times 20$ pixels at 
160~\micron\ (Ge:Ga device). 
The nominal fields of view are $5' \times 5'$ at 24 
and 70~\micron\ and $0\farcm75 \times 5'$ at 160~\micron. 
The resolution of the instrument is diffraction limited to 6$''$, 
18$''$, and 40$''$ at 24, 70, and 160~\micron, respectively. 
The operating modes of all the observations were 
``scan map'', the telescope scanning the sky at a constant rate.

\begin{figure*}
\includegraphics[width=\columnwidth]{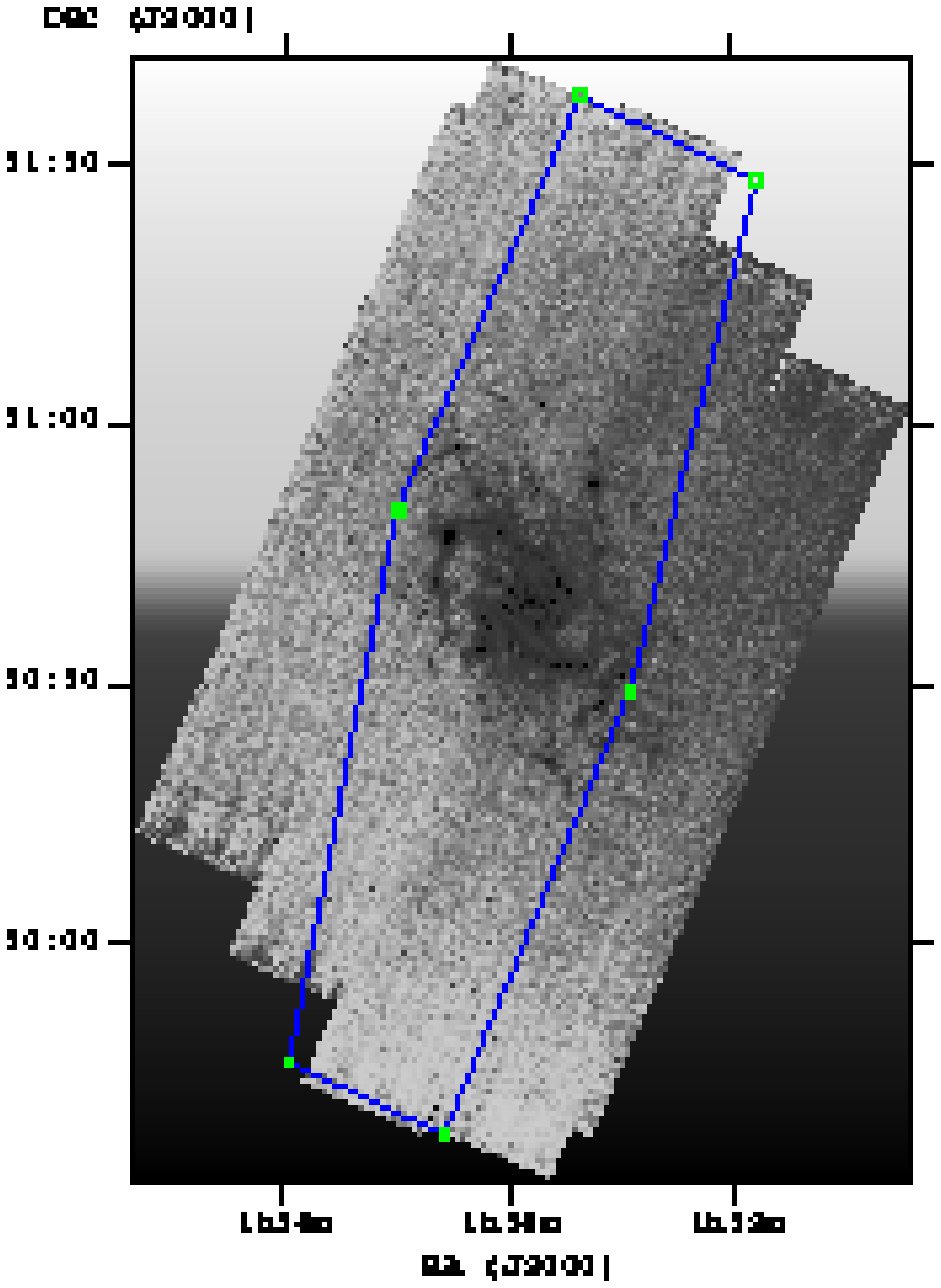}
\includegraphics[width=\columnwidth]{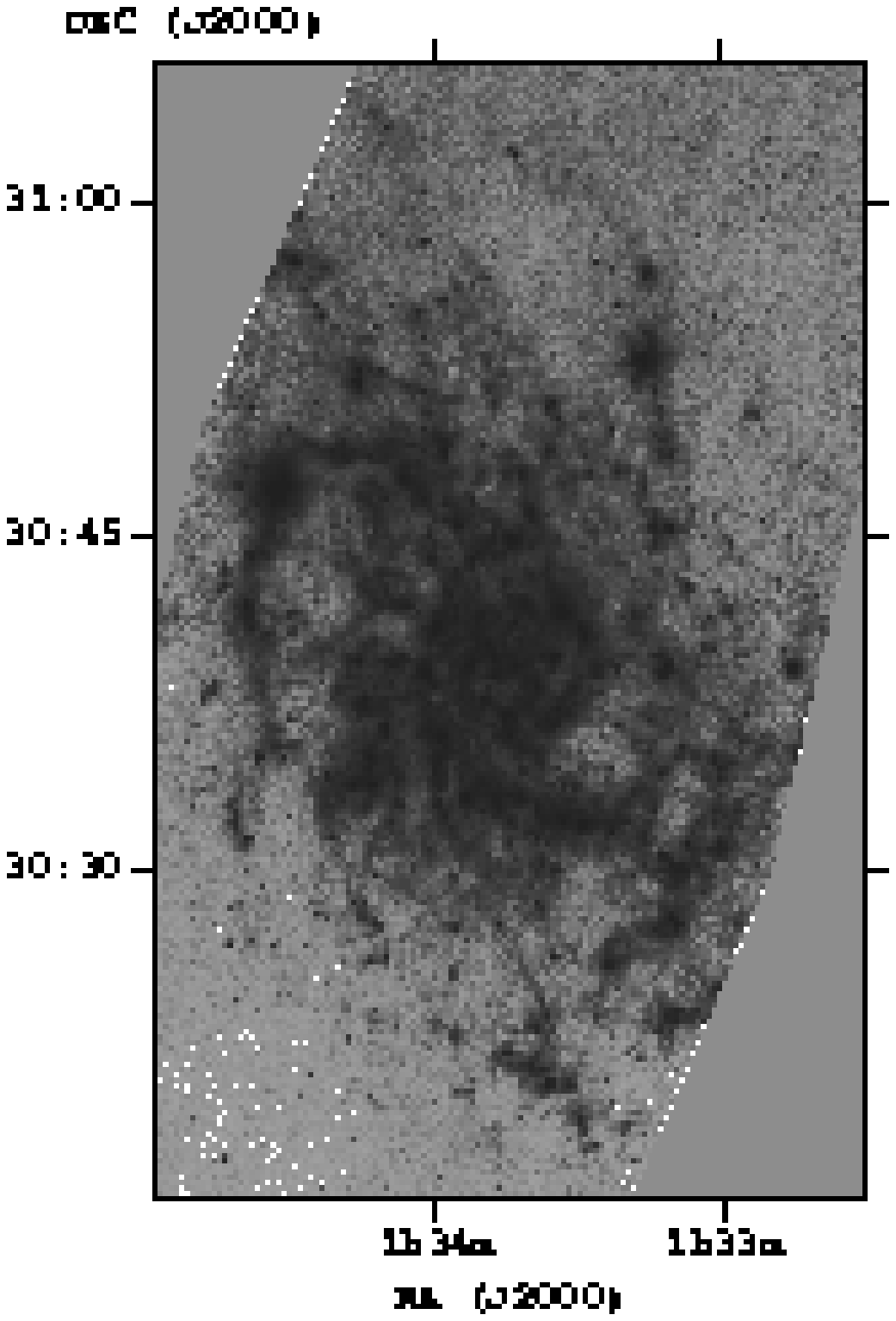}
\includegraphics[width=\columnwidth]{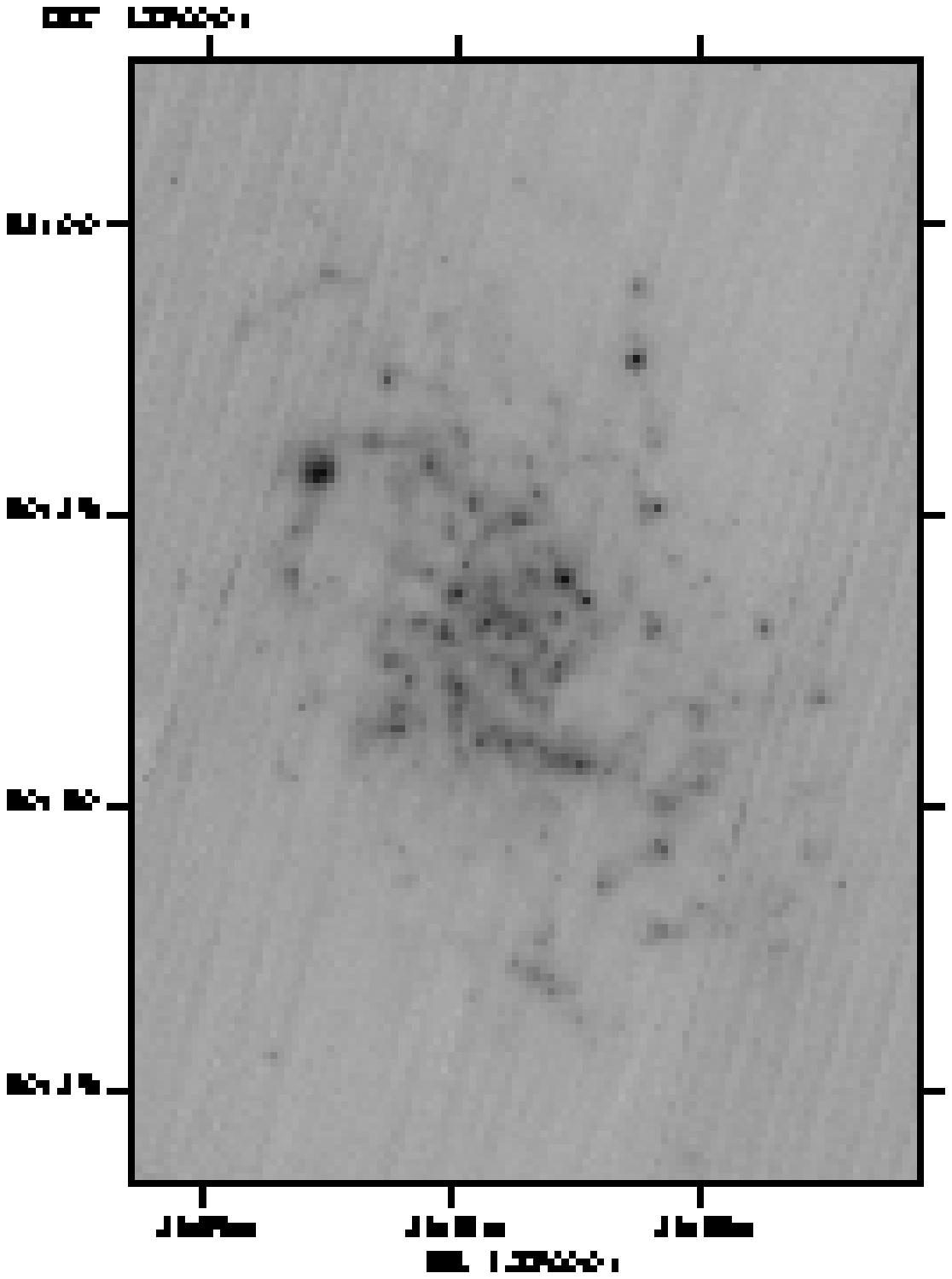}
\includegraphics[width=\columnwidth]{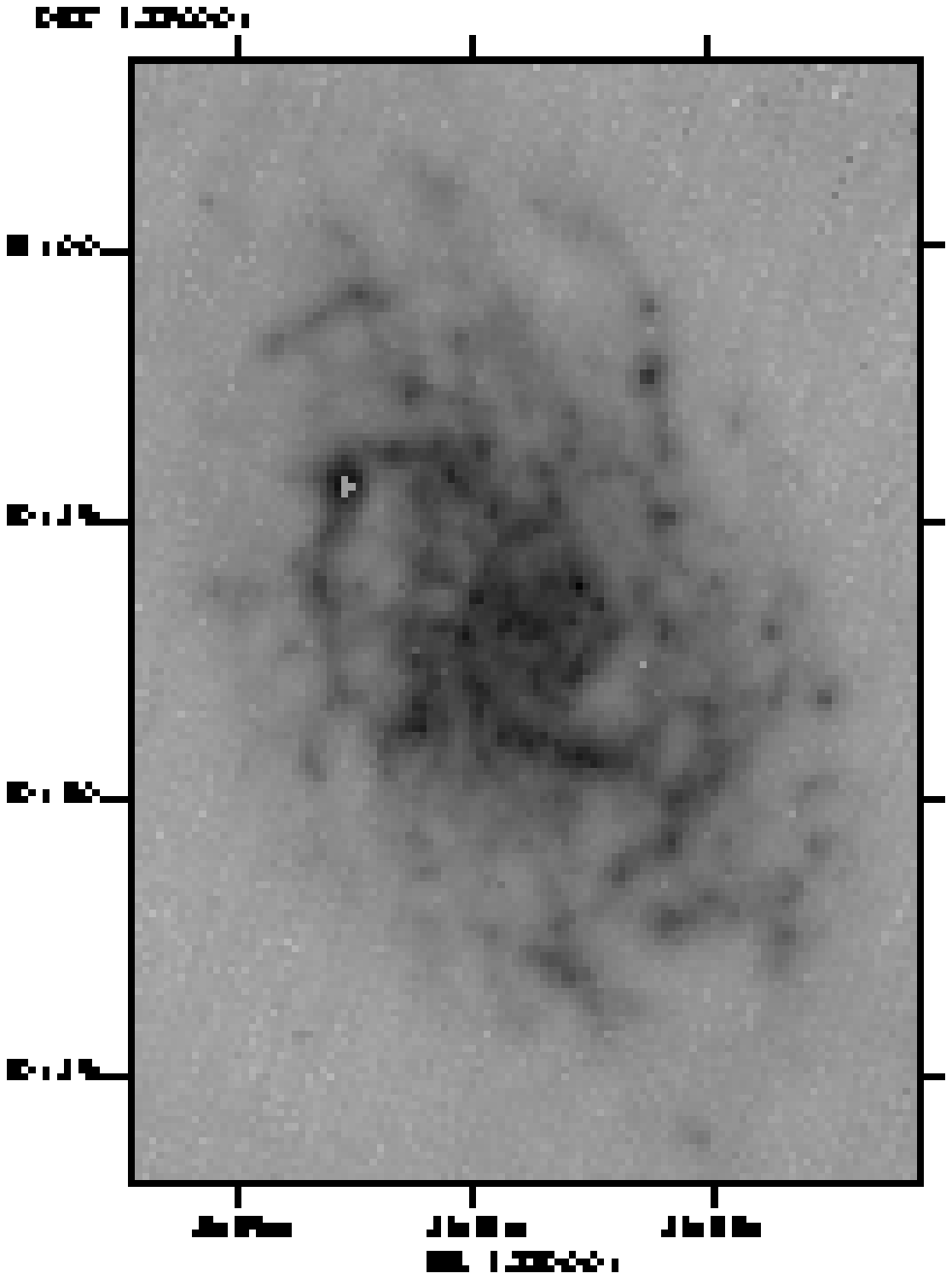}
\caption{MIPS images of \mm. The upper panels display the 24~\micron\ images: 
the full image (one single epoch) is on the left. The inner region enclosed by 
the continuous line will be further investigated in the present paper, and is shown in the upper right panel after overlapping the data from two different epochs. 
The bottom left panel shows the 70~\micron\ channel, 
and the lower right the 160~\micron\ channel. 
The images are about $30' \times 50'$, north is up and 
east is left. }
\label{fig:MIPS}
\end{figure*}

As for IRAC, we retrieved eight MIPS datasets from the archive
(AORs 3647744, 3648000, 3648256, 3648512, 3648768, 3649024, 3649280 and 3649536).
The 24~\micron\ BCDs were created by the SSC pipeline version S16.1.0. 
The exposure times of the DCEs were 3.67 s (except for the 
first frames of each series, where the integration time per 
pixel is lower and which consequently were not used in 
the further reduction steps). 
Since the observations were acquired over two epochs,
the respective mosaics were assembled separately using Mopex. 
For consistence with the IRAC reduction, 
the pixel size was chosen to be 1\farcs20. 
The first epoch includes 4199 frames, and the second one includes 4500 frames. This latter map is displayed in Fig.~\ref{fig:MIPS}. 
The mean redundancy per sky position is 22 
(varying from 10 to 40), in each mosaic. 
This high level of redundancy removes any spurious effects such as 
cosmic-ray hits and provides more reliable measurements. 
In the present article, as we plan to focus on point sources 
(\hii\ regions, SNRs and PNe) located near the centre of the galaxy, 
we chose to concentrate, for each epoch, on the two central AORs. 
These two images were then aligned with the reference of 
the coordinates of 8 point sources and combined together 
by keeping the average of the pixel values. 
A small background gradient was still visible in the 
combined image and was removed by fitting a surface to selected 
regions (IRAF task: imsurfit). The zone defined by the superposition of the two central AORs 
of each epoch is marked on the 24~\micron\ image (see Fig.~\ref{fig:MIPS}). 
The resolution measured on the final image is $6''$.

The 70 and 160~\micron\ MIPS observations were acquired with
the same AORs as the 24~\micron\ observations. 
The version S16.1.0 of the SSC pipeline was also used to create the BCDs. 
The integration times of the DCEs is 4.19 s, both for the 
70 and 160~\micron\ observations. We used Mopex to create the 
final mosaics (again the pixel sizes were chosen to be $1\farcs2$), 
including 8750 individual frames at each of the two wavelength channels. 
The mean redundancy is about 20 (from 8 to 42) and 6 
(from 2 to 10) at 70 and 160~\micron, respectively. 
The final images show some striping along the 
scan direction that is a residual instrumental artifacts due 
to the time-dependent responsivity of the Ge:Ga detectors 
\citep{2004ApJS..154..253H, 2004ApJS..154..259H}. 
The final resolution is about $16''$ and $40''$ for 
the 70 and 160~\micron\ images, respectively. 
The final MIPS images (24, 70, and 160~\micron) are 
shown in Fig.~\ref{fig:MIPS}.

\subsection{H$\alpha$ image} \label{sub:Ha}

We also include in our analysis an optical narrow-band H$\alpha$ image, 
a standard star formation indicator. 
The \mm\ H$\alpha$ emission was observed at the 
Kitt Peak National Observatory (NOAO). 
The dimensions of the CCD are $2048 \times 2048$ 
with pixels of $2\farcs03$ and	
a total field of view of about $70' \times 70'$. 
More details of the observations and reduction process can be found  
in \citet{2000ApJ...541..597H}; 
the estimated [N {\sc ii}] contamination accounts for 
at most 5\% of the measured flux. 
The total H$\alpha$ luminosity of \mm\ is about 
$7 \times 10^6 L_\odot$ and is dominated by 
\hii\ regions \citep{1997AJ....113..236D}.
The optical extinction in \mm\ is found to be globally low, \av$<0.4$~mag.
Towards \hii\ regions \av$\sim 1$~mag on average even though it can 
reach values as high as 2~mag towards high surface brightness cores
\citep[e.g.][]{1980ApL....21....1I,1997AJ....113..236D}. 
The moderate values of extinction imply that H$\alpha$ can be used 
as an SF tracer globally, but towards individual sources an
extinction correction can become necessary.

\section{Large scale structure of stars and dust}\label{sec:largeScale}

In the present paper, we concentrate on the inner region of \mm\ as defined by the \spitzer\ inner AORs which is outlined by the continuous line in Figs.~\ref{fig:IRAC} and \ref{fig:MIPS} for the IRAC and MIPS data, respectively. 

In order to help the interpretation of \spitzer\ data, 
we first review briefly which are the main emission 
phenomena that dominate each IRAC and MIPS waveband.

\subsection{Origin of the emission detected by \spitzer}

The stars (mainly Red Giant Branch and/or Asymptotic Giant Branch, according to the age
of the population) account for most of the 
emission in the 3.6~\micron\ IRAC band. 
A mix of stars and very hot dust continuum contribute to the emission 
in the 4.5~\micron\ band. 
Nevertheless, in discrete sources with negligible continuum (see Sect. \ref{sec:knownSources}),
the emission seen in both these bands can have a 
strong component of ionic and molecular lines
\citep[e.g.,][]{2006AJ....131.1479R,2006astro.ph..7598W}.
In the 5.8~\micron\ IRAC band, the stellar contribution 
decreases and the dust continuum component gets stronger.
The longest-wavelength 8.0~\micron\ IRAC band still traces hot dust
continuum but in most cases, the polycyclic aromatic hydrocarbons (PAH) 
dominates.
If there are no PAHs and low continuum, however, 
in discrete sources there can be a
sizable contribution from two H$_2$ lines and the
[Ar III] at 8.9\,\micron\ \citep[e.g.,][]{2004ApJS..154..296H}. 

The PAH emission features (mainly concentrated at 3.3, 
6.2, 7.7, 8.6, and 11.3~\micron) are the 
optically active vibrational modes of PAH molecules 
\citep{1984A&A...137L...5L,1989ApJS...71..733A}, 
resulting from internal conversions of energy following 
absorption of an optical or UV photon. 
The IRAC bands 3.6, 5.8, and 8.0~\micron\ contain 
PAH emission features at 3.3, 6.2, and 7.7 \& 8.6~\micron, 
respectively \citep{2003ARA&A..41..241D, 2004ApJS..154..322C}. 
The 4.5~\micron\ channel is the only one 
with no PAH features, although 
it contains the bright hydrogen recombination line 
Br$\alpha$ (4.05~\micron).

We investigate the far-infrared (FIR) emission with the
MIPS images. 
The 24~\micron\  emission is contributed partly by
thermal emission of classic (large) dust grains and partly by 
very small grains (VSGs). Indeed, VSGs were introduced by 
\citet{1990A&A...237..215D} as the carrier of the emission 
in the 25~\micron\ IRAS band. 
The 70~\micron\ emission traces cooler dust, 
and the 160~\micron\ one shows the diffuse cold dust component.

\subsection{Structure of infrared emission in \mm}

Figure \ref{fig:IRAC} shows that
the stellar component is smoothly distributed everywhere in the 
disk of \mm. 
The warm and hot dust traced by the 5.8~\micron\ 
channel smoothly follow the morphological patterns (spiral 
arms and flocculent structures) of \mm\ with a 
higher concentration in the inner part of the galaxy. 
Compared with the other components, the 
8.0~\micron\ contribution shows most clumpy distribution; 
evidently
PAHs are preferentially located along the spiral arms and flocculent 
structures, rather than being uniformly distributed in the ISM. 
The very centre of the galaxy gathers all the 
three contributions: stars, dust and PAHs.

\subsection{The colour images}

To help investigate the large scale structure of \mm\ as seen by
\spitzer\ we made two sets of colour images: one degraded to the worst 
resolution of the IRAC channels, 
and a second set degraded to the resolution of the MIPS 24~\micron\ image.
We approximated the PRF with a Gaussian and convolved both sets to
the resolution of the worst image in the set (8~\micron\ and 24~\micron,
respectively).

\begin{figure*}
\includegraphics[width=\columnwidth]{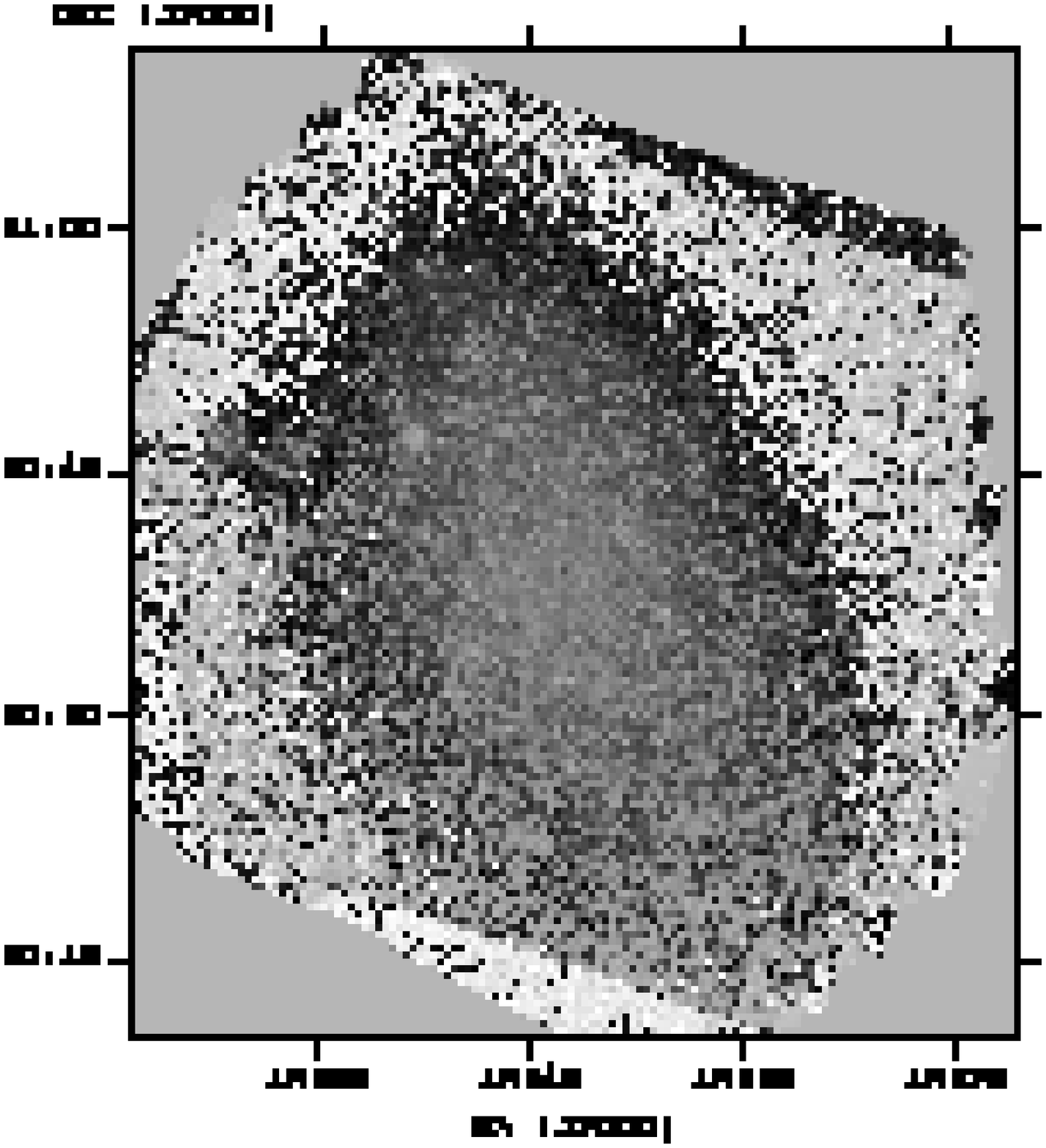}
\includegraphics[width=\columnwidth]{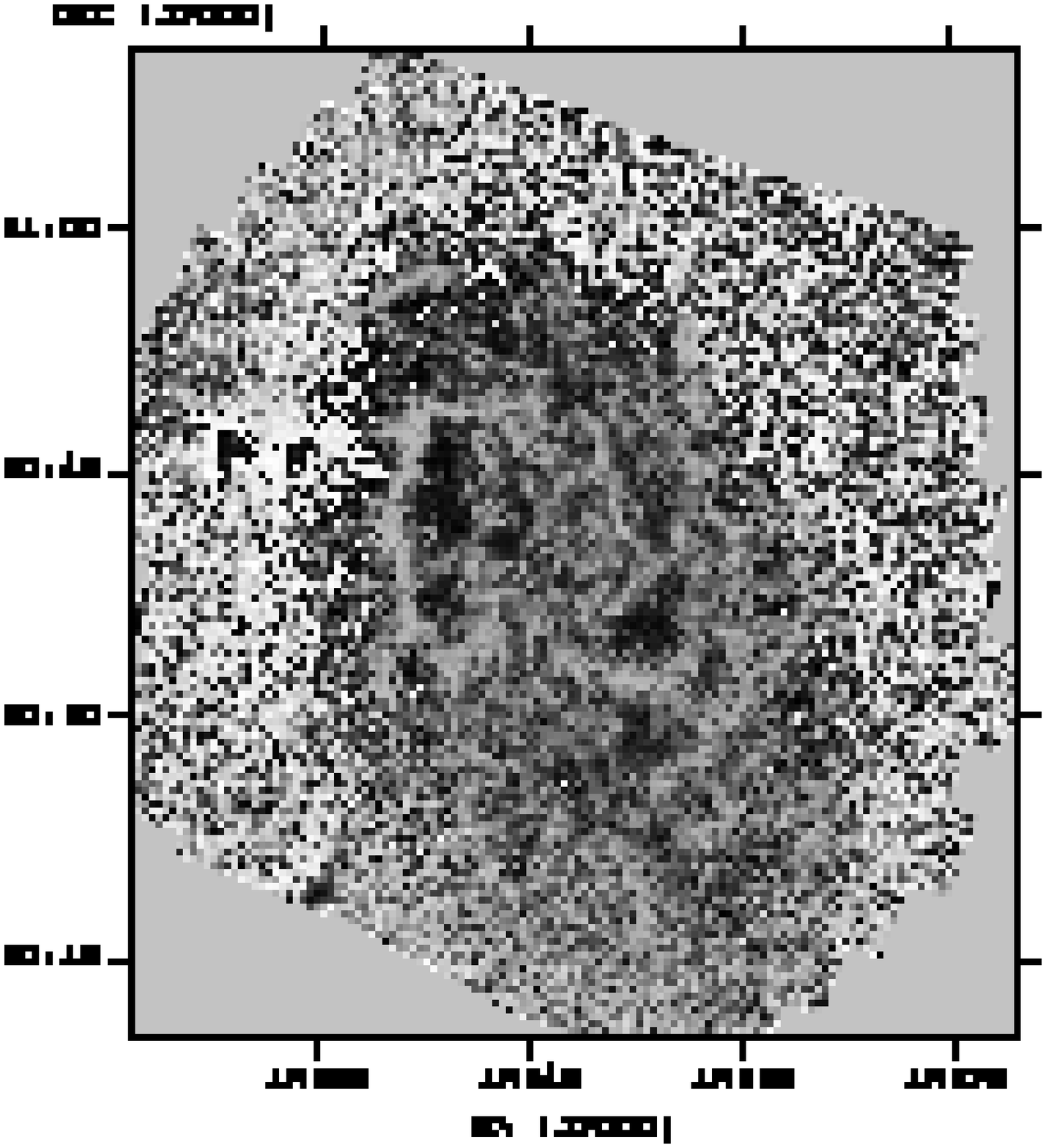}
\caption{Colour images representing the ratio of the 3.6 over 4.5~\micron\ 
images (left) and 4.5 over 8.0~\micron\ (right).
These images have a common resolution roughly equivalent to that in the 8.0~\micron\
image.
White corresponds to low colour ratios, and black to high ones.
}
\label{fig:col}
\end{figure*}

Fig.~\ref{fig:col} shows the 3.6/4.5 ratio (left panel) and 4.5/8.0 ratio 
(right) of \mm.
The 3.6/4.5 shows a remarkably constant ratio ($\sim$1.7) over almost the entire galaxy,
confirming the statement by \citet{2004ApJS..154..229P}, 
who find that for each galaxy, the [3.6]-[4.5] colour
is nearly constant with radius and consistent with 
stellar photospheric emission. 
The 5.8/8.0 ratio (not shown in Fig.~\ref{fig:col}) is also 
approximately constant ($\sim$0.5), similarly to other galaxies
such as NGC\,300 by \citet{2004ApJS..154..253H},
even though it might not be a general feature.
In both the 5.8 and 8.0~\micron\ bands there is an increase of 
the diffuse emission relative to the 3.6 and 
4.5~\micron\ bands, as in our Galactic plane.
The 8.0/24 colour (not shown) also does not show any evident radial gradient or
prominent structure and is practically constant across the disk.
Similarly to the 5.8/8.0 ratio, this result points out the overall 
constancy in the ratio of PAHs and warm dust emission in this galaxy.

On the other hand, the 4.5/8.0 images (right panel of Fig.~\ref{fig:col}) and
the 3.6/5.8 ratio (not shown) both retain the structural
pattern of \mm, tracing the inner features, the two main spiral 
arms and the filamentary structure of the disk.
In the first case, this is probably due to the predominant PAH
features in the spiral arms and filaments;
in the second, the ISM as reflected in the 5.8~\micron\ band
is emerging relative to the stars at 3.6~\micron.
The 8.0~\micron\ emission along the spiral arms and along filaments surrounding
the dimmer star forming regions at large galactocentric distances
become prominent.

\begin{figure}
\centering
\includegraphics[width=\columnwidth]{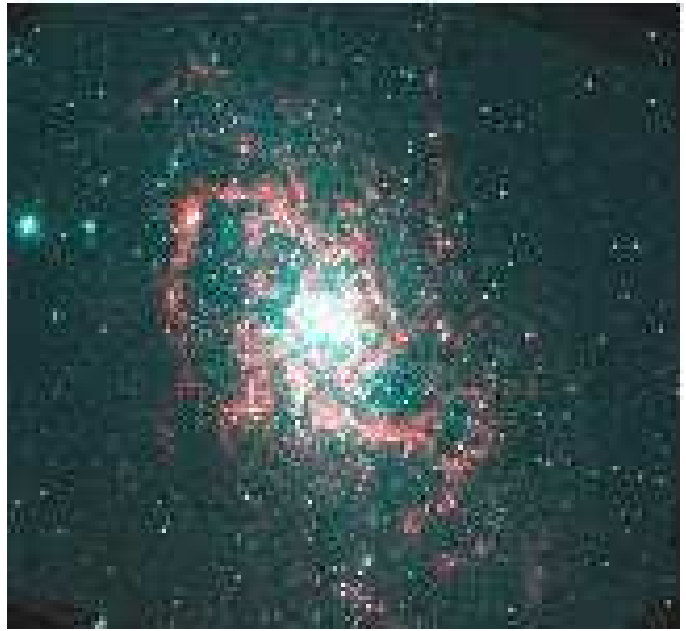}
\caption{\mm\ three colour image, composite of the central region 
of the complete IRAC data. Blue is a weighted mean of 3.6 and 4.5~\micron,
representing the evolved stellar population. 
Green depicts the 5.8~\micron\ light after removal 
of stellar emission, and red 
traces the 8.0~\micron\ 
channel also corrected for stellar emission 
[{\it See the electronic edition of the Journal for a 
colour version of this figure.}].}
\label{fig:m33_composite}
\end{figure}

We also constructed an IRAC three colour image of \mm\ by
subtracting the stellar component in the shorter-wavelength images
from the 5.8 and 8.0~\micron\ ones.
To remove the stellar contribution from the 5.8 and 
8.0~\micron\ images, we used the recipe by 
\citet{2004ApJS..154..253H,2004ApJS..154..229P,
2005ApJ...633..871C,2006astro.ph..5605P}. 
The 3.6 and 4.5~\micron\ images 
were combined assuming colours appropriate for an M0 III 
star ([3.6]-[4.5]=-0.15 in Vega mag).
The final combination is shown in 
Figure~\ref{fig:m33_composite} and displays 
the stars in blue, 
the star-subtracted 5.8~\micron\ in green and the 
star-subtracted 8.0~\micron\ in red. 
The dominance of the 8~\micron\ emission in the spiral
arms is clearly revealed by Fig.~\ref{fig:m33_composite}.
The underlying stellar disk is also evident;
the excess of stellar light in the centre has been pointed out by
previous near-infrared ground observations \citep{1994ApJ...434..536R},
and is related to the possible presence of a bulge or bar component.

\subsection{Distribution of cool dust}

Over the last fifteen years, many studies have been 
devoted to the FIR emission of \mm\ using IRAS and ISO. 
\citet{1990ApJ...358..418R} observed \mm\ 
with IRAS at 12, 25, 60, 100~\micron\ and found a 
striking correlation between the spatial IR structure and 
the \hii\ regions. 
They estimated that 75\% of the integrated IR 
emission arises from stars younger than a few times 10$^8$ yr. 
The close correspondence between the H$\alpha$ 
image and the FIR (60 and 100~\micron) 
morphology was also noted by \citet{1997AJ....113..236D}. 
\citet{2003A&A...407..137H} studied \mm\ with 
ISOPHOT at 60, 100 and 170~\micron, and concluded
that the FIR emission is composed of a 
superposition of two components: i) a 
warm one, found in spiral arms and SF regions, 
heated by UV radiation from OB stars, contributed by dust 
at about 45~K; and ii) a cold one around 16~K which is smoothly distributed 
over the disk and heated by diffuse interstellar radiation.

More recently, \citet{2004ApJS..154..259H} used MIPS observations 
of \mm\ to investigate the origin of the FIR emission,
and found that the emission in the 24 and 70~\micron\ 
MIPS images follows closely the structure of the ionised gas, 
suggesting heating by hot, ionising stars. 
The overlays of Fig.~\ref{fig:m33_halpha-c24} (see Sect.~\ref{sec:compHalpha}) show 
the link between 24 and H$\alpha$, i.e. between the hot
dust emission and the formation sites of massive stars 
\citep[see also][]{2007A&A...466..509T}. 
This link is also evident by comparing the H$\alpha$ and 
70~\micron\ emission maps.
Moreover,
comparing thermal and non-thermal radio images of \mm\ with MIPS
suggests that the 24 and 70~\micron\ emission is more
closely associated with the thermal radio component than with
the non-thermal one \citep{2007A&A...466..509T}.
Both the 24 and the 70~\micron\ emission originates in fact
from warm dust associated with massive star formation.
We discuss in more detail the connection
between 24~\micron\ discrete sources and 
H$\alpha$ knots 
below.

A more diffuse component emission at 160~\micron\, is evident
when comparing it with the 24~\micron\ map. Cold dust heated 
by a diffuse interstellar radiation field clearly contributes
to the longer wavelength MIPS band.

\subsection{Comparison with H$\alpha$} \label{sec:compHalpha}

\begin{figure}
\includegraphics[width=\columnwidth, angle=-90]{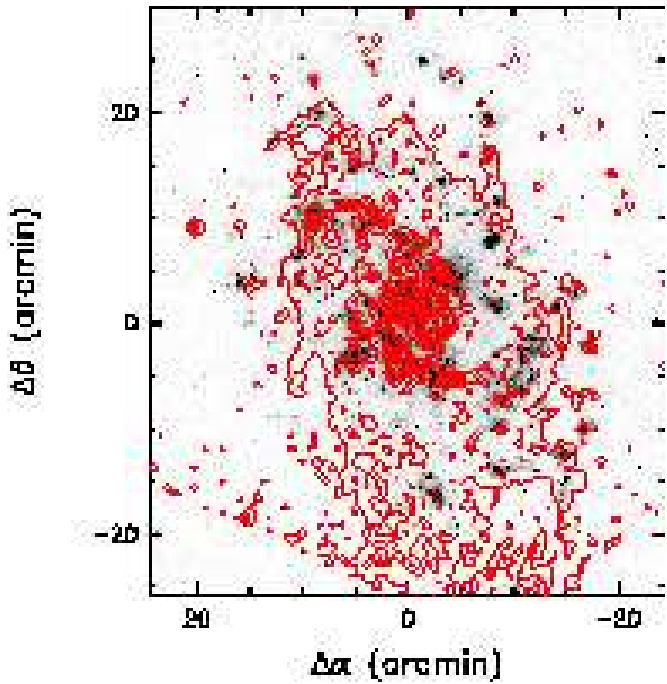}
\caption{H$\alpha$ image of \mm\ with 
the over-plot of 8.0~\micron\ contours.}
\label{fig:m33_halpha-c80}
\end{figure}

\begin{figure}
\includegraphics[width=\columnwidth, angle=-90]{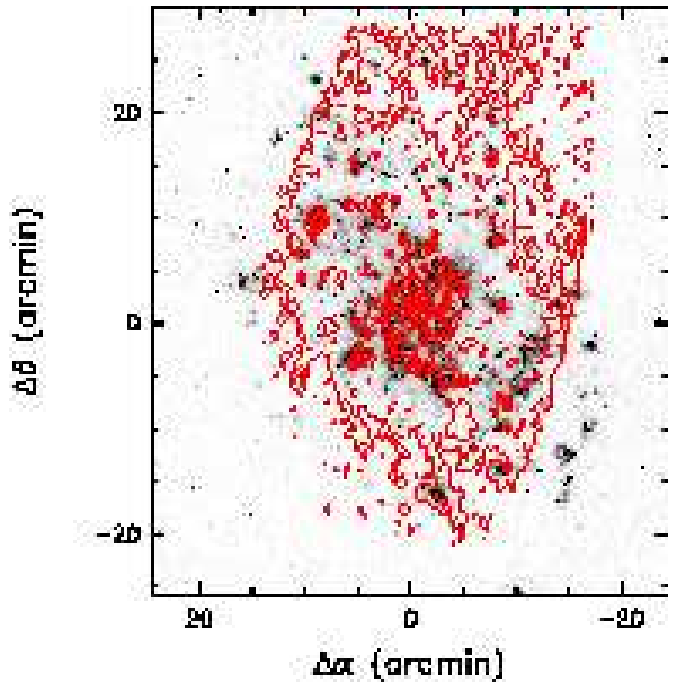}
\caption{H$\alpha$ image of \mm\ with the 
over-plot of 24~\micron\ contours.}
\label{fig:m33_halpha-c24}
\end{figure}

In Fig.~\ref{fig:m33_halpha-c80} and Fig.~\ref{fig:m33_halpha-c24}, 
we show the 8.0 and 24~\micron\ contours on the H$\alpha$ image. 
Other than the two brightest \hii\ regions (NGC\,604 and IC\,133), 
the highest levels of 24~\micron\ 
emission are concentrated in the inner part of the galaxy, and 
the 24~\micron\ emission mainly follows the spiral arms. 
Notably, not all the H$\alpha$ bright spots have 
bright 24~\micron\ counterparts. Moreover, the 
centre of the 24~\micron\ emitting sites can be shifted with respect 
to the centre of the H$\alpha$ emission sites. 
This might indicate that some of the optical emission from 
the \hii\ regions is obscured by dust or may 
be due to different intrinsic structures related to the evolution of 
\hii\ regions. Of particular interest are the 24~\micron\ 
emission spots not associated with H$\alpha$ emission, 
which could trace deeply embedded \hii\ regions, 
or a different population of objects. 
In Fig.~\ref{fig:m33_halpha-c80}, we can 
see the same general trends with a concentration of the 
8~\micron\ emission in the centre and along the 
spiral arms of the galaxy. 
At larger radii, 
the 8~\micron\ emission becomes more diffuse respect to
the 24~\micron\ component which matches better the bright H$\alpha$ 
knots. At 8~\micron, the southern arm 
shows a much more concentrated emission than the northern arm. A
similar trend has also been found for the molecular gas:
giant molecular clouds and diffuse molecular gas emission follow 
the pattern of the arm in the south
as traced by H$\alpha$ much more closely than in the north
\citep{2003ApJS..149..343E,2004ApJ...602..723H}.

\section{The IR emission of nebulae in \mm} \label{sec:knownSources}

We used six catalogues of three classes of known-type sources in \mm: 
three catalogues of \hii\ regions selected at different wavelengths, 
two catalogues of SNRs, and one catalogue of PNe:

\begin{enumerate}
\item IR \hii\ regions: 28 sources \hfill 
\citet{2002ApJ...566..880G,2002ApJ...568..679W}
\item Radio \hii\ regions: 40 sources \hfill \citet{1999ApJS..120..247G}
\item Optical \hii\ regions: 78 sources \hfill \citet{2007arXiv0705.3116M}
\item Optically selected SNRs: 98 sources \hfill \citet{1998ApJS..117...89G}
\item Radio selected SNRs: 30 sources \hfill 
\citet{1995ApJ...445..173D,1999ApJS..120..247G}
\item PNe: 152 sources \hfill \citet{2004ApJ...614..167C}
\end{enumerate}

\noindent
The properties of these will be compared to our catalogue
of 24~\micron -selected sources in Sect.~\ref{sec:S24}.

We have selected our sample of \hii\ regions from 
three different catalogues, 
which include  \hii\ regions of different morphologies, 
locations within the galaxy, and brightness.
Our IR sample was obtained by merging the sources
observed by \citet{2002ApJ...568..679W}
and \citet{2002ApJ...566..880G}.
It includes in total 28 sources
located in three chains, to the east, west, 
and south of the galaxy nucleus. 
Our radio sample comprises 40 \hii\ regions with 20-6~cm spectral
index ($\nu^\alpha$) $\alpha  >$ -0.2 as
catalogued by \citet{1999ApJS..120..247G},
and extends out to 20$'$ galactocentric radii.
A list of 78 optically selected \hii\ regions,
from narrow band images and with known optical spectra,
was kindly provided to us by L.~Magrini \citep{2007arXiv0705.3116M}. 
Such regions are in general much fainter than the IR or radio
selected regions and are located preferentially 
at large galactocentric radii.

To probe dust emission during the final stages of stellar evolution, 
we selected two samples of SNRs, one defined optically 
and one defined in the radio. 
The optical sample \citep{1998ApJS..117...89G} contains
98 SNR candidates selected
primarily on the basis of optical emission-line ratios. 
The radio sample of SNRs contains 30 (radio-bright) sources
appearing both in \citet{1995ApJ...445..173D} and \citet{1999ApJS..120..247G}.
Their selection criteria are effective at finding middle-aged remnants. 
In fact, since their SNRs have diameters in 
the range 20 - 48 pc, most of the SNRs should be 
in the Sedov-Taylor expansion stage. 
The contamination in their sample is estimated to be 
less than 1, but there are 20 SNRs embedded 
within \hii\ regions. 
Still, several remnants are associated with nearby \hii\ 
regions and some are embedded in them. 

Finally, we adopt the catalogued PNe of \citet{2004ApJ...614..167C}. 
As a result of a photometric and spectroscopic survey, 
they identified 152 PNe candidates in \mm. 
By comparison with \citet{2001A&A...367..498M}, 
the contamination by \hii\ regions or SNRs 
is estimated at $\sim$22\%.

\subsection{Photometric methodology} \label{sub:Photometry}

We have six source lists of known type, 
for a total of more than 400 objects 
distributed over the entire disk of \mm.
Our aim is to use the IRAC/MIPS-24 colours of these 
as empirical diagnostics to infer the dust properties
in various kinds of nebulae. This is done by comparing 
their IR colours with the theoretical colour
diagrams (see Sect.~\ref{sec:diagnostic}). 

To these source lists we then add the 24-$\mu$m selected  
sample and we shall infer the nature of these sources
by comparing their IR colours with those of known-type of sources
(see Section \ref{sec:S24}). 
We are also interested in the H$\alpha$ 
emission of our objects, in particular for the 24-$\mu$m selected  
sample and therefore we need to perform 
photometry on the H$\alpha$ image. 
Therefore, we must perform photometry in a total of 
six bands (4 IRAC, MIPS-24, and H$\alpha$). 
We merge the IR photometry, and consider 
only those sources which have measurable emission in all 
the IR bands examined, in order to place 
them on our diagnostic colour diagrams.

The proximity of \mm\ combined with the superb resolution 
of \spitzer\ makes photometry of individual sources in the 
galaxy challenging. Isolating sources is difficult, 
because of the extremely crowded fields, especially the inner 
disk and circumnuclear regions. 
Moreover, we expect that many of the objects 
(e.g., \hii\ regions and SNRs) will be extended, 
so that we cannot accurately perform point-source 
fitting and subtraction with point-response functions. 
In addition, many classes of sources are not 
prodigious emitters in the mid-infrared (MIR), which further 
complicates the automatization of the photometric procedures. 
Hence, we investigated many photometric schemes before 
finally arriving at a reliable solution.

Following \citet{2005ApJ...633..871C}, 
we first attempted aperture photometry of the sources in our 
lists. The initial centres were taken from the 
source lists described in Sect. \ref{sec:knownSources}, 
and the centre of the virtual aperture was then 
allowed to vary up to a certain distance from the initial position. 
While this is a viable solution in other situations, 
it turned out not to be feasible in \mm. 
First, the definition of the fiducial aperture should 
depend on source extent, but this differs among 
the source classes, so it was not clear 
how to define the limiting aperture. 
This was an especially vexing problem for the wavelengths in 
which a given source does not have measurable emission, 
since the automatic centring routines did not converge. 
Second, the determination of the background was extremely 
problematic because of the variable degree of crowding 
and the variable degree of diffuse emission. 
It was also unclear how to define either ``regional'' 
or ``global'' backgrounds in the case of \mm. 
Thus, we were forced to find another solution.

We then attempted elliptical cross-section Gaussian 
fitting of sources \citep[e.g.,][]{1995AJ....110.2739C}. 
The fitting box was fixed to a series of dimensions (from 25 to 10 pixels), 
and the centre, the orientation, the width, 
and a constant background level were left free to vary. 
As before, the initial centres were taken from the source lists. 
The larger fitting boxes allowed the centre to wander too 
much from the initial position. Hence, the source 
identified by the fitting routine frequently did not correspond 
to the source in the list, although reasonably accurate 
backgrounds were obtained when checked against the 
source surroundings. 
The smaller fitting boxes better constrained the centre, 
but in crowded fields gave a background which was much too high. 
This resulted in gross underestimations of the total flux, 
especially when the source was rather faint 
and comparable to the estimated background. 
An additional problem encountered with Gaussian fitting was 
its unsuitability for multiple sources. Many of 
the sources in our lists are in truth ``clusters'' 
of smaller sources when examined with the \spitzer\ 
spatial resolution and at wavelengths different from those in which 
the sources were defined. 
These problems taken together led us to distrust this 
method for a large fraction of sources in our lists.

To overcome these shortcomings, we finally devised a 
``hybrid'' solution, which optimally combines the 
photometry from Gaussian fitting and that from isophotal photometry. 
This last was accomplished with SExtractor 
\citep{1996A&AS..117..393B} which is able 
to deal with very large images automatically, and handles reasonably 
accurately a wide variety of object shapes and sizes. 
The idea was to calculate the total flux 
integrated over an irregularly shaped aperture containing flux levels above 
some isophotal limit. 
The hybrid method consists of three steps: 
(1)~perform Gaussian fitting of all sources as described 
in the previous paragraph; 
(2)~perform SExtractor photometry as described below; 
(3)~merge the two sets of photometry according to an 
algorithm which decides which photometric measurement to prefer, 
in order to have one measurement per source.

For (1), we adopted a fitting box of 10 pixels; 
this was the best compromise between centre wandering 
and background level, as the photometry 
was typically centred on a position relatively close to the 
initial one, although with a potentially spuriously low 
flux because of the background determination.

In practice, for (2), we used the ASSOCIATIVE feature of 
SExtractor, and input our source lists with initial positions. 
The initial SExtractor photometry consisted in those sources 
extracted at a 10$\sigma$ 
level above the background, at a distance within 
15 pixels ($\equiv$\,\dr) from the initial position. 
These two values resulted to be the best compromise 
between source association and source identification. 
Lower values of signal-to-noise ratio ($<10\sigma$) 
resulted in more sources being identified, but were in 
the mean farther away.
The background for the SExtractor photometry was defined to 
be the local median of the image in 
a 64-pixel box centred on the source.

Step (3), the merging of the Gaussian fitting and 
SExtractor photometry, was accomplished by examining the radial distance 
of each measurement from the initial position. 
When the SExtractor centre was within \dr\ of 
5 pixels (in radius), and as long as it was within a 
factor of two of the Gaussian fitting centre, it was 
preferred over the Gaussian method. 
We experimented with a few values of \dr\ 
and a few values of the relative ratio of the radial distances, 
and concluded that this gave the best results overall. 
Indeed, it appears that this hybrid approach overcomes the 
problems of underestimating source fluxes because of background 
estimates, and successfully calculates the total flux even 
when sources are in fact multiple. 
Indeed, the SExtractor photometry and the Gaussian 
fitting results are fairly well correlated except for low 
fluxes with high background where Gaussian fitting tends to 
underestimate the true flux, and high fluxes for 
extended or clustered sources.

\subsection{Detection rates}

\begin{figure}
\centering
\includegraphics[width=\columnwidth]{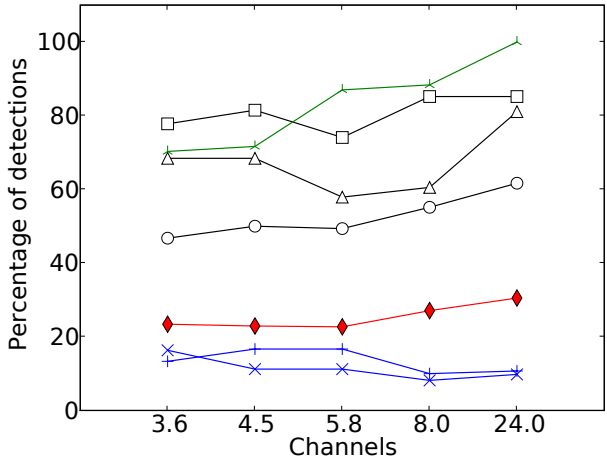}
\caption{Detection rate 
in the different wavebands. 
Open symbols represent the \hii\ regions selected in 
IR (squares), radio (triangles) and optical (circles). 
The (blue) crosses represent the optically selected SNRs and the 
(blue) plus signs the radio selected SNRs . 
The filled (red) diamonds represent the PNe. 
The (green) tripods represent our 24~\micron\ sources.}
\label{fig:fracDet}
\end{figure}

The detection statistics are illustrated in  
Fig.~\ref{fig:fracDet} and Table~\ref{tab:detec}. 
The matching radius \drmax\, for the coincidence of
sources, was determined by examining the scatter
on the diagnostic diagrams, see Sect. \ref{sec:diagnostic}.
In the end we adopted a value \drmax\,=\,3 pixels
(about 14 pc), which is sufficiently large to 
accommodate possible astrometric uncertainties.

\begin{table}
\begin{center}
\caption{Detection rate in all four IRAC and 
IRAC+MIPS-24 bands.} \label{tab:detec}
\begin{tabular}{c c c c | c c c}
\hline \hline
     & IRAC & Total & \% & IRAC+MIPS-24 & Total & \%\\
\hline
ISO  & 17 &  27 & 61\% & 15 &  27 & 54\% \\
Hra  & 18 &  38 & 45\% & 18 &  37 & 45\% \\
\hii & 27 &  77 & 35\% & 19 &  60 & 32\% \\
SNo  &  3 &  98 &  3\% &  2 &  92 &  2\% \\
SNR  &  1 &  30 &  3\% &  1 &  28 &  3\% \\
PNe  &  9 & 141 &  6\% &  4 & 118 &  3\% \\
24Ss  &306 & 515 & 59\% &306 & 515 & 59\% \\
\hline
\end{tabular}
\end{center}
\end{table}

\begin{figure}
\centering
\includegraphics[width=\columnwidth]{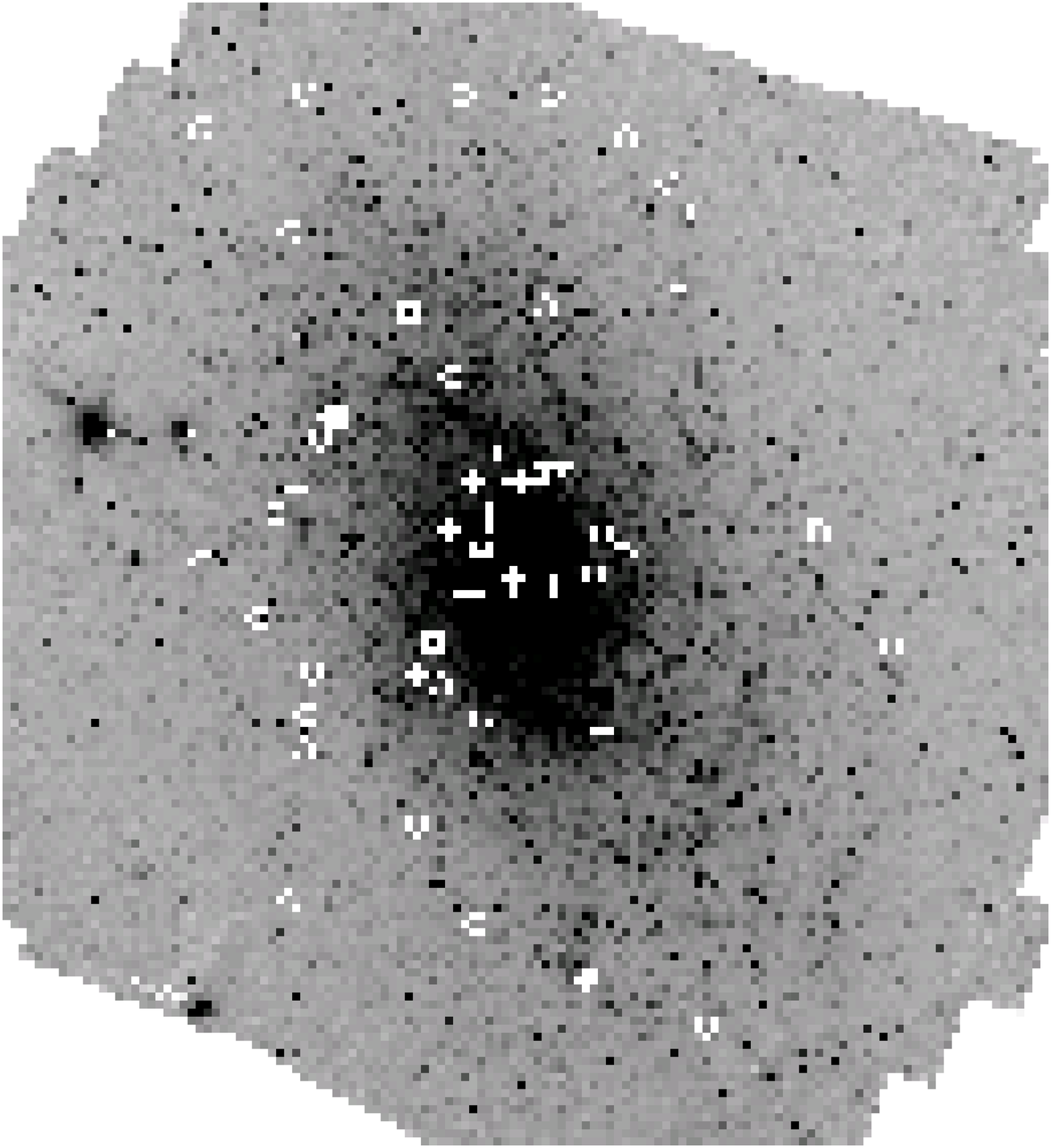}
\includegraphics[width=\columnwidth]{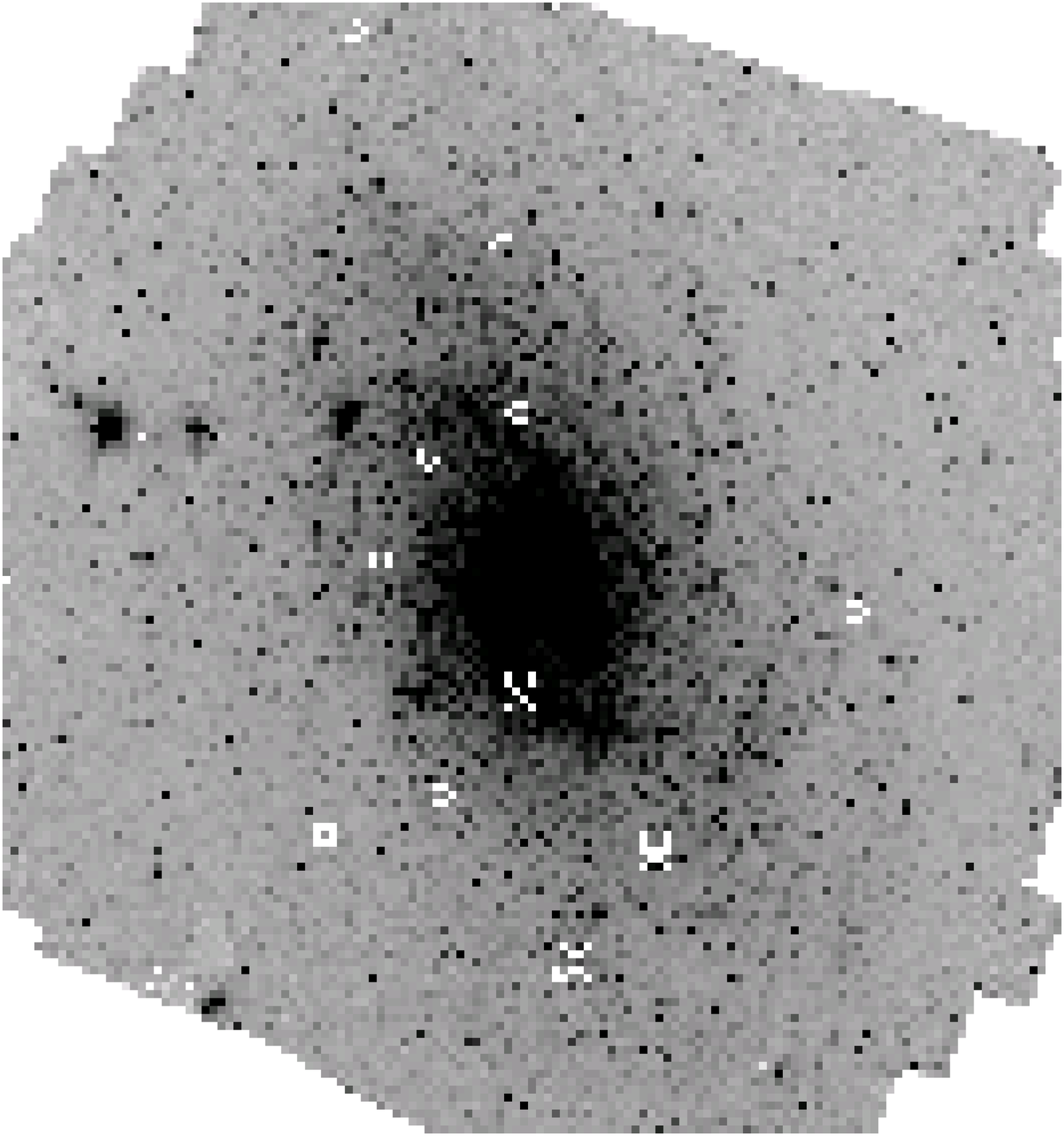}
\caption{The position of catalogued sources 
with photometry in all four IRAC bands superimposed on the 4.5~\micron\ 
IRAC image. The upper panel shows the IR selected 
(open squares), radio selected (pluses) and optically 
selected (open circles) \hii\ regions. 
The lower panel shows the locations of the PNe 
(open circles), optically selected (crosses) and radio 
selected (pluses) SNRs. The images are about 1$^\circ$ 
on a side, north is up and east to the left 
.}
\label{fig:locationKnown}
\end{figure}

The positions of catalogued sources with photometry in 
all four IRAC bands are shown in Fig.~\ref{fig:locationKnown}, 
on the IRAC 4.5~\micron\ image of \mm.  
The upper panel refers to \hii\ regions and we note that
the optically selected sample is mainly in the outer regions 
of the galaxy, while the IR and radio selected \hii\ 
regions are in the inner parts of \mm. 
The lower panel shows the locations of PNe and SNRs. 
The single detection of the radio-selected SNR coincides with 
an optically-selected one.

\subsubsection{\hii\ regions}

The detection rate of \hii\ regions 
depends on the waveband and on the catalogue. 
As expected, the brighter \hii\ regions 
(radio and IR sample) are detected at a 
higher rate 
than the faint ones (optically selected catalogue). 
The detection rate for bright regions increases even more if we
loosen the strict requirement about the centre displacement.
For large \hii\ regions  in fact
the centre of the IR emission can be displaced by 
more than 3 pixels from the nominal position, 
and hence not accepted as coincident. 
The lower resolution in the 24~\micron\ band 
makes objects rounder, and the Gaussian fitting is often
more accurate than
in the 5.8 and 8.0~\micron\ band. 
This partly accounts for the higher detection rate 
in the 24~\micron\ MIPS wavelength. 
But, there is a general tendency for the 
\hii\ regions to be less visible in the shortest 
wavelengths (3.6 and 4.5~\micron) than in the 8.0 
and 24.0~\micron\ channels. 
The detection rate at 5.8~\micron\ always shows a drop, with respect to 
the adjacent channels (see Fig.~\ref{fig:fracDet}). 

\subsubsection{Supernovae remnants}

We found a low rate of IR counterparts for the SNRs identified in 
optical and in radio. 
The percentage of detections is about 15\% for both samples.
Four SNRs were detected in at least one band: 
two were detected at all wavelengths, and other two only at 4.5~\micron.
The four detected SNRs have radii of $\sim$ 8 -- 16 pc,
while the two undetected ones have radii greater than 20
pc and may be of a lower surface brightness. 
The low number of SNR detections is consistent with previous IRAS studies 
trying to find Galactic SNR counterparts in the IR:
17\% \citep{1989ApJS...70..181A} and 18\% \citep{1992ApJS...81..715S}.
Also, using IRAC, \citet{2006AJ....131.1479R} found only 18 IR 
counterparts out of 95 SNRs in the inner Galaxy. 
Very few Galactic SNRs were detected by MSX, the Midcourse Space Experiment \citep{2006astro.ph.10569C}.
In the Large Magellanic Cloud, \citet{2006ApJ...642L.141B} found
no emission from  Type Ia SNRs and \citet{2006astro.ph..7598W} did not 
detect the two largest SNRs of their sample in any of the IRAC or MIPS wavebands.

In our samples, the detection rate is also about 30\% less
in the 8.0 and 24~\micron\ bands with respect to the
shorter IRAC wavelengths. This might be due to the destructive
effect of the blast wave on the
smaller grains; PAHs can already be razed
in slow shocks and they are never detected in dense,
shocked clumps \citep{2002ApJ...564..302R}.
The emission seen at 3.6 and 4.5~\micron\ is likely due
to ionic and molecular lines, such as Br$\alpha$ 4.1~\micron\
\citep{2006AJ....131.1479R,2006astro.ph..7598W}.
The environment of the SNRs should have a strong
influence on their IR detection rate: remnants interacting
with a denser external medium should be more easily detectable. 
If PAHs in the surrounding ISM are destroyed
by nearby supernova events \citep{2004ApJS..154..275G},
they will not emit at 8.0~\micron\ (nor
contribute to 3.6 and 5.8~\micron),
which could explain the quite low rate of detection
of a remnant in all 4 bands.

\subsubsection{Planetary nebulae}

The detection rate of PNe is about 25\%
for the three shorter IRAC wavelengths and then rises steeply to
reach more than 50\% in the 24~\micron\ band.
This is consistent with the result by
\citet{2005MNRAS.362.1199C} who detect PAH
emission in less than half of a sample of 43 Galactic PNe.
In addition, \citet{2004ApJS..154..296H} claim that the
IRAC colours of PNe are red; 
despite typically negligible continua, they are bright
in the 8.0~\micron\ band probably because of a 
sizable contribution from the two H$_2$ lines and an
[Ar III] line at 8.99~\micron.

\subsection{Spectral energy distributions}

\begin{figure}
\includegraphics[width=\columnwidth]{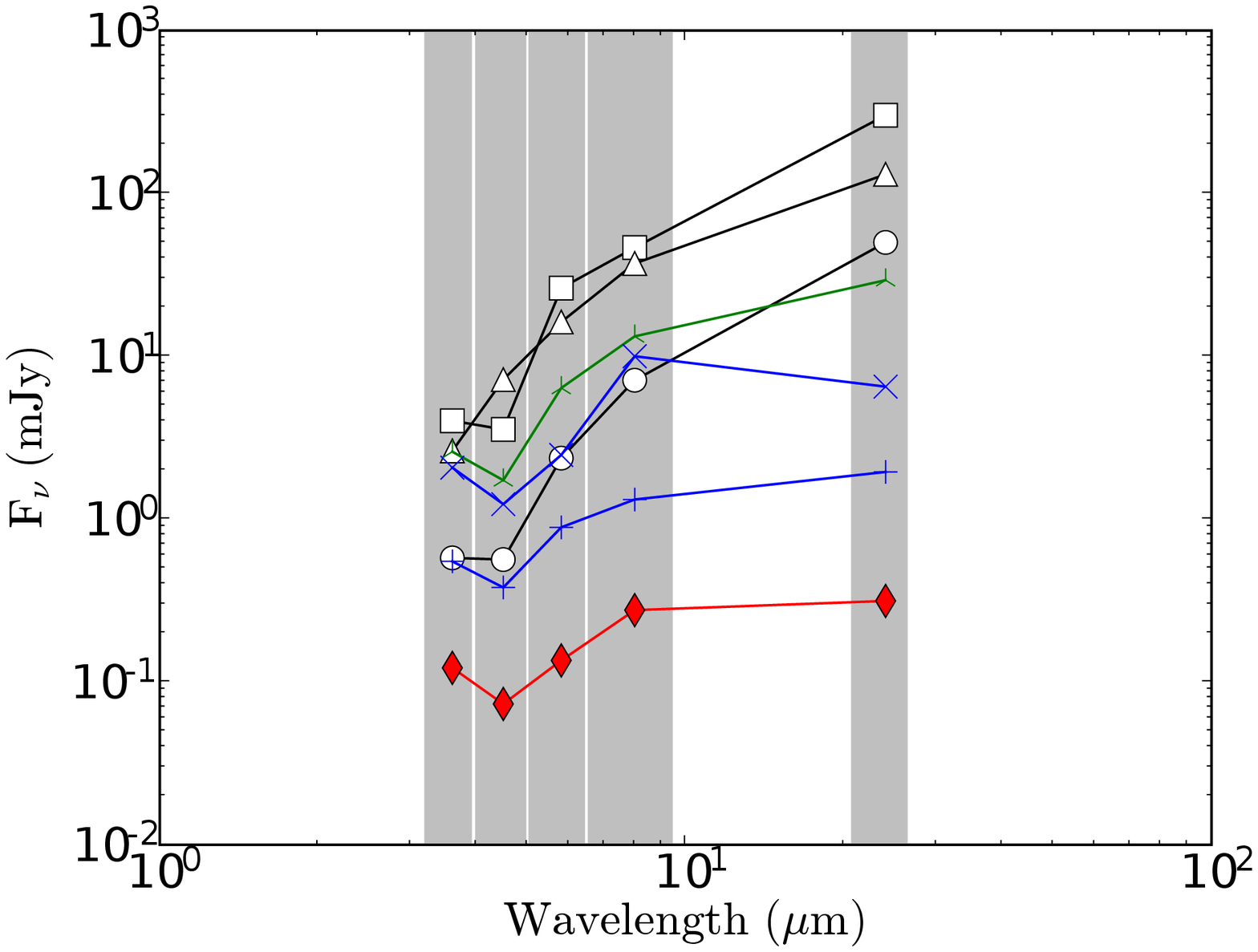}
\caption{Mean IR SEDs of the different kind of sources. 
Open symbols represent the \hii\ regions selected in 
IR (squares), radio (triangles) and optical (circles). 
The (blue) crosses represent the SNRs optically selected and the 
(blue) pluses the SNRs radio selected. The filled (red) 
diamonds represent the PNe. The (green) tripods represent 
the 24~\micron\ unknown sources. The grey vertical stripes
depict the widths of the IRAC and MIPS-24 filters.}
\label{fig:sed-mean}
\end{figure}

The average IR Spectral Energy Distribution (SED)
for our different lists of objects is shown in
Fig.~\ref{fig:sed-mean}. 
\hii\ regions, either IR or radio selected, display
the highest fluxes, about two order of magnitudes
above the faintest in the list, that is the PNe.
The \hii\ optical sample instead, largely consisting
of quite faint regions, sits at lower values,
in between the two samples of SNRs.
The steep rise between 8.0 and 24~\micron\ of all
the \hii\ regions samples, implies that \hii\
regions are invariably associated with warm dust.
SNRs and PNe show milder slopes, their SEDs being almost flat, 
or even declining as for optically selected SNRs. This last feature
reveals changes of the ISM characteristics related to SN events.

In \hii\ regions, the drop between 3.6 and 4.5~\micron\ 
can be accounted for by the photospheric contribution of young stars 
in the two shortest wavelength IRAC channels
or perhaps line emission. 
Candidate lines in the 3.6~\micron\ 
band include [Fe {\sc iii}] 3.229~\micron, 
and [Co {\sc iii}] 3.492~\micron\ \citep{2007astro.ph..2117G}, but 
line emission can be important in other bands as well
as long as there is little continuum emission.
This drop has also been detected by 
\citet{2004ApJS..154..328M,2004ApJS..154..275G,
2005ApJ...620..731J,2006astro.ph..7598W,2007astro.ph..2117G}. 
The absence of the 4.5~\micron\ drop 
in the means of radio selected \hii's 
is due primarily to a single object with a high 4.5~\micron\ flux.

\section{Diagnostic diagrams} \label{sec:diagnostic}

\subsection{Theoretical diagnostic diagrams for IRAC and IRAC/MIPS flux ratios}

To better understand the IRAC/MIPS colour diagrams of 
discrete sources, we assembled a series of models with different physical 
parameters, as described more in detail below. The IR colours 
are derived by convolving the model spectra with the IRAC and 
MIPS-24 filter response curves, as provided by the 
\spitzer\ Science Center web page. 
The colours are here computed as ratio of flux densities 
(in Jy, i.e. per unit frequency).

\subsubsection{\hii\ regions with Cloudy}

\hii\ regions are possibly the most 
prominent class of sources in our images. 
We modelled them with version 05.07 of Cloudy, 
last described by 
\citet{1998PASP..110..761F} \citep[see also][]{2003ARA&A..41..517F}. 
The central star was taken to be a log\,{\it g}=4 Kurucz model 
\citep{1979ApJS...40....1K} at T$_{\rm eff}$\,=\,37\,000\,K, 
as modified by Rubin to better account for the stellar flux in 
the model atmospheres between 41\,eV and 54\,eV. 
In all cases, we assumed a stellar ionising luminosity of 
$10^{49}$\,sec$^{-1}$, spherical geometry, constant gas density, 
a covering factor of 0.005, and a metal abundance of 50\% solar.

Grains were ``turned on'' in the Cloudy models so that heating and 
cooling of the grains and the gas are calculated 
in a self-consistent way. A standard PAH distribution is included 
in the grain population, with treatment 
of photoelectric heating and collisional processes 
\citep[see ][]{1998PASP..110..761F}. 
We did not explicitly introduce a photo-dissociation 
region (PDR) in the calculations.

We accounted for the range in typical \hii\ region 
parameters by varying the outer radius and the total 
hydrogen density so as to roughly follow the well-known 
size-density $n_e\,\sim\,D^{-1}$ relation for \hii\ 
regions \citep{1999PASP..111.1049G,2001ApJ...549..979K}. 
Four models were calculated with the following pairs in size-density 
space with density given first in cm$^{-3}$ 
followed by outer radius in pc: (1000,2); (100,10); (20,30); (10,60).

\begin{figure}
\centering
\includegraphics[width=\columnwidth]{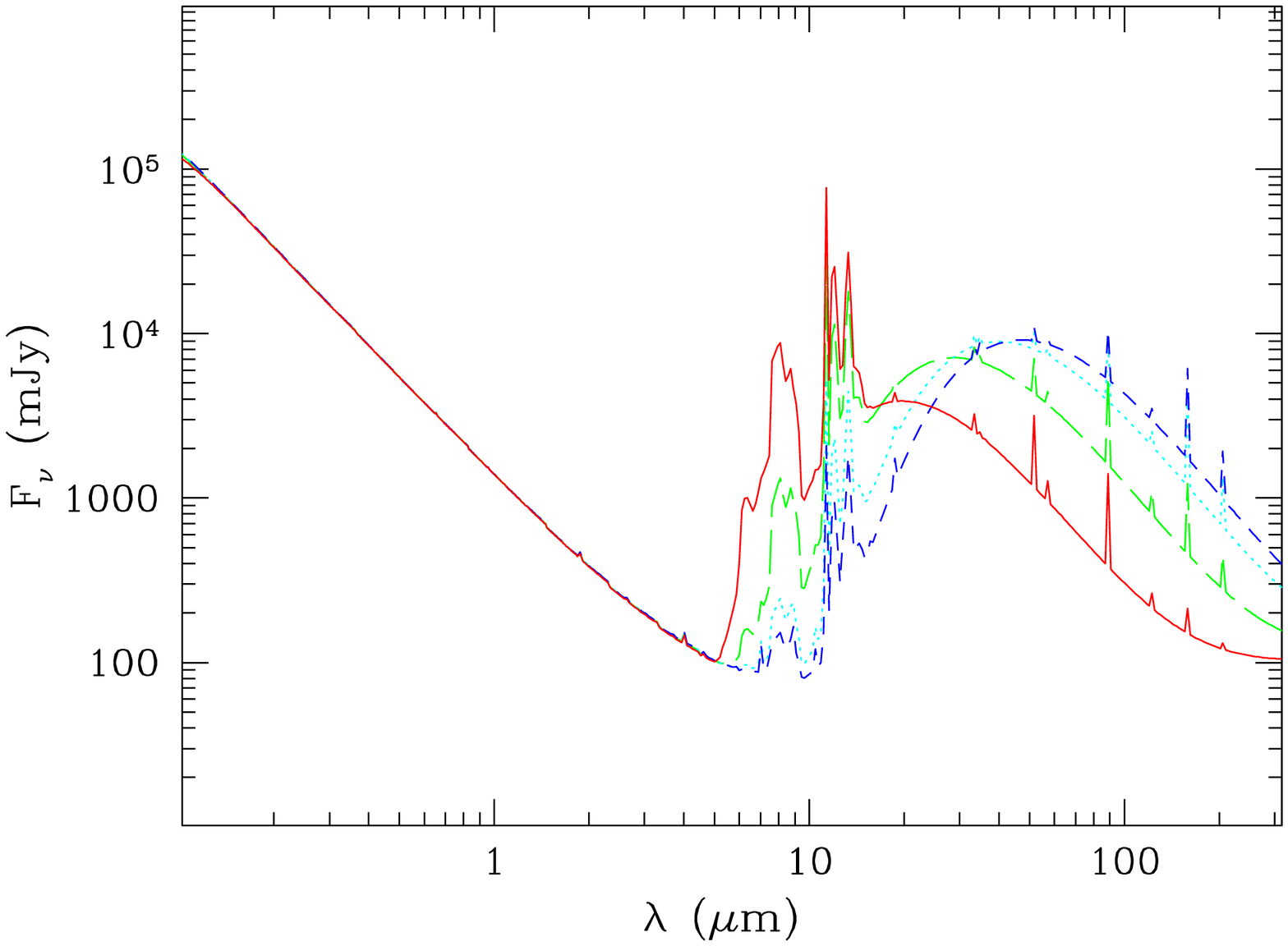}
\caption{Spectra obtained with Cloudy code, for the four models. 
The parameter pairs in size-density space with density given 
first in cm$^{-3}$ followed by outer radius in pc 
are represented by the (blue) short-dashed line for (100,10); 
the (cyan) dotted line (20,30); the (green) long-dashed 
line (10,60); and the (red) solid line (1000,2).}
\label{fig:cloudy_spec}
\end{figure}

The resulting spectra are shown in Fig.~\ref{fig:cloudy_spec}, where the 
(blue) short-dashed line shows (100,10); the (cyan) 
dotted line (20,30); the (green) long-dashed line (10,60); 
and the (red) solid line (1000,2). 
It can be seen from the figure that varying the density by 
a factor of 100 drastically changes the PAH emission 
and its dominance relative to the continuum. 
Since the PAHs in most \hii\ regions originate in the surrounding PDR 
rather than in the ionised region itself 
\citep[e.g][]{2004ApJS..154..315W,2004ApJS..154..322C,2006ApJ...639..788D}, 
the simulations may not be truly representative of \hii\ 
regions unless a PDR is absent. However, they clearly span a 
wide range in IR peak wavelength and PAH strength, 
so should be adequate for helping to interpret the 
IRAC/MIPS24 diagnostic diagrams.

\subsubsection{Embedded \hii\ regions with \dusty}

The \hii\ regions we have modelled with Cloudy are basically optically 
thin, because we have not performed multiple iterations on the 
solution. Moreover, even with iteration, optical depth 
effects in the dust itself are not taken into account in 
the Cloudy formalism. Hence, to better model compact and 
ultra-compact \hii\ regions which are expected to be embedded in 
an opaque dust cocoon, we have run a 
series of models with \dusty\ \citep{1997MNRAS.287..799I}. 
\dusty\ solves semi-analytically the radiative transfer physics in a 
dust envelope by exploiting the self-similar 
nature of the problem. The dust is located in a shell
external to the \hii\ region itself. 
The program allows for a variety of 
spherical shell configurations, and provides a realistic spread 
of dust temperatures as the 
radiation field changes with distance from the source. 
At the same time, this code suffers from some limiting 
assumptions, such as spherical symmetry, no treatment of 
stochastic processes for PAH emission, 
and the decoupling of the gas from dust. 
Nevertheless, for our purposes, it is a powerful 
tool with which to investigate the dust continuum emission from 
young dust-enshrouded stars and star clusters.

With \dusty, it is necessary to specify the temperature 
at the inner radius of the dust shell, \tin which,
for a given incident radiation field, 
determines the inner radius of the region, $R_{\rm in}$. 
The radial distribution of 
the dust is described through the ratio of the outer radius 
to the inner one, \yout\ ($\equiv\,R_{\rm out}/R_{\rm in}$), and 
the power-law exponent $p$ describing the radial fall-off $R^p$. 
In all cases, we adopted a standard ``MRN'' size 
distribution \citep{1977ApJ...217..425M}, and a 
standard dust composition \citep{1984ApJ...285...89D}. Finally, 
we must specify the radial optical depth  \av\ which, 
for a given geometry, fixes the dust mass. 
The \dusty\ models were calculated for 10 values 
of \av, ranging logarithmically from 0.1 to 100 
(specifically 0.1, 0.215, 0.464 for each decade in \av). 
We adopted single-zone models, with only one stratum in the shell.

Two temperatures were input for \tin , 300\,K 
and 600\,K, which correspond roughly to the range observed in 
extragalactic \hii\ regions \citep{2005A&A...434..849H}. 
Because of the non-negligible thickness of the dust shell, such 
relatively high temperatures at its inner rim are 
also required to explain the {\it average} 
temperatures observed in Galactic \hii\ regions 
\citep{2004MNRAS.355..899C,2007AJ....133..639G}. 
We fixed the power-law exponent $p$ to $-0.2$ as a 
compromise between uniform regions and those with a steeper fall-off. 
This exponent strongly affects the amount of cool dust in 
the shell, but does not significantly influence the MIR 
portion of the spectrum that we are examining with 
the IRAC/MIPS-24 diagnostic diagrams. 
The thickness ratio \yout\ varied 
from 10 to 10\,000; \yout\ also controls the amount of cool dust 
in the shell as thicker shells have proportionately more cool dust.

To model the exciting star cluster input to \dusty, 
we incorporated the Starburst 99 simulations 
\citep[SB99,][]{1999ApJS..123....3L}. 
Because we are interested in {\it young, dusty} star 
clusters, a single age of 3\,Myr was used. SB99 
takes into account the nebular continuum emission which can dominate 
the near-infrared (2\,$\mu$m -- 4\,$\mu$m, NIR) 
spectral range, and thus could in principle significantly influence our 
IRAC colour diagrams (see below). 
In practice, the slope of the input spectrum of a 3\,Myr 
starburst is quite similar to the star 
used for the Cloudy simulations, at least down to 0.1\,$\mu$m, 
so we expect few systematic differences in the dusty 
star clusters modelled with \dusty\ relative to the Cloudy models. 
The metallicity of the input spectrum was taken 
to be 20\% solar; this was to better account for possible 
lower-metallicity embedded sources in the outer disk. 
However, as illustrated below, this choice of metallicity does 
not significantly affect the resulting IRAC/MIPS-24 colours.

\begin{figure}
\centering
\includegraphics[width=0.49\columnwidth]{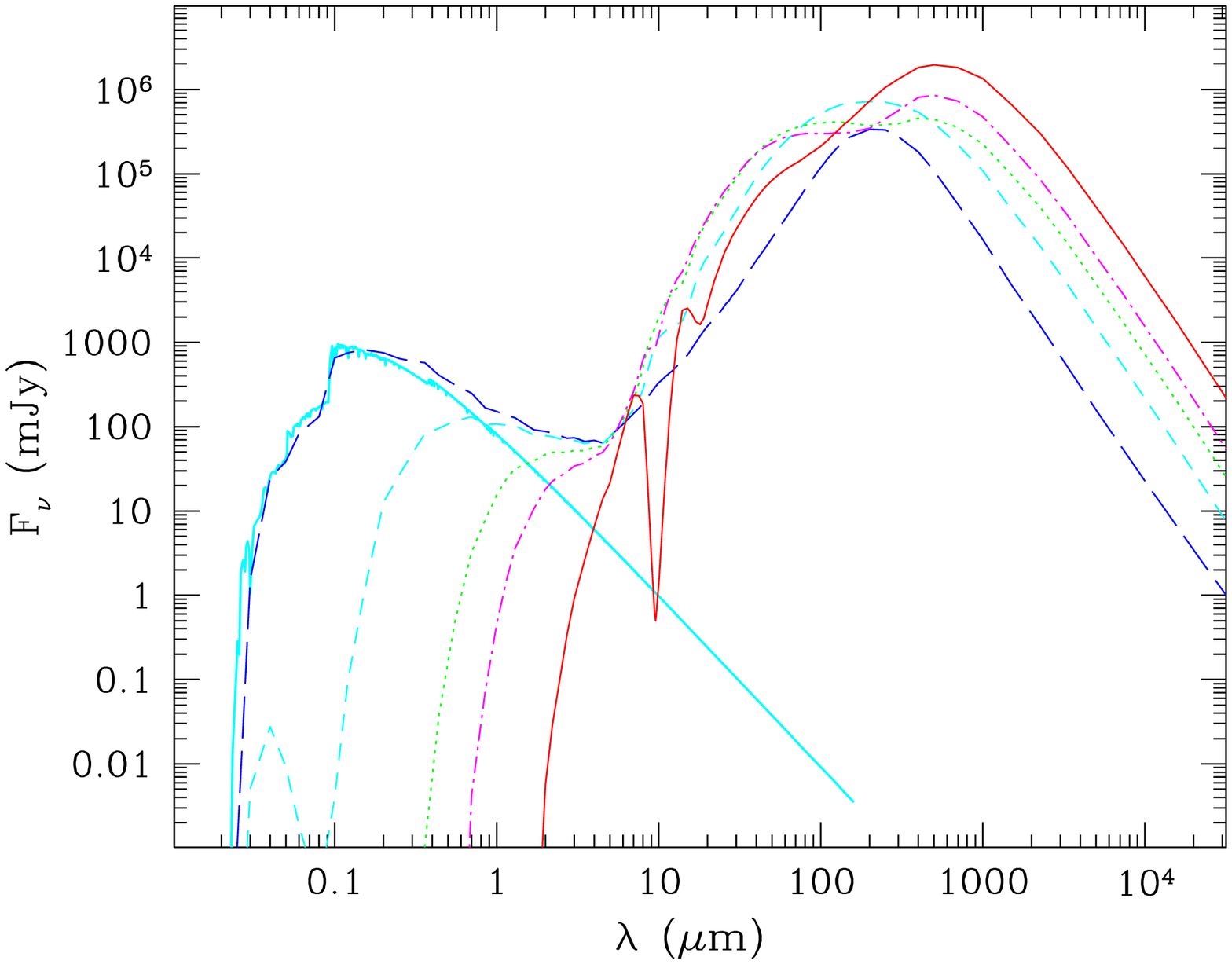}
\includegraphics[width=0.49\columnwidth]{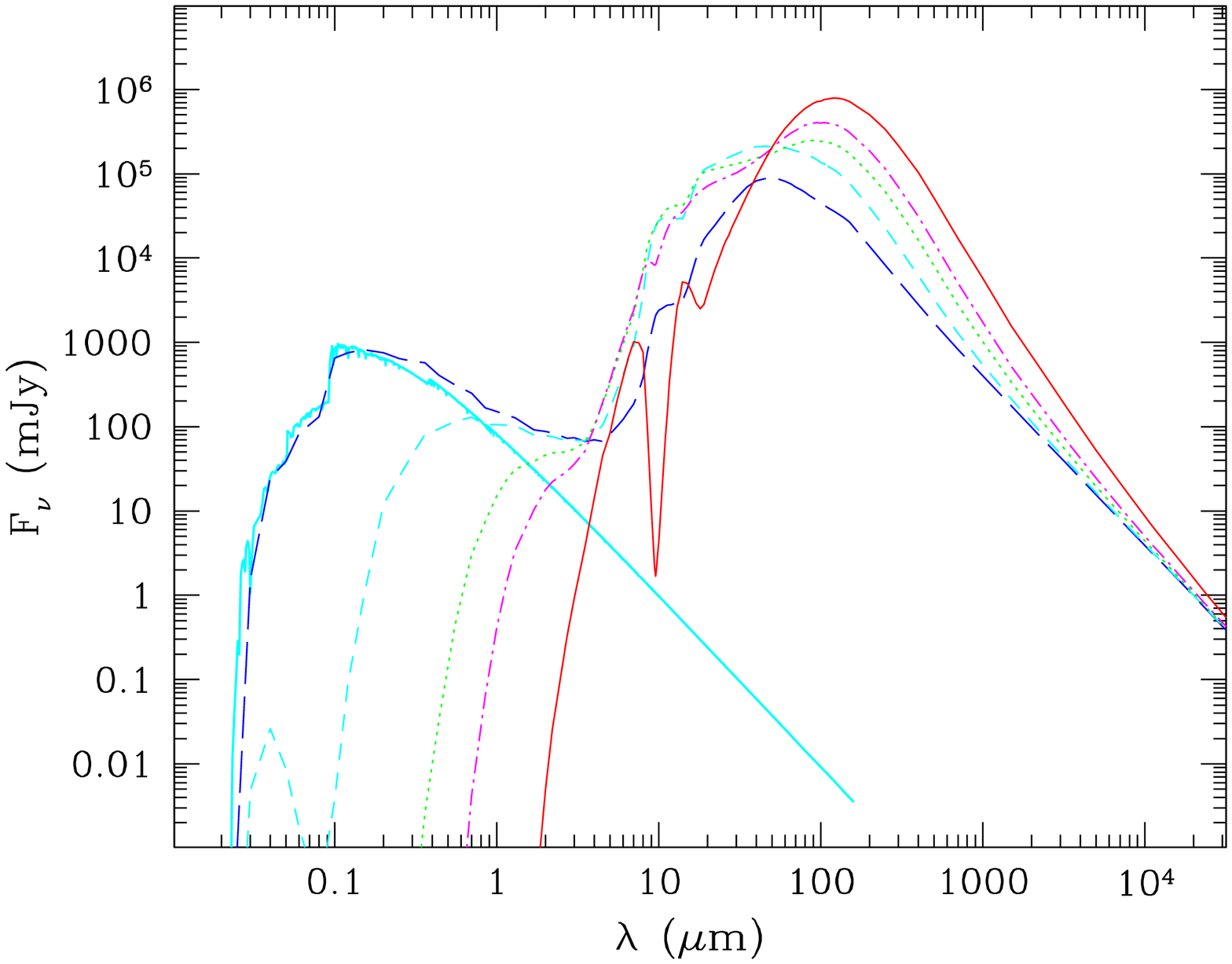}
\includegraphics[width=0.49\columnwidth]{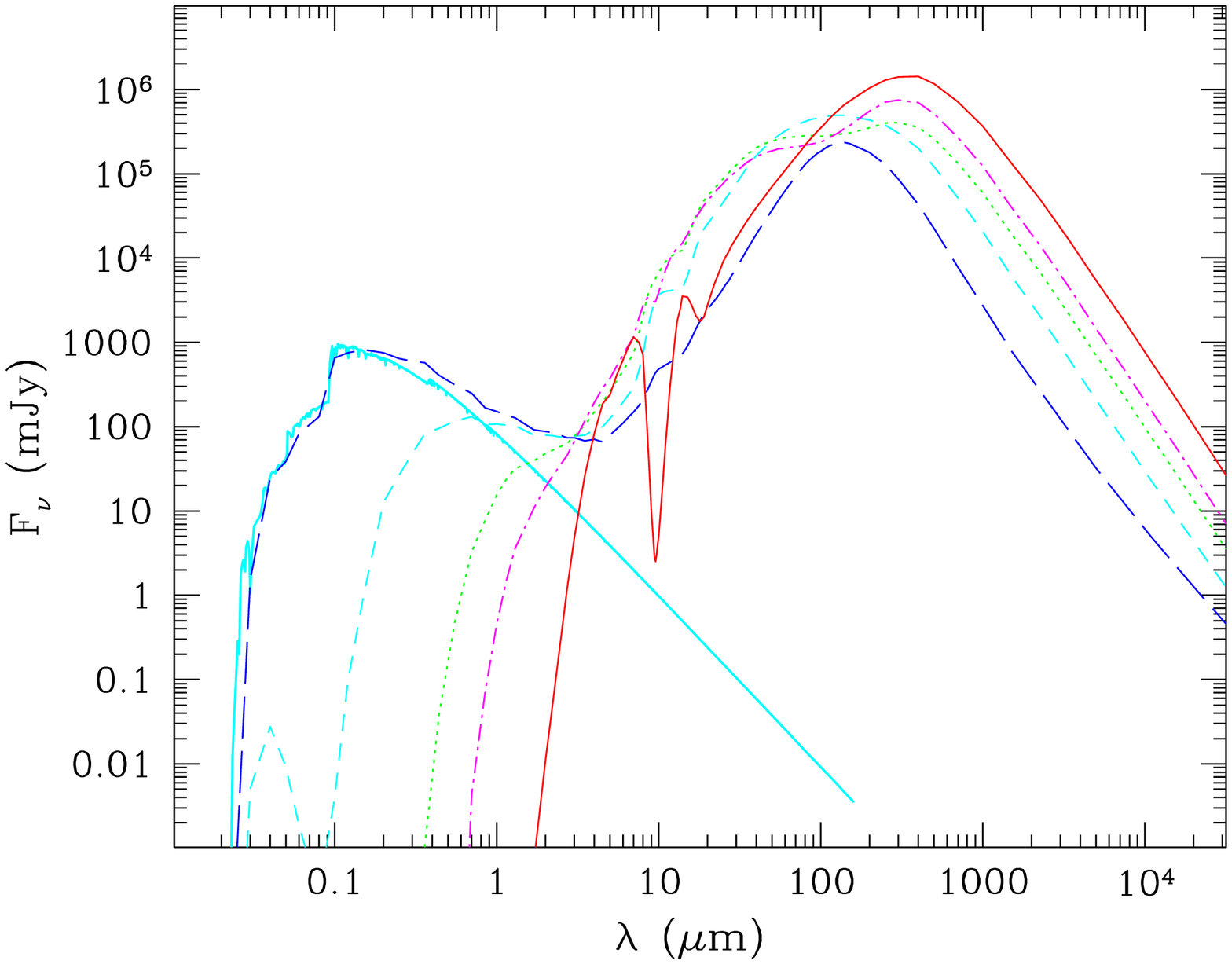}
\includegraphics[width=0.49\columnwidth]{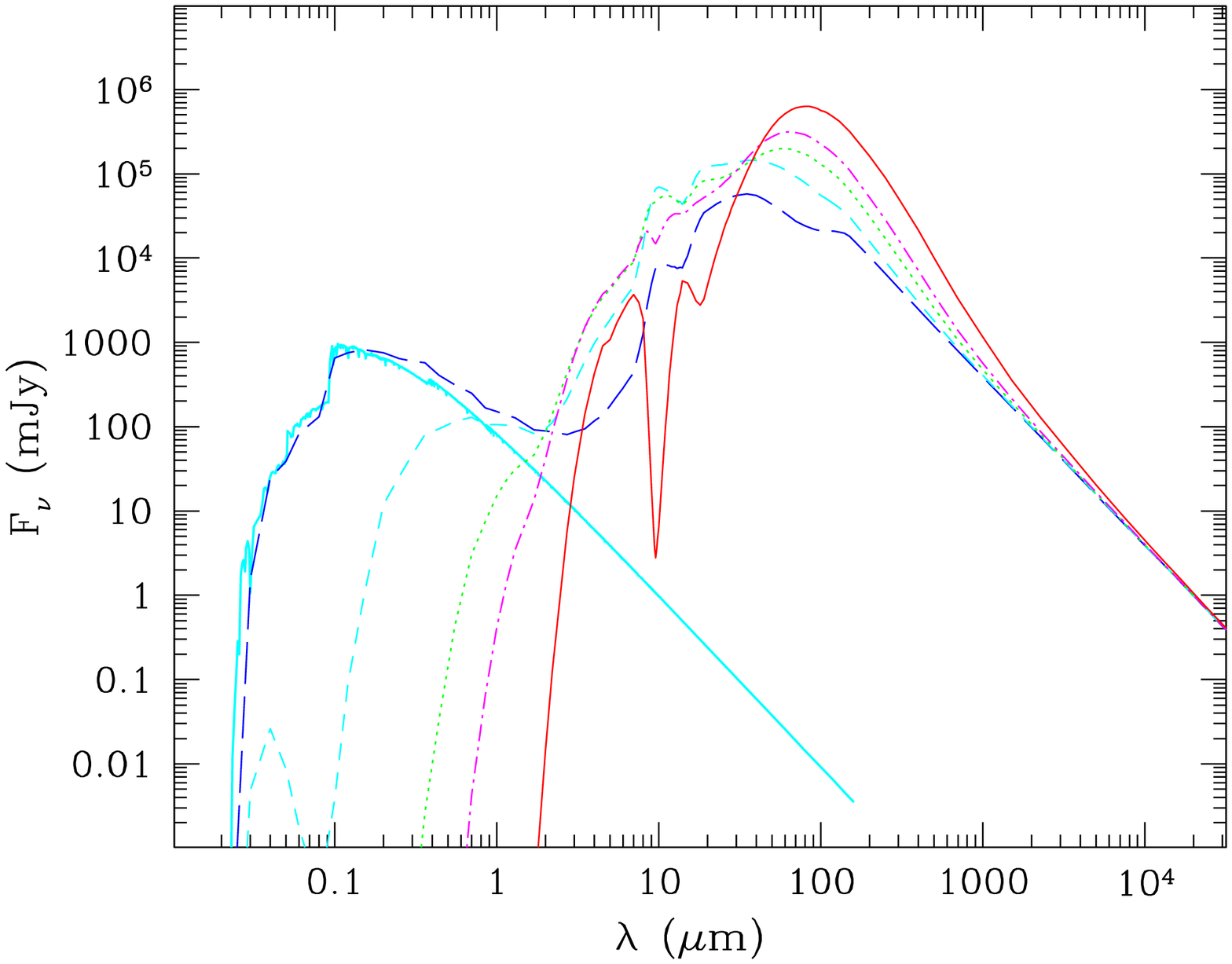}
\caption{The resulting \dusty\ spectra. We plot only 5 
values of \av, namely 0.1, 2.15, 10.0, 21.5, and 100 
(shown as long-dashed, short-dashed, dotted, 
dot-dashed, and solid lines, respectively). Also shown (as a heavy 
solid line which ends at $\lambda\sim\,100\,\mu$m) 
is the input stellar spectrum of a 3\,Myr starburst from SB99.
The upper panels show \tin =\,300\,K, and the lower ones \tin =\,600\,K.
The left panels show thick shells (\rout =\,10000), and 
in the right panels thin ones (\rout =\,100).
}
\label{fig:dusty_spec}
\end{figure}

We show examples of the resulting \dusty\ spectra in 
Fig. \ref{fig:dusty_spec}. Here we plot only 5 values 
of \av, namely 0.1, 2.15, 10.0, 21.5, and 100 
(shown as long-dashed, short-dashed, dotted, 
dot-dashed, and solid lines, respectively). Also shown 
(as a heavy solid line which ends at $\lambda\sim\,100\,\mu$m) 
is the input stellar spectrum of a 3\,Myr starburst from SB99. 
The figure clearly shows the importance of nebular continuum 
emission and the onset of warm dust 
in the spectral region sampled by IRAC.

\subsubsection{Blackbodies}

To illustrate the similarity between the emission from  stellar 
populations and 
blackbodies in the IRAC bands, we also calculated blackbody 
spectra at temperatures ranging from 100\,000\,K to 300\,K. 
In the IRAC/MIPS spectral region, there 
is virtually no difference among the blackbodies from 
100\,000\,K to 3500\,K. As expected, their colours are 
also very similar to the SB99 simulations of the 
stellar populations in a 3\,Myr cluster.

\subsubsection{Interstellar medium with PAH features}

An example of the spectrum of the diffuse ISM at 
high Galactic latitudes was also included in the diagnostics. 
The spectrum was taken from Figure 8 of \citet{2001ApJ...554..778L}, 
and the digital version was kindly provided by B. Draine. 
In the IRAC/MIPS-24 spectral range, the diffuse high-Galactic 
latitude ISM shows extremely strong PAH emission, 
superimposed on a rising continuum.

\subsubsection{Reflection nebula NGC\,7023}

NGC\,7023 is a well-studied PDR that, 
because of its proximity, is an excellent source to 
study the physical processes occurring in such regions. It is 
a reflection nebula illuminated by a B3Ve star, 
and is located in a cavity of the molecular cloud. 
This cavity hosts a dense PDR with its peak 
located away from the star itself \citep{2000A&A...354.1053F}. We 
obtained the short-wavelength portion of 
the ISO spectrum of this object from the uniform database published by 
\citet{2003ApJS..147..379S}. 
Like the diffuse ISM, NGC\,7023 also shows relatively strong 
PAH features superimposed on a rising continuum.

\subsubsection{Results}

\begin{figure}
\includegraphics[width=\columnwidth]{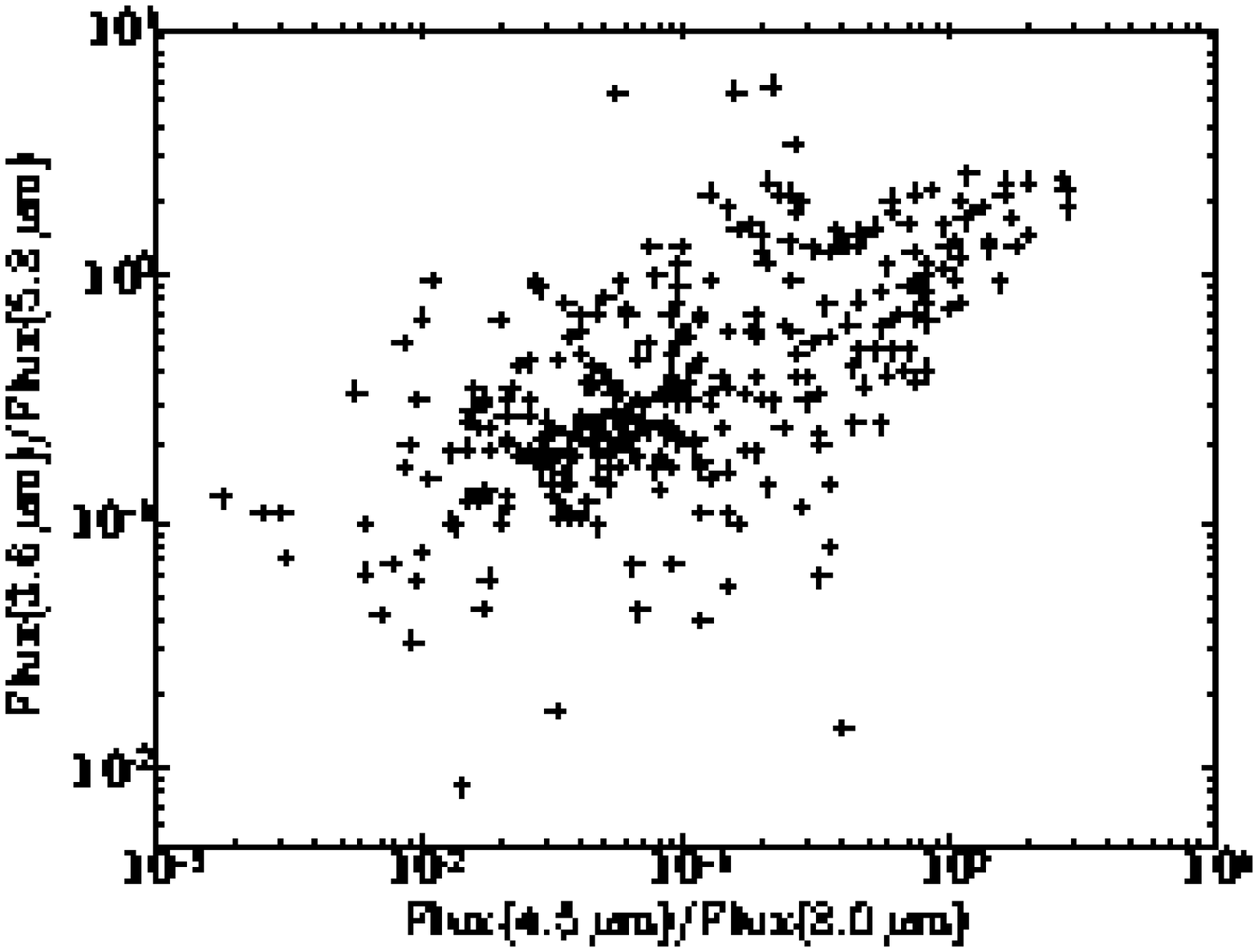}
\includegraphics[width=\columnwidth]{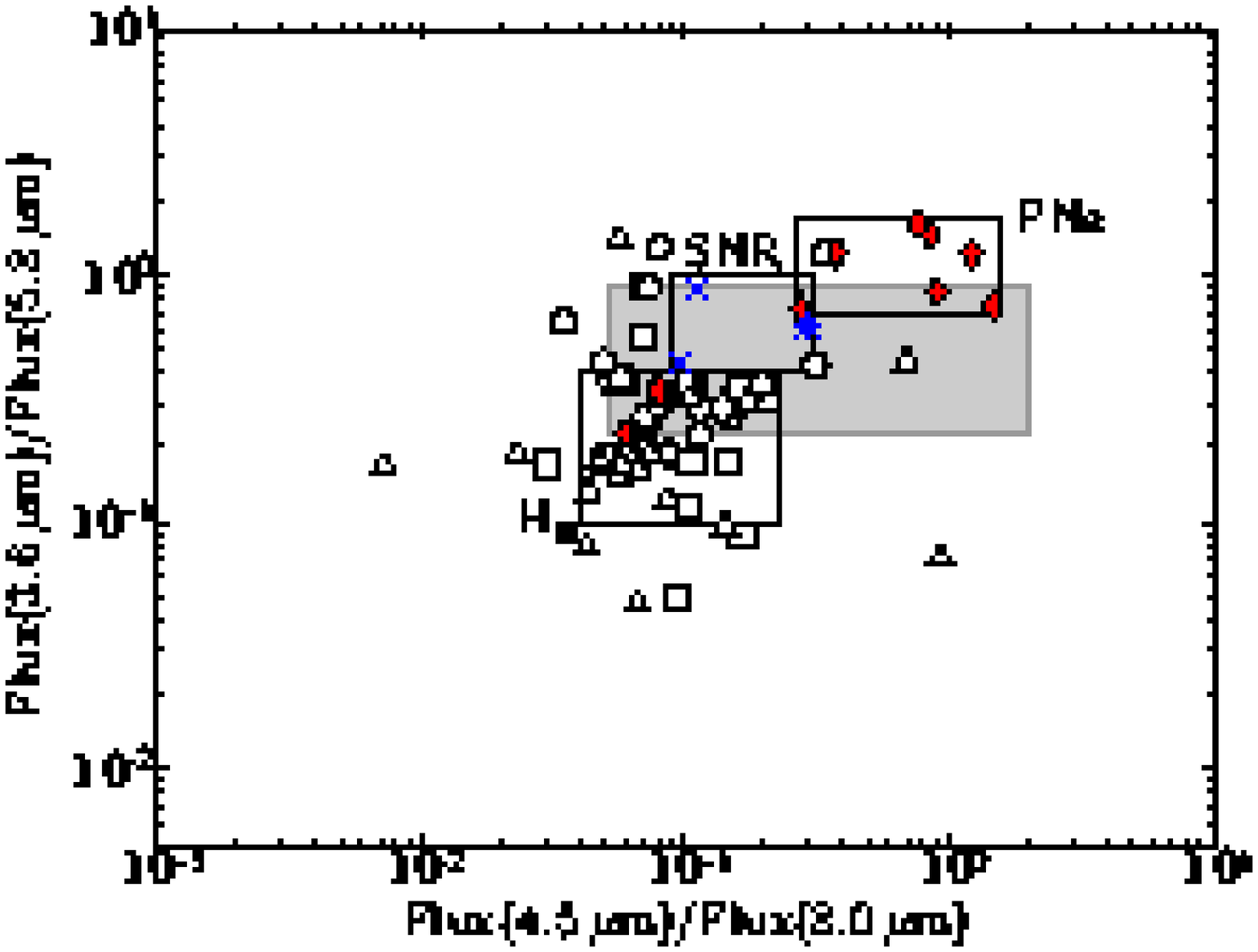}
\includegraphics[width=\columnwidth]{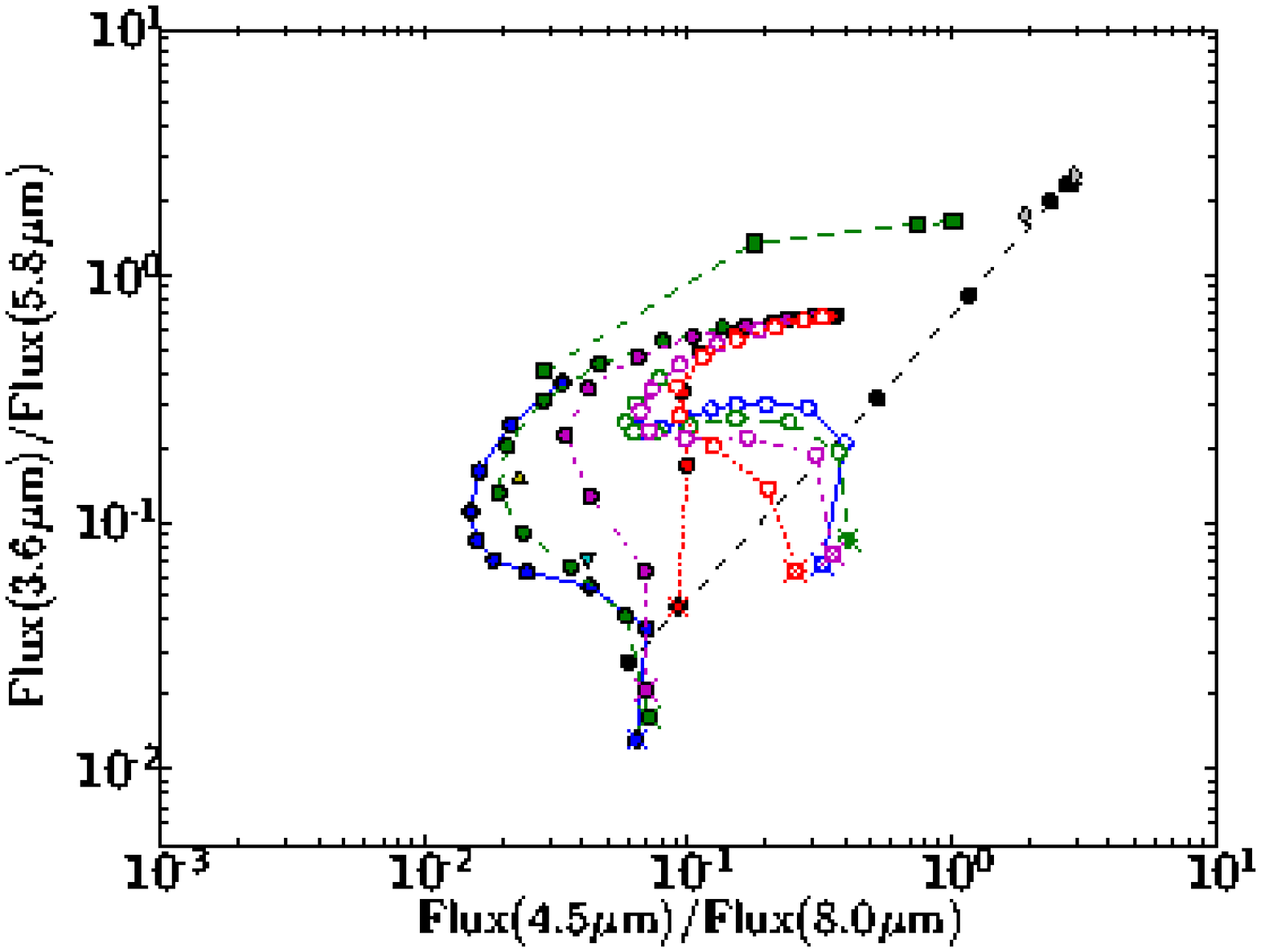}
\caption{{\bf Bottom panel:} Blackbodies are depicted by 
filled (black) circles. NGC\,7023 is represented by a (yellow) 
triangle, the ISM by Draine by an inverted (cyan) triangle, 
the SB99 simulations by (grey) diamonds. 
The Cloudy \hii\ region is represented by filled (green) squares. 
Finally, the \dusty\ models are represented by filled (\tin\ = 300 K) 
and empty (\tin\ = 600 K) circles, 
for a variety of shell thickness \yout\ = 10 (solid), 
100 (dashed), 1000 (dash-dot) and 10\,000 (dotted line). 
Different values of \av\ (logarithmically from 0.1 to 100~mag)
for a given \tin\ and \yout\ are connected by the different lines;
the highest value of \av\ is marked by $\times$.
{\bf Middle panel:} The open symbols represent the \hii\ 
regions selected in IR (squares), radio (triangles) and optical (circles). 
The (blue) crosses represent the SNRs optically selected and the (blue) 
plus the SNRs radio selected. The filled (red) 
diamonds represent the PNe. 
The open rectangles delineate the zones occupied by the majority of sources of a given type. 
The grey rectangle delimits the zone of molecular shocked SNRs \citep{2006AJ....131.1479R}. 
{\bf Top panel:} The 24~\micron\ sources are depicted 
by (black) pluses.}
\label{fig:ccDiag1}
\end{figure}

\begin{figure}
\includegraphics[width=\columnwidth]{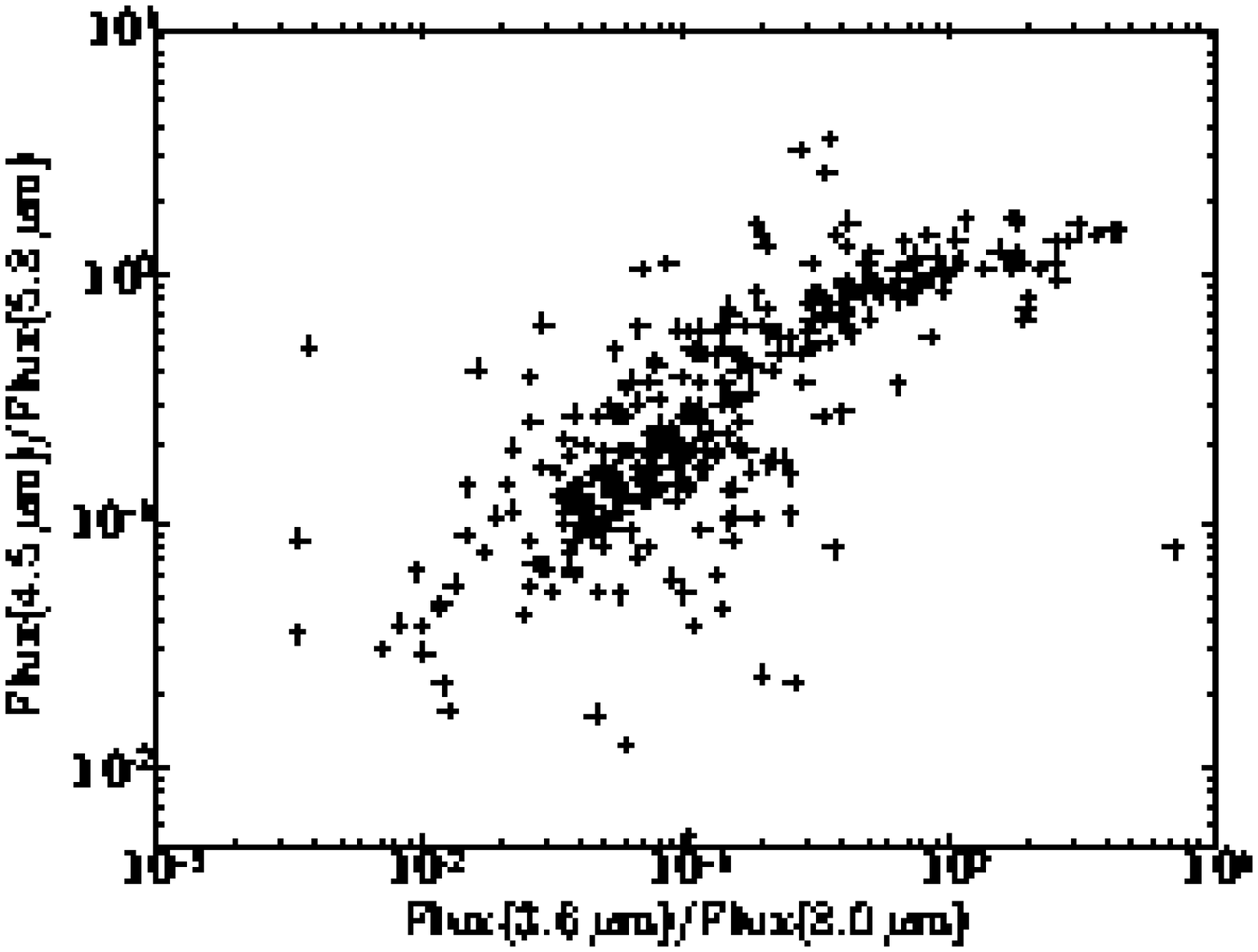}
\includegraphics[width=\columnwidth]{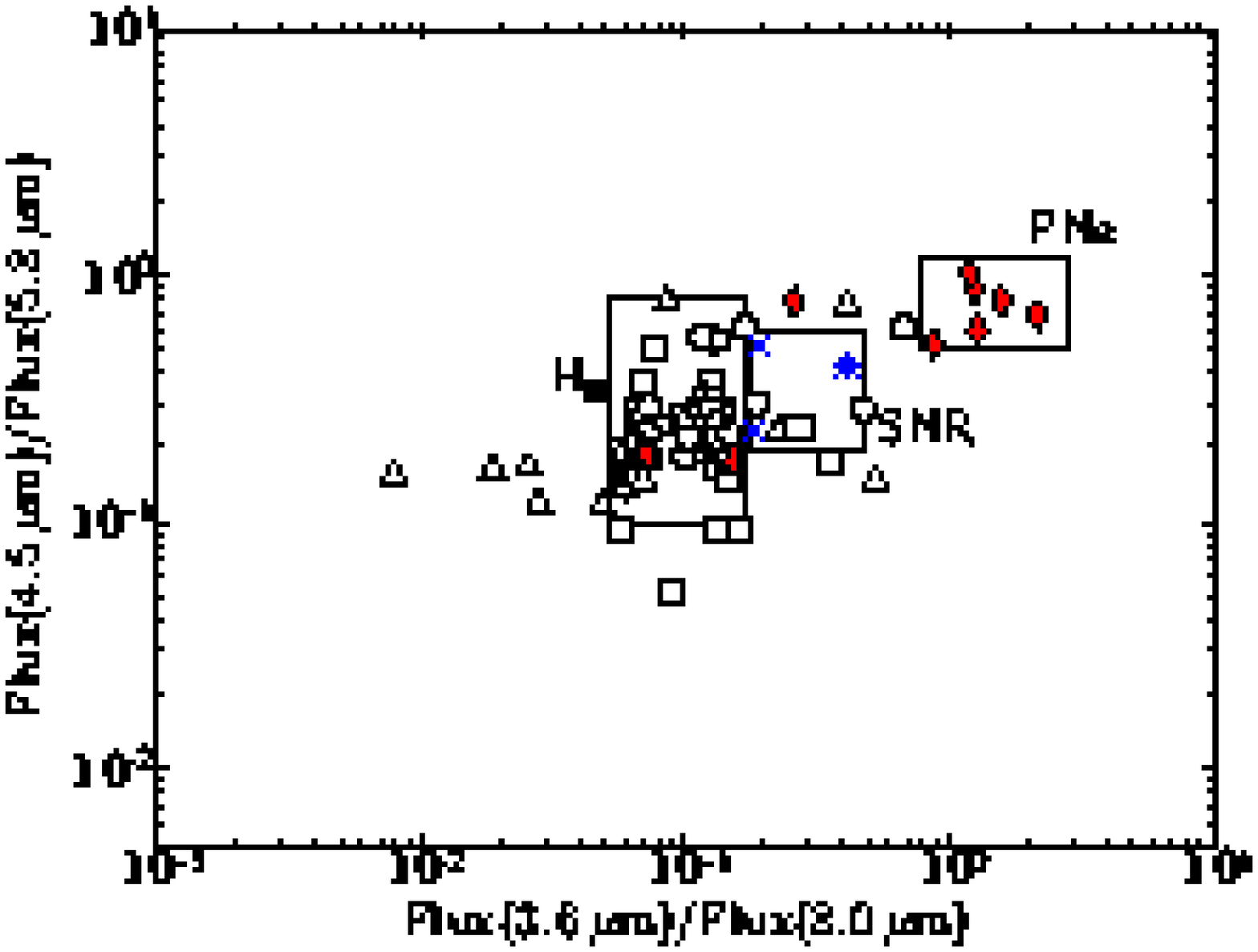}
\includegraphics[width=\columnwidth]{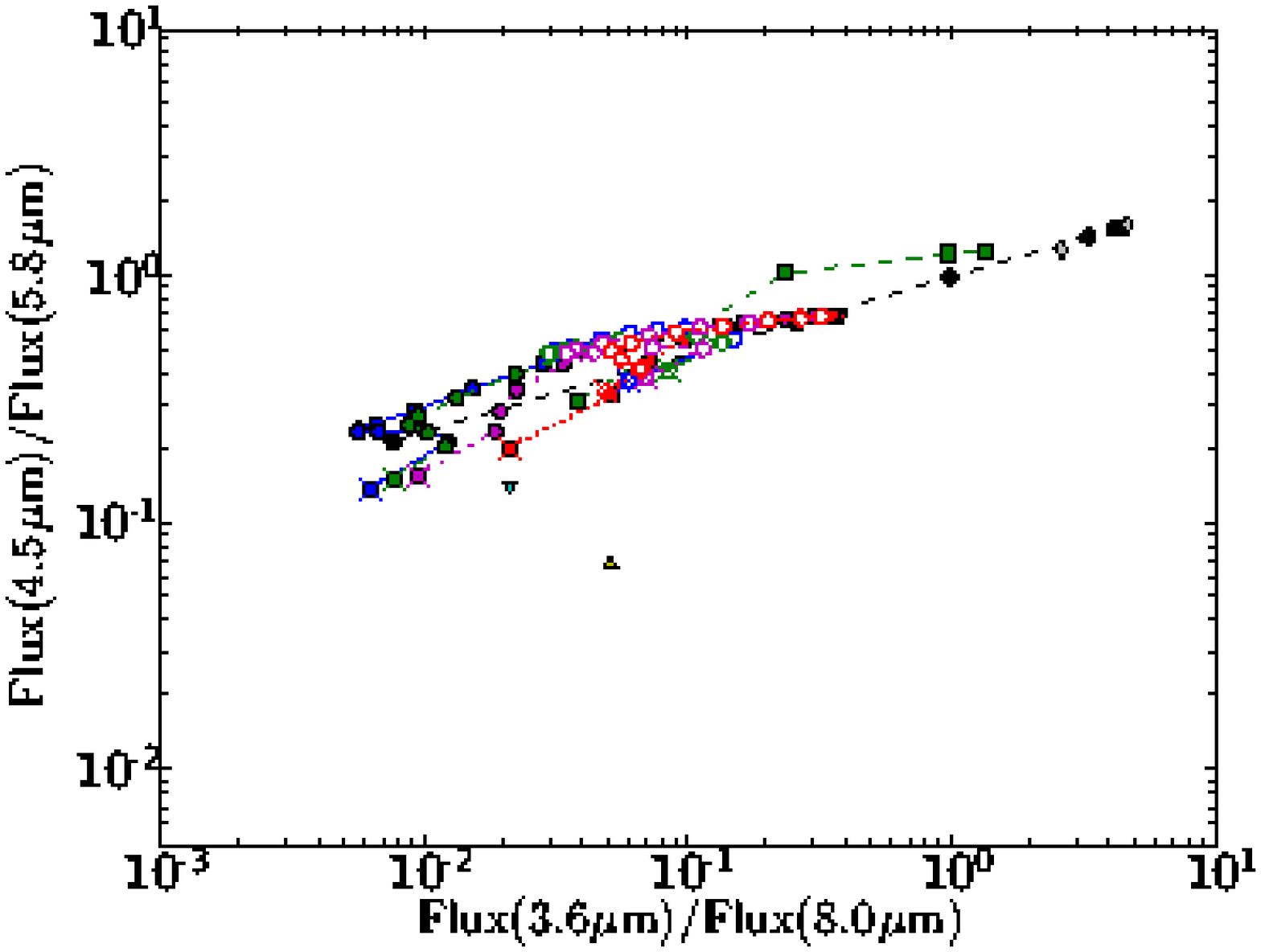}
\caption{Same as Fig.~\ref{fig:ccDiag1}, but for different IRAC flux ratios.}
\label{fig:ccDiag2}
\end{figure}

\begin{figure}
\includegraphics[width=\columnwidth]{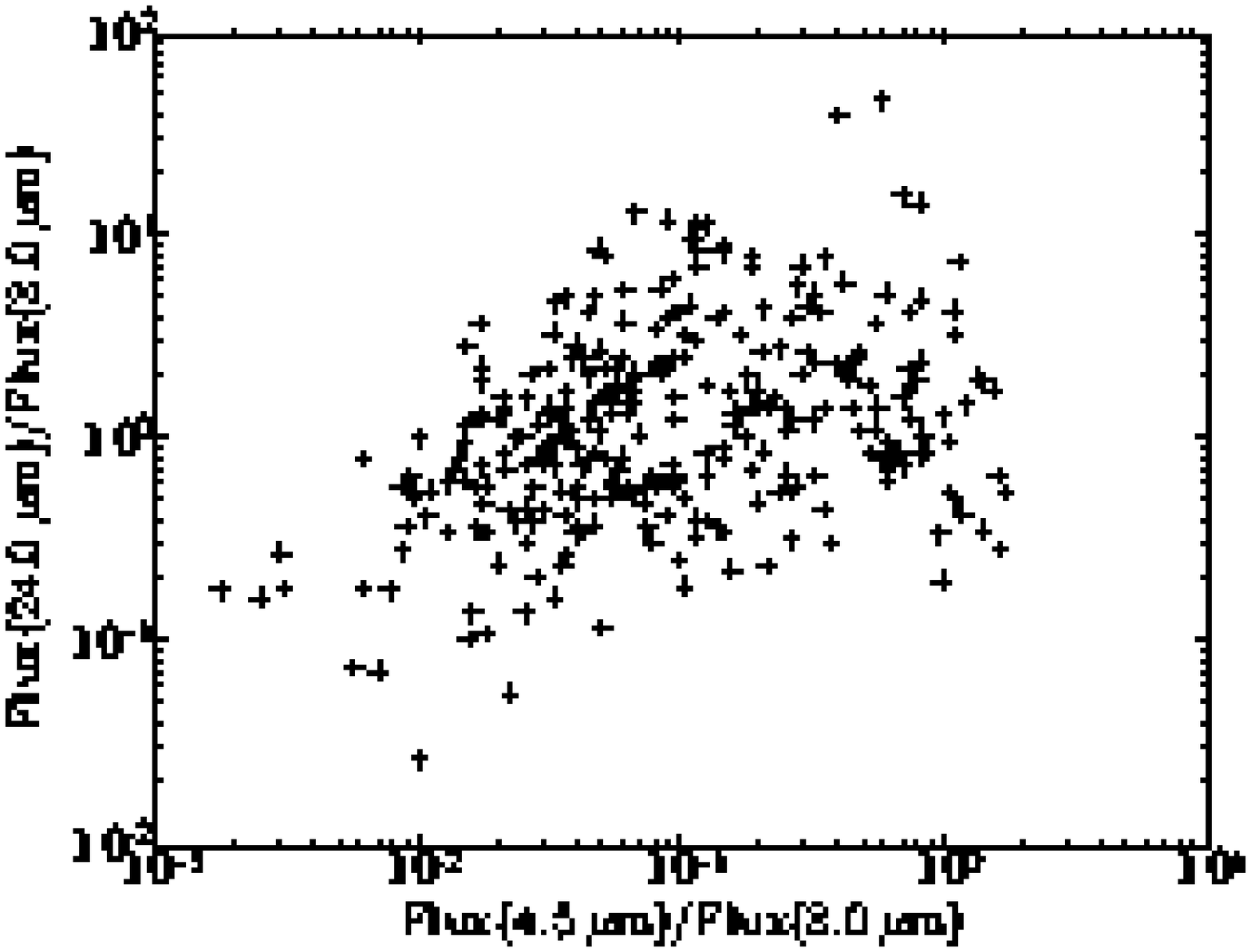}
\includegraphics[width=\columnwidth]{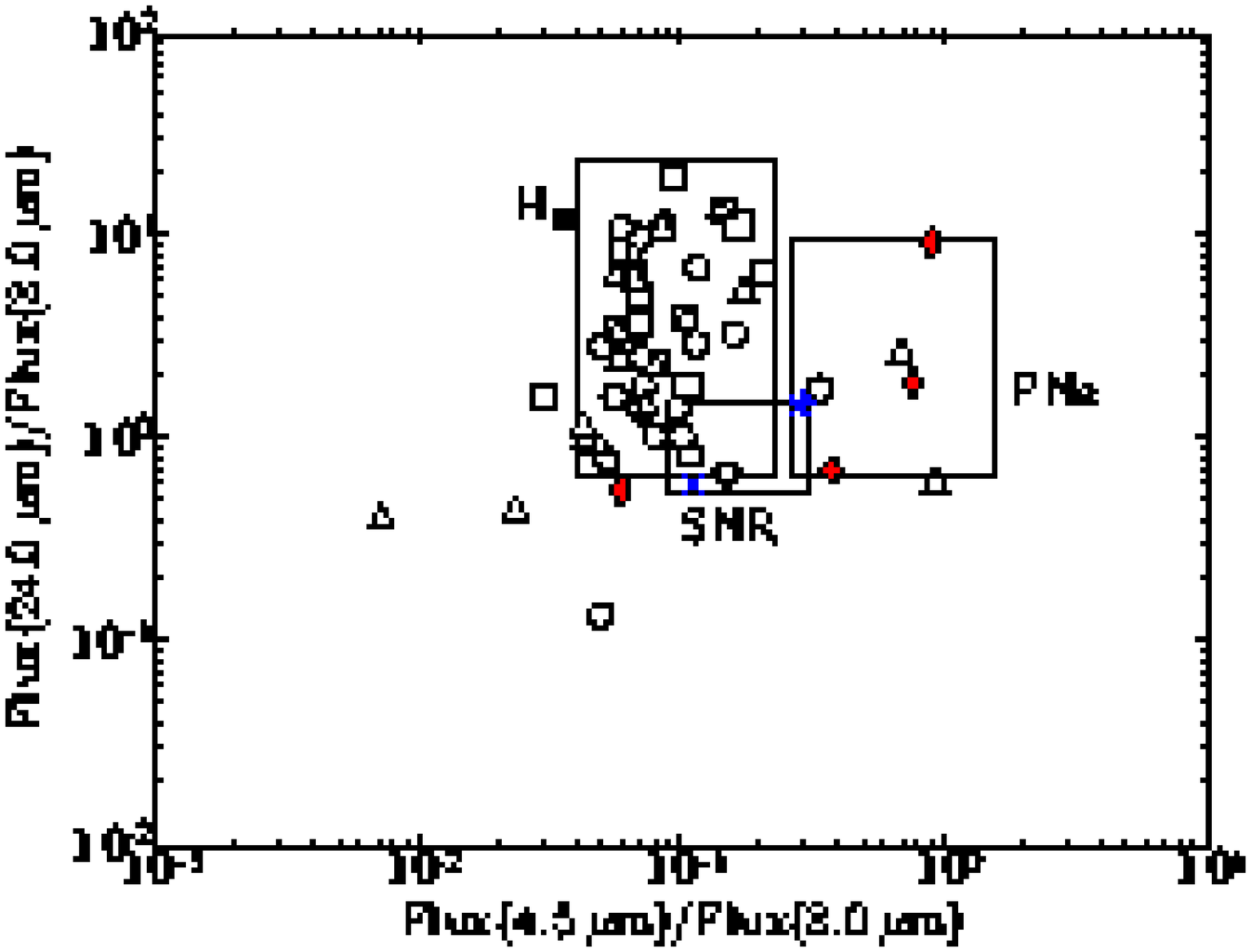}
\includegraphics[width=\columnwidth]{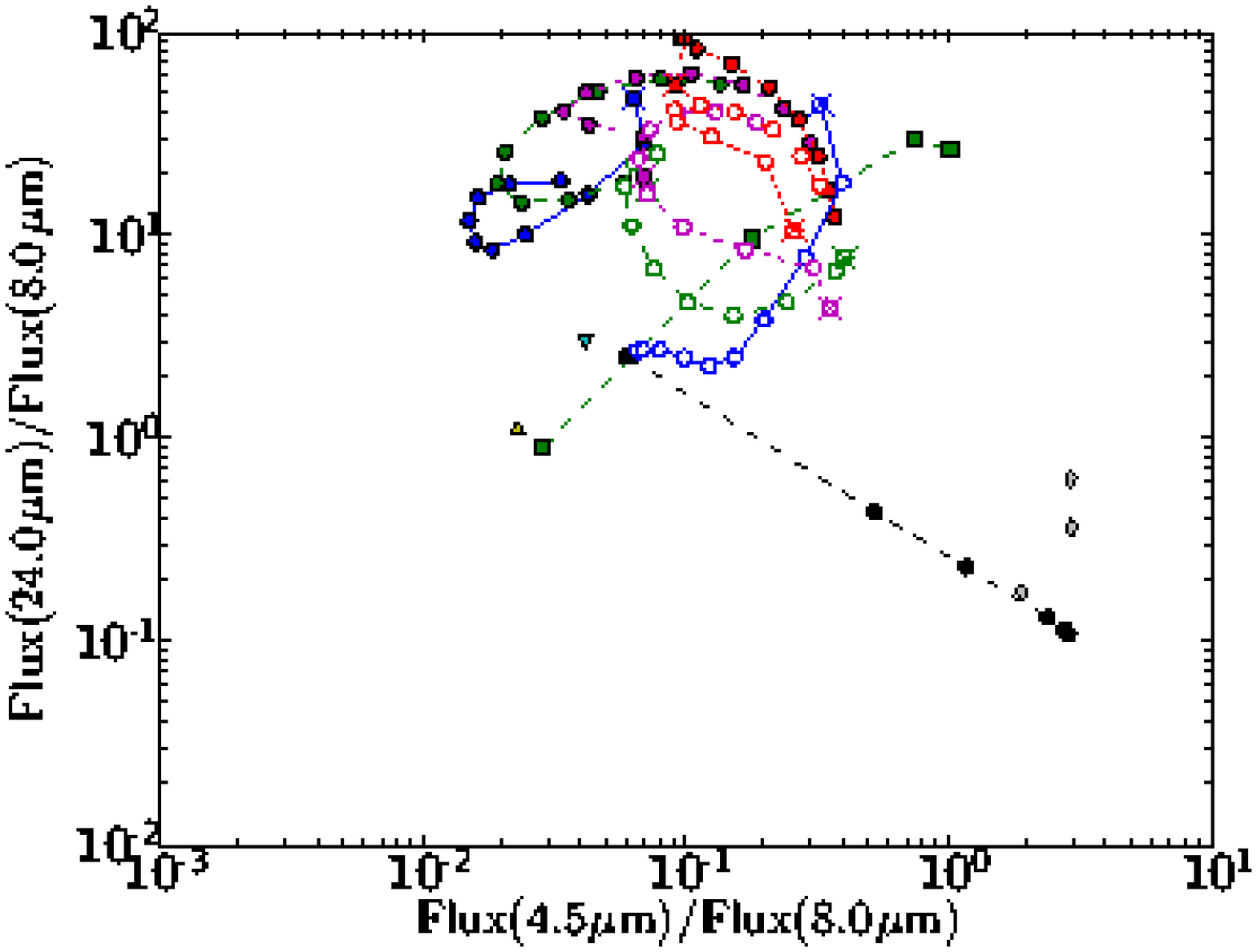}
\caption{Same as Fig.~\ref{fig:ccDiag1}, but now including
the MIPS 24~\micron\ flux ratio.}
\label{fig:ccDiag3}
\end{figure}

The IRAC and IRAC/MIPS-24 diagnostic diagrams relative to
the models described above are shown in the bottom panels 
of Fig.~\ref{fig:ccDiag1}, 
Fig.~\ref{fig:ccDiag2}, and Fig.~\ref{fig:ccDiag3}. 
Blackbodies are depicted by filled black circles. NGC\,7023 
is represented by a (yellow) triangle, 
the diffuse high-latitude ISM by an inverted (cyan) triangle, 
the SB99 simulations by (grey) diamonds. 
The Cloudy \hii\ region is represented by filled 
green squares. Finally, the \dusty\ models are 
represented by filled (\tin\ = 300 K) and empty 
(\tin\ = 600 K) circles, for a variety of shell thickness 
\yout\ = 10 (solid), 100 (dashed), 1000 (dash-dot) 
and 10\,000 (dotted line). 

There are several interesting 
features that can be noticed in the plots. 
First, a PAH-dominated spectrum such as the diffuse 
ISM, the small dense Cloudy \hii\ region, and the 
reflection nebula have virtually identical IRAC colours to some of 
the \tin\,=300\,K \dusty\ models. The implication here 
is that strong PAH features can be easily 
confused with a rising continuum in the MIR.  
The only way to resolve the ambiguity is through detailed 
spectral information {\it or by using the 
24\,$\mu$m filter in conjunction with the IRAC 8\,$\mu$m one}. 
Indeed, our plots show that only 2 
IRAC filters, together with MIPS-24, are sufficient 
to separate strong continuum emission in the MIR 
from strong PAH features. 
Second, Figure~\ref{fig:ccDiag1} shows that 
strongly embedded sources (e.g., \av\,$\sim 100$) 
occupy a well-defined region in the IRAC colour diagram. This 
may potentially be useful for singling out ultra-compact dusty 
\hii\ regions from our 24\,$\mu$m source catalogue. 
Third, stars (roughly equivalent to 3500\,K blackbodies) 
occupy a well-defined region in both diagrams, and should 
therefore be easily separable from other classes of objects.

On the negative side, the \dusty\ model predictions in the IRAC/MIPS-24 
diagram are not easily distinguishable from Cloudy predictions. 
Moreover, even the highly embedded clusters do not occupy 
a distinct region in the colour diagrams. 
This makes the use of IRAC bands necessary, in order 
to better separate high extinction regions from optically thin ones.

\subsection{Observational diagnostic diagrams}

A comparison of the IRAC colours of the
models$+$templates vs. observations of the different
kinds of nebulae of Sect.~\ref{sec:knownSources}
is shown in Figs.~\ref{fig:ccDiag1} and \ref{fig:ccDiag2}. 
Diagnostic diagrams using IRAC colours were presented 
by \citet{2006AJ....131.1479R} (focusing on SNRs 
emission mechanisms) and by 
\citet{2007MNRAS.374..979C}.
We present two versions of the IRAC diagram, in order
to better assess the diagnostic value of the different ratios.
We also propose in Fig.~\ref{fig:ccDiag3}
a new diagnostic diagram involving the 24~\micron\ 
MIPS channel which allows us to disentangle some 
degeneracies present when using only IRAC colours. 
In the next Section, 
will use both the theoretical colour diagnostics and the observed 
colours of known-type objects to investigate the nature of
sources selected at 24~\micron.

\subsubsection{\hii\ regions}

In the first IRAC colour diagram
(see the middle panels of Figs.~\ref{fig:ccDiag1} and \ref{fig:ccDiag2}), 
we find a concentration of \hii\ regions in 
the zone dominated by PAH emission \citep[see][]{2006AJ....131.1479R}. 
In the same zone, lie the reflection nebulae NGC\,7023 
and the diffuse Galactic ISM as well. 
The dispersion in the colours of 
\hii\ regions makes it difficult to separate them 
from SNRs and PNe. 
Statistically, the \hii\ regions 
are more concentrated near the regions defined by low values of 
the two shortest IRAC channels and/or equivalently 
high values of the two longer IRAC wavelengths. 
The most intense PAH bands (6.2, 7.7, 8.6~\micron) are 
located in the 5.8 and 8.0~\micron\ channels 
and this may explain the colours of \hii\ regions.


We compared the IRAC colours of our catalogued \hii\ 
regions with the diffuse \hii\ regions studied by 
\citet{2006astro.ph.10569C}. 
Upon conversion to their magnitude scale \citep{2004ApJS..154..229P},
we find the IRAC-colour distributions to be mutually consistent.

\begin{figure}
\includegraphics[width=\columnwidth]{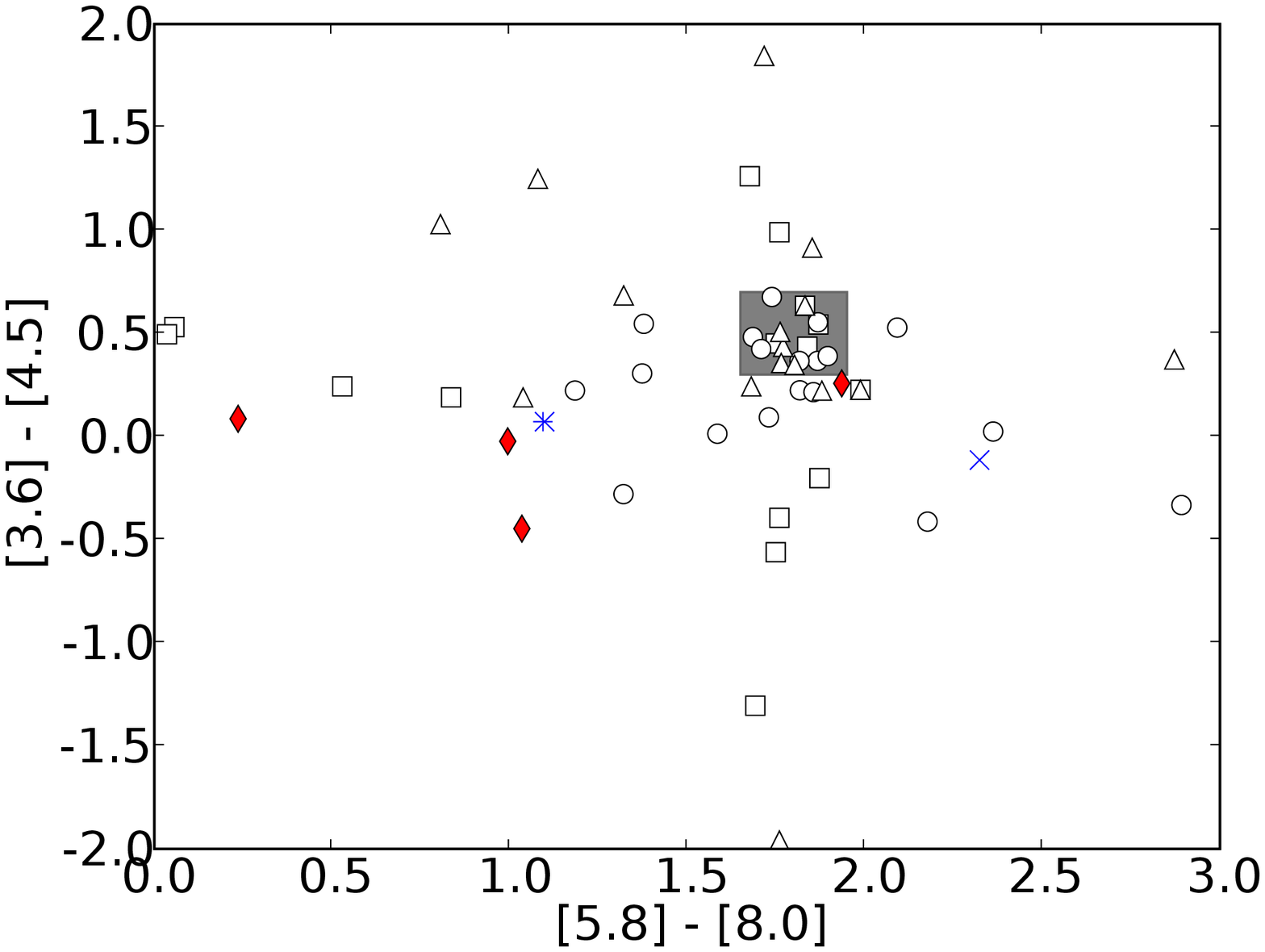}
\caption{Diagnostic diagrams in magnitude, for comparison with 
\citet{2006astro.ph.10569C}. The 
open symbols represent the \hii\ regions selected 
in IR (squares), radio (triangles) and optical (circles). 
The (blue) crosses represent the SNRs optically selected and the 
(blue) plus the SNRs radio selected. 
The filled (red) diamonds represent the PNe. 
Two radio \hii\ regions are lying outside the 
limits of the displayed plot.}
\label{fig:ccDiagMIPSKnown-Hora}
\end{figure}

\subsubsection{Supernovae remnants}

As a word of caution, since (especially young) SNRs are
often associated with \hii\ regions, they are often hard
to discern and to measure photometrically given the
crowding and confusion in the field.
In the colour diagrams,
the few detected SNRs lie between
the \hii\ regions and the PNe, generally closer to the 
\hii\ regions.

Since the colours of pure synchrotron emission are never 
observed, there are several alternative emission mechanisms  
for the IR radiation that have been proposed by 
\citet{2006AJ....131.1479R} such as 
shocked molecular lines, shocked ionic lines, 
and PDR PAH emission. About half of their 
SNRs have colours consistent with molecular shocks, 
17\% are consistent with ionic shocks, 
and 22\% with PAH emission from unshocked ISM. 
The others are thought to represent a mixture of 
shock types. All our 4 SNRs appear to 
belong to the class
of shocked molecular emission (see the -grey- filled rectangle in 
the middle panel of Fig.~\ref{fig:ccDiag1}), 
that is shocked molecular gas cooling mainly via 
MIR emission lines. 
Also, optically, SNRs appear as shock-heated emission 
nebulae \citep{1999ApJS..120..247G}, confirming the prevalence 
of the two shock mechanisms described by \citet{2006AJ....131.1479R}. 
The colours of our SNRs are also close to 
the colours of some \hii\ regions. 
Note that the colours of the SNRs could depend 
of the epoch at which the SNR is observed 
(for instance 
because of the onset or decrease of shocks as they expand into the ISM).

\subsubsection{Planetary nebulae}

With few exceptions, catalogued PNe are confined to a region
characterised by low values of the 5.8
and 8.0~\micron\ emission, the bandpasses where PAHs are 
believed to contribute substantially to the IR emission.
Indeed, PNe spectra usually show an absence or very weak presence of 
PAH, as noted by \citet{2002A&A...387..301B} 
with {\it ISO} 
and by \citet{2004ApJS..154..271B} and with \citet{2006ApJ...652L..29B} \spitzer.
In the 4.5~\micron\ band, the [Mg iv] 4.485~\micron\ and 
[A vi] 4.530~\micron\ lines could give an 
additional contribution for high-excitation PNe 
\citep{2005ApJ...627..446C}. 
Thus, in absence of detailed spectral information 
\spitzer\ colours can provide some diagnostics to help
to identify PNe candidates
\citep[see also ][]{2004ApJS..154..296H}. 

\section{The 24 \micron\ sources in \mm} \label{sec:S24}

\spitzer\ images provide a unique
opportunity to compile vast catalogues of MIR discrete
sources down to stellar luminosities for galaxies 
in the Local Group.
In the present section we describe and discuss some features of 
the catalogue of  24~\micron\ sources (24Ss) 
which we have generated for \mm.
In the first subsection, we describe the source selection
criterion and flux measurements; 
in the second subsection, we investigate their nature
in light of both theoretical and observational diagnostic diagrams; 
in the last subsection we present their luminosity function.

\subsection{24 \micron\ source catalogue}

To select and measure individual sources from the 24 \micron\ image, 
we used the SExtractor software \citep{1996A&AS..117..393B}. 
Using a convolution with a Gaussian filter we claim a 
detection when the source area is larger 
than 10 pixels and the peak signal-to-noise ratio 
is at least 10 times the local background. Guided by the 
galaxy pattern traced by the 24~\micron\ diffuse emission,
we removed spurious sources such as foreground 
stars or artifacts due to map edge effects. 
The final catalogue comprises 515 objects.
The brightest sources at 24~\micron\ 
are located in the central region and along the 
spiral arms. Fainter sources are more smoothly distributed but still 
follow the flocculent pattern of the galaxy.

\subsection{Nature of the 24~\micron\ sources}

Perhaps not surprisingly, sources in our 24~\micron\ catalogue turn out to
have a higher detection rate than 
any of the other kinds of objects considered 
in the previous sections. 
This is particularly marked at 5.8~\micron\ 
where the detection rate for the 24~\micron\ 
sample is much higher than for \hii\ regions (see Fig.~\ref{fig:fracDet})
although, as seen in  Fig.~\ref{fig:sed-mean},
the averaged SED of the 24Ss is similar to the one of 
\hii\ regions.

The position of the 24Ss 
in the colour diagrams are illustrated  in the top panels 
of Figs.~\ref{fig:ccDiag1}, \ref{fig:ccDiag2} \& \ref{fig:ccDiag3}.
In Fig.~\ref{fig:ccDiag1}, the region with the highest 
concentration matches well
the one of \hii\ regions. However the distribution 
of 24~\micron\ sources extends into the regions populated  
by SNRs and PNe. 
The comparison with the theoretical models indicates 
a paucity of sources with high extinction. However, as
Fig.~\ref{fig:ccDiag1} shows,
neither 
a pure blackbody, Cloudy, or the two 
\dusty\ models can alone reproduce the 
colour distribution of the 24~\micron\ sources. 
Fig.~\ref{fig:ccDiag2} shows a clear departure of the 24Ss
sample from the expected narrow 4.5/5.8 colour dispersion. 
Such departure is particularly marked in the red
tail of the distribution where
both 24Ss and known-type sources tend to have colours 
similar to the PDR observed in the reflection nebulae NGC\,7023.

The hybrid IRAC-24~\micron\ MIPS colour diagnostics
of the selected sources are illustrated in Fig.~\ref{fig:ccDiag3}. 
No 24S displays a 24/8.0~\micron\ flux ratio 
as high as the \dusty\ models with shell inner temperature
\tin\ = 300~K; 
the warmer models (\tin\ = 600~K) seem 
more appropriate to describe their physical conditions.
The range spanned by the 24/8.0 ratio in the 24~\micron\ 
sources is large but similar to the one covered by the 
catalogued \hii\ regions, indicating that in both samples 
the contribution from PAH features may vary widely.
So, 24Ss with both a low 4.5/8.0 and a low
24/8.0 ratio could well be \hii\ regions with high PAH emission 
(see, for instance, the location of the radio-selected \hii\ regions). 
The Cloudy \hii\ region simulations 
seem to match well the colours of the 24Ss, 
in spite of their large scatter. On the contrary, 
the observations rule out most stellar 
colours: both pure blackbodies and the SB99 templates. 
However, as above,
the general agreement between any single model 
and the 24Ss distribution in the colour diagrams is unsatisfactory;
such a trend is probably due to a variable PAH component
which is not considered in most of our templates
(but see loci of NGC\,7023 and the ISM). 
In the next section, we will select a subsample of 
24Ss associated with the H$\alpha$ emission. 
These sources, likely to be optically visible \hii\ 
regions, achieve a better agreement with
low-extinction models. 

\subsection{Luminosity function of 24~\micron\ sources} \label{sec:S24_sub:lum}

The 24~\micron\ emission of \mm\ within an ellipse of 
major axis 40$'$ and axial ratio of 0.62 is 40.9~Jy. 
At a distance $D = 840$~kpc, such ellipse corresponds
to a circle at galactocentric distance of 4.8~kpc
inclined by 52\degr\ with respect to our line of sight.
The 24~\micron\ integrated emission is in good agreement with the 
value reported by \citet{2004ApJS..154..259H} (see their Fig.~2) 
and implies a 24~\micron\ luminosity 
$( L = \nu F_\nu \times 4 \pi D^2)$  of $4.3 \times 10^{41}$~ergs~s$^{-1}$. 
The contribution of the 515 discrete 
24Ss amounts to 
$1.4 \times 10^{41}$~ergs~s$^{-1}$, 
hence the diffuse component accounts for 
two thirds of the 24~\micron\ emission.

\begin{figure}
\includegraphics[width=\columnwidth]{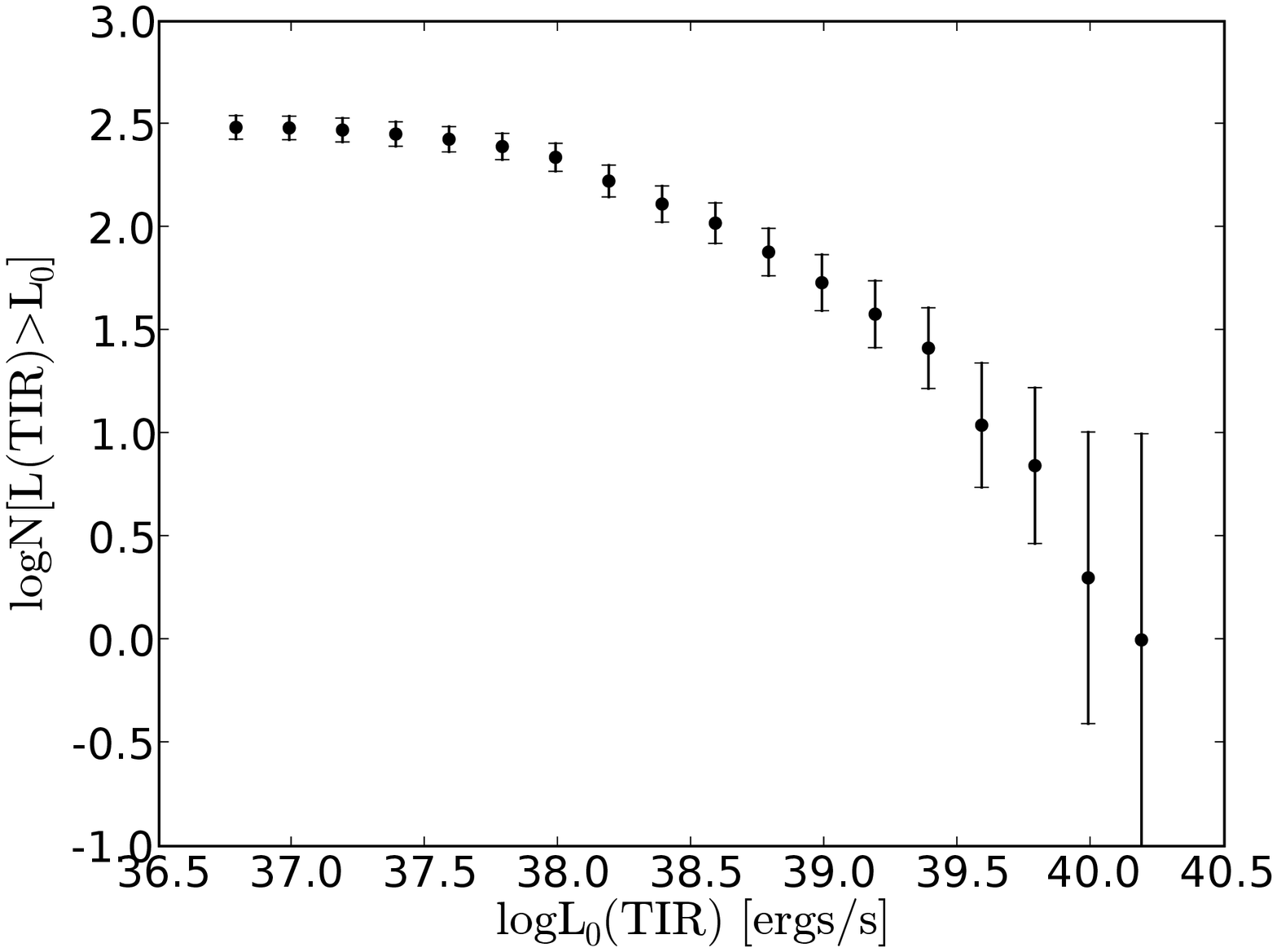}
\includegraphics[width=\columnwidth]{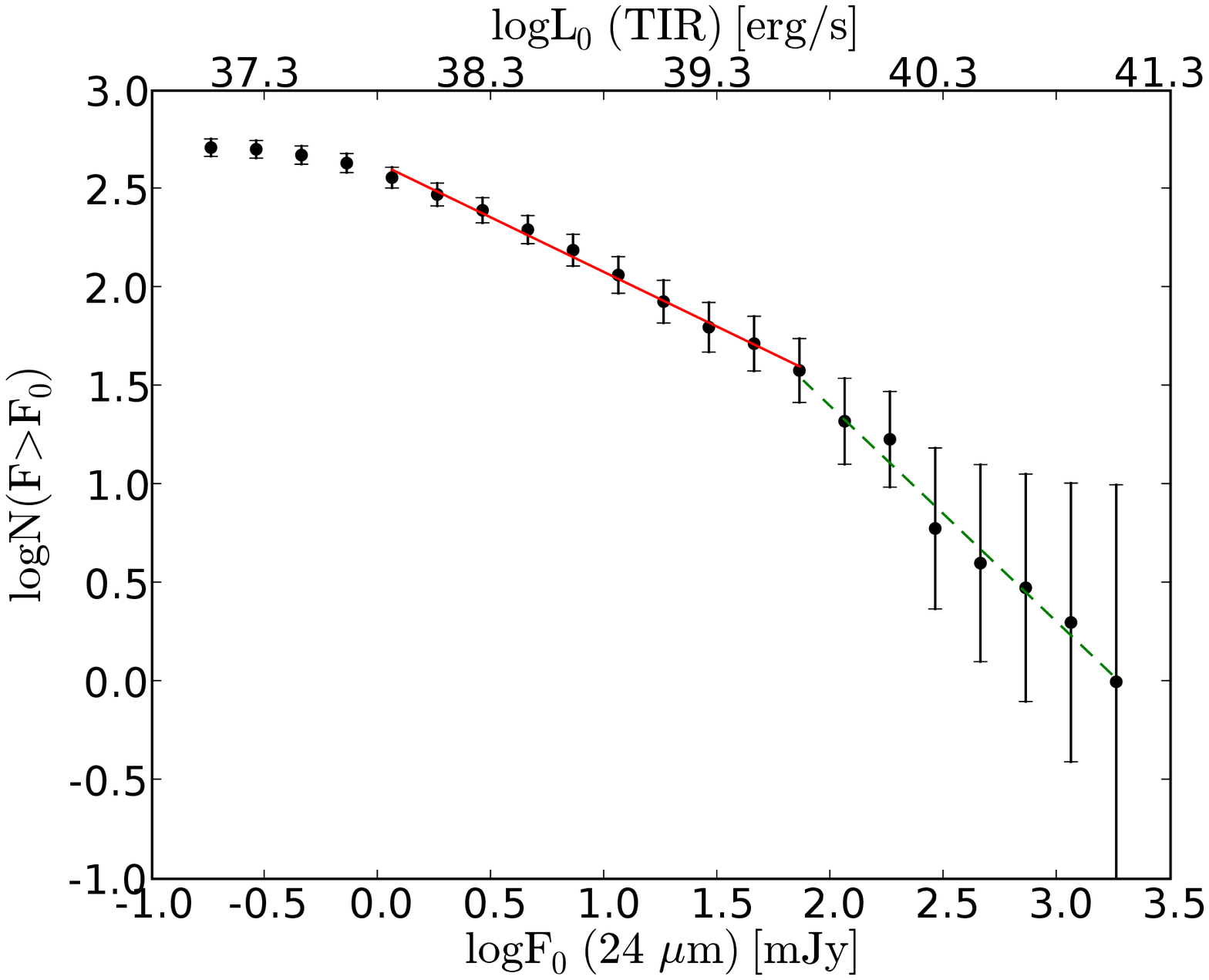}
\caption{{\bf Bottom panel:} Cumulative distribution 
of the 24~\micron\ sources. The 24~\micron\ 
fluxes are reported F$_\nu$(24~\micron) (bottom axis), 
as well as the TIR luminosity L$_\nu$(TIR) (top axis). 
The linear best fits are depicted by a (red) 
solid line for the F$_\nu$(24~\micron) 
fluxes between 1 and 70~mJy, and by a (green) 
dashed line for the F$_\nu$(24~\micron) fluxes 
greater than 70~mJy. {\bf Top panel:} Cumulative distribution of 
the 24~\micron\ sources detected also at 8.0~\micron, 
the TIR luminosity is calculated using for each source the effective 
8~\micron/24~\micron\ flux ratio.}
\label{fig:lumFunc}
\end{figure}

The cumulative luminosity function of 
the 515 discrete sources catalogued at 24~\micron\ is
shown in Fig.~\ref{fig:lumFunc}. 
In this figure, the abscissas are reported as flux density at
24~\micron\ in mJy and also as total IR luminosity (TIR).
To estimate the TIR  luminosity, we follow the prescription of 
\citet{2005ApJ...633..871C} 
(see their Eq.~1). The conversion involves the 24~\micron\ 
and the 8~\micron\ measurements.
Since we do not have a detection at 8~\micron\ 
for all 24Ss, we compute the average 
8~\micron/24~\micron\ flux ratio  for 
the sources detected in both bands: the average is close 
to unity and therefore, in the lower panel of Fig.~\ref{fig:lumFunc}, 
which includes all 24Ss, we use $\log$L(TIR)~=~$\log$L(24~\micron)+0.908. 
As a check, in the upper panel of Fig.~\ref{fig:lumFunc}, 
we plot the cumulative luminosity 
function only for the sources detected in both bands, using 
for each source the measured 8~\micron/24~\micron\ 
flux ratio: the shape and meaning (see below) of the
two curves is quite compatible. 
The main effect, when using a constant average flux ratio, is 
a slight expansion of the range in L(TIR) by
about 0.4 dex, together with a brightening by the same factor.

From the flattening of the faint end of the distribution, 
we estimate the completeness limit of the
catalogue around 1~mJy, or an L(TIR) of $5 \times 10^{37}$ ergs~s$^{-1}$
on the averaged scale. That is the bolometric luminosity of a B1.5\,V star
\citep{2000asqu.book.....C} and so our listing should be quite
complete even for faint obscured \hii\ regions. 
Apart from the levelling off at low fluxes,
as often observed for  \hii\ regions, open clusters, and associations
\citep{1997ApJ...476..144M}, the luminosity function
displays a double slope behaviour, markedly steeper 
at the high luminosity tail.
The change of slope, in the simplest scenarios,
represents the change of regime between poor and
rich clusters, where rich means numerous enough
to represent fairly the high-mass IMF \citep{1998AJ....115.1543O}.
In this framework, the transition point between the two
regimes marks the luminosity of the single brightest
star; below this value, the observed statistics is modified by
the sampling variance.
We find that the transition point occurs around F$_\nu$(24~\micron) = 70~mJy,
that is, in the averaged scale, an L(TIR)~$\simeq 5\times10^{39}$ ergs~s$^{-1}$
which, still bolometrically, corresponds to an O3\,V star
\citep{1996ApJ...460..914V}. This implies: first, that most of the bright
24Ss are in fact luminous young stellar clusters and, second,
that our L(TIR) scale cannot be grossly mistaken.
From F$_\nu$(24~\micron) = 1 
to F$_\nu$(24~\micron) = 70~mJy we fit a power index $-0.55 \pm 0.01$ 
(plain -red- line) and, for larger values 
an index $-1.09 \pm 0.08$ 
(dotted -green- line). 
It is easily shown that an index about unity, in the log-log 
cumulative distribution, implies for the rich galactic clusters  
a distribution $N(L) dL \propto L^{-2} dL$ or, equivalently,
$N(n_*) dn_* \propto n_*^{-2} dn_*$, $n_*$ being the number of
stellar members. Again, this is precisely the slope observed
in the luminosity function of \hii\ regions in late spirals
\citep{1991ApJ...370..526C}, as well as the slope determined for
Galactic clusters and associations, e.g. 
\citet{1997ApJ...476..144M}. 

The brightest source reported in 
Fig.~\ref{fig:lumFunc}, at F$_\nu$(24~\micron)~$\simeq 1.8$~Jy,
would be roughly equivalent to an open cluster with around 
$3.5 \times 10^7$~L$_\odot$; assuming that all members 
($0.06$ to $90$~M$_\odot$) are on the main sequence,
this makes a cluster of some $10^5$~M$_\odot$, more than $10^6$ members,
and more than 1000 ionising ($>15$~M$_\odot$) stars.
On the other side, the faintest detected discrete 24Ss
have F$_\nu$(24~\micron) of 0.1~-~0.2~mJy,
that is L(TIR)~$\simeq 10^{37}$~ergs~s$^{-1}$.
This is about the bolometric luminosity of zero age main sequence
B2 stars \citep{1973AJ.....78..929P}; so it is possible that these
objects are also (the faintest) \hii\ regions and
our luminosity distribution appears to precisely match the whole range
expected for ionised regions and their complexes.
Anyway at low luminosity, especially below $10^4$~L$_\odot$, 
the situation may be more intricate since also evolved stars
can contribute substantially,
more or less depending on the history of the region, 
to the population of IR sources; carbon giants, for example,
at the distance of \mm\ are expected to emit an average
F$_\nu$(24~\micron)~$\simeq 0.25$~mJy \citep{2007MNRAS.376..313G}.
The nature of the 24Ss will be further explored, 
by investigating their spatial location within the galaxy and IR colours,
in a forthcoming paper.

\section{Star Formation} \label{sec:24Ha}

\subsection{Global star formation rates} \label{SFR_global}

In regions of moderate visual extinction,
H$\alpha$ line emission is expected to be a reliable indicator
of recent star formation. \mm\ has globally a low extinction
rate and therefore we can assess the performance of MIR emission 
as a star formation tracer, by comparing the SFR inferred from
MIR emission with that derived from H$\alpha$ emission. 

H$\alpha$ fluxes were measured on the map of 
\citet{2000ApJ...541..597H} cited earlier (Sect.~\ref{sub:Ha})
and converted to SFR following \citet{1998ARA&A..36..189K} 
We find a total SFR from H$\alpha$, 
within 5~kpc from the centre, 
of 0.22~M$_\odot$ yr$^{-1}$ if 
no extinction corrections are taken into account. 
On average, this is a reasonable assumption 
since \citet{2007astro.ph..1897T} 
found a perfect 3.6~cm-H$\alpha$ correlation 
for the thermally emitting regions in \mm.
In any case, given that the diffuse H$\alpha$ emission
amounts to 40\% of the 
total \citep{2000ApJ...541..597H}, an extinction correction
of individual sources  
by \av\,=\,1 on average 
(as claimed by \citealt{1997AJ....113..236D} and 
\citealt{1980ApL....21....1I}), would increase the 
SFR in the same area to 0.3~M$_\odot$ yr$^{-1}$.

To estimate the SFR from the IR emission, we used two independent methods: 
we first estimate the TIR luminosity within a radius of 5~kpc
using the data at 8.0 and 24~\micron. 
With a flux of 47.3~Jy,
the integrated 8.0~\micron\ luminosity L$_\nu$(8.0~\micron) 
is $4.0 \times 10^{28}$~ergs~s$^{-1}$~Hz$^{-1}$, corresponding to
$\nu L_\nu(8)\,=\,1.5\times10^{42}$~ergs~s$^{-1}$.
This is slightly higher than 
L$_\nu$(24~\micron)$=3.5 \times 10^{28}$~ergs~s$^{-1}$~Hz$^{-1}$ 
(or $\nu L_\nu(24)\,=\,4.4\times10^{41}$~ergs~s$^{-1}$).
Such a difference in the 8 and 24\,\micron\ flux (47.3 vs. 40.9~Jy)
is consistent with the IR SEDs measured in 
nearby galaxies \citep{2005ApJ...633..857D}, 
in which IR fluxes at 8.0~\micron\ often exceed those at 24~\micron. 
Using these values, Eq.~1 in \citet{2005ApJ...633..871C} yields 
an L(TIR) =  $1.0 \times 10^9$~L$_\odot$.
The IR SFR is usually computed from the FIR (40-120~\micron) luminosity. 
\citet{2000ApJ...533..682C} found that the ratio 
L$_{\rm IR}$(1-1000~\micron)/L$_{\rm IR}$(40-120~\micron) 
is about 1.75. In our case this translates into a 
FIR luminosity of $5.8 \times 10^8$~L$_\odot$ 
within a radius of 5~kpc in \mm. 
This is in good agreement with the value of $6.5 \times 10^8$~L$_\odot$
found from the total IRAS fluxes reported by \citet{1990ApJ...358..418R},
especially considering that we are including only the central 10~kpc in
our estimate.
Following the precepts of \citet{1996A&A...306...61B} for late spirals,
that is SFR~[M$_\odot$~yr$^{-1}$]~=~$8 \times 10^{-44}$~L(FIR)
with uncertainties of around 50\%, 
we obtain a SFR equal to 
$0.18^{+0.18}_{-0.07}$~M$_\odot$ yr$^{-1}$. 
Within the uncertainties this estimate is compatible with those
derived from H$\alpha$ emission (with and without extinction corrections).
We obtain the same value from the TIR--SFR  relation by \citet{1998ARA&A..36..189K},
taking into account the luminosity conversion (TIR vs. FIR). 

The second method to derive the SFR from the IR emission involves the linear 
relation between radio and IR luminosities, and SFR 
\citep{2001ApJ...554..803Y}. \citet{2005ApJ...632L..79W} derived SFR 
directly from the 24~\micron\ luminosity (see their Eq.~1): in our case, 
this translates into a SFR of 0.17~M$_\odot$~yr$^{-1}$. 
Using the 8~\micron\ luminosity (their Eq.~2) gives a higher
value of SFR, 0.26~M$_\odot$~yr$^{-1}$, even after subtracting
the stellar contribution to the 8~\micron\ flux, $\sim$10\%. 
Although within the errors of the 24~\micron\ and FIR results
\citep[from][]{1990ApJ...358..418R}, the 8~\micron\ value is $\la$40\% higher
than either the 24~\micron\ or the H$\alpha$ one.
A similar trend was found by \citet{2005ApJ...633..871C} in which
the 8~\micron\ SFR estimate can be significantly larger 
in weakly ionized regions such as the disk of \mm\ taken globally.
This is because the 8~\micron\ emission is produced by more than one
mechanism, and thus does not trace SFR as accurately as other IR wavelengths.
 
In general then both MIR prescriptions for computing the 
SFR reach a satisfactory agreement with the 
H$\alpha$-derived value 
and, beyond confirming the adopted SFR computing scheme, implies: 
a) that actively star forming sites are the main contributors to
 the MIR emission;
b) that mean extinction is low also in the inner \hii\ regions; 
c) that highly obscured sources do not contribute
significantly to the MIR emission. 

\subsection{Star formation rates in 24~\micron\ sources} \label{SFR_24Ss}

We then searched for H$\alpha$ emission at the location of the 24Ss. 
This was done in order to compare locally the SFRs from H$\alpha$
with those from the MIR, but also to check whether sources with 
and without an H$\alpha$ counterpart have on average the 
same IR colours. On the narrow-band (continuum 
subtracted) H$\alpha$ image, we performed the 
photometry of the 24Ss using the same technique 
described in Sect.~\ref{sub:Photometry}. 
We find that
293 of the 24Ss are also detected in H$\alpha$ 
(within 3\farcs6).
An inspection of the H$\alpha$ image shows that
most bright 24Ss with no H$\alpha$ counterpart
actually lie near an \hii\ region, with
an offset which, though larger than the photometric errors,
could still be accommodated in case of asymmetries 
of the H$\alpha$ shell.
This is the case, for example, of some sources along the spiral arms.
Some faint 24Ss sources might instead really be absent in the H$\alpha$ map. 
Indeed, the percentage 
of 24Ss undetected in H$\alpha$ gets higher close to the completeness 
limit of our survey. We will analyse this finding 
in more detail in a forthcoming paper. 
We note however that the luminosity function of the 24Ss with H$\alpha$ 
counterpart is totally compatible with the one for the whole 
sample presented in Sect.~\ref{sec:S24_sub:lum}. This is because contamination
of the 24~\micron\ sample by non star forming regions happens mostly below the 
completeness limit of the derived luminosity function.\\

\begin{figure}
\includegraphics[width=\columnwidth]{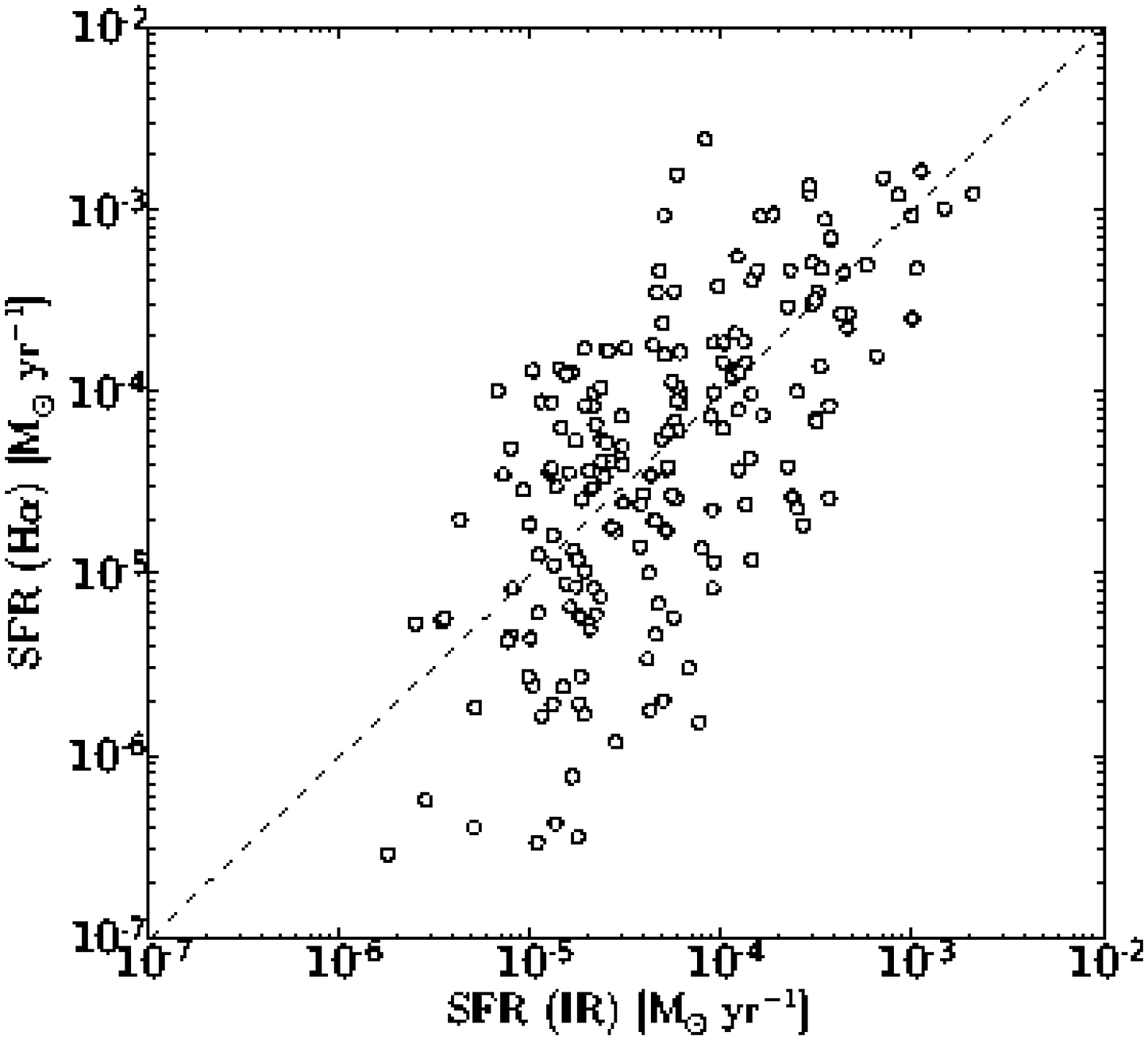}
\caption{Comparison of the SFR calculated from 
the H$\alpha$ luminosity and from the IR emission 
based on the 8.0 and 24~\micron\ emissions.
The dashed line shows the SFR(H$\alpha$)~=~SFR(IR) relation.} 
\label{fig:compSFR}
\end{figure}

To estimate the SFR in discrete sources, we adopted the same conversion 
methods used for the global SFRs and,
in order to estimate the TIR luminosity following
Calzetti's formula, we restrict our sample of 293 sources 
to the 186 sources which are also detected in the IRAC bands. 
The resulting SFRs are compared in 
Fig.~\ref{fig:compSFR}, where the SFR 
is derived from H$\alpha$ without any average or individual
extinction  corrections and the FIR fluxes have been converted
to SFR as in the previous subsection: 
SFR~[M$_\odot$ yr$^{-1}$]~=~$8 \times 10^{-44}$~L(FIR). 
The scatter about the line of slope unity is large,
roughly a factor of 5 ($1\sigma$),
and it is difficult to assess the presence of any
systematic deviation.
The scatter itself, given its magnitude, cannot be imputed to
extinction only. 
Most of the variance is likely linked to the uncertainties
intrinsic to the methods used to infer SFRs. A very similar plot can be obtained 
using the SFR estimated only from the 24~\micron\ emission (see previous subsection) 
of the discrete sources and is consequently not shown here.

\begin{figure}
\includegraphics[width=\columnwidth]{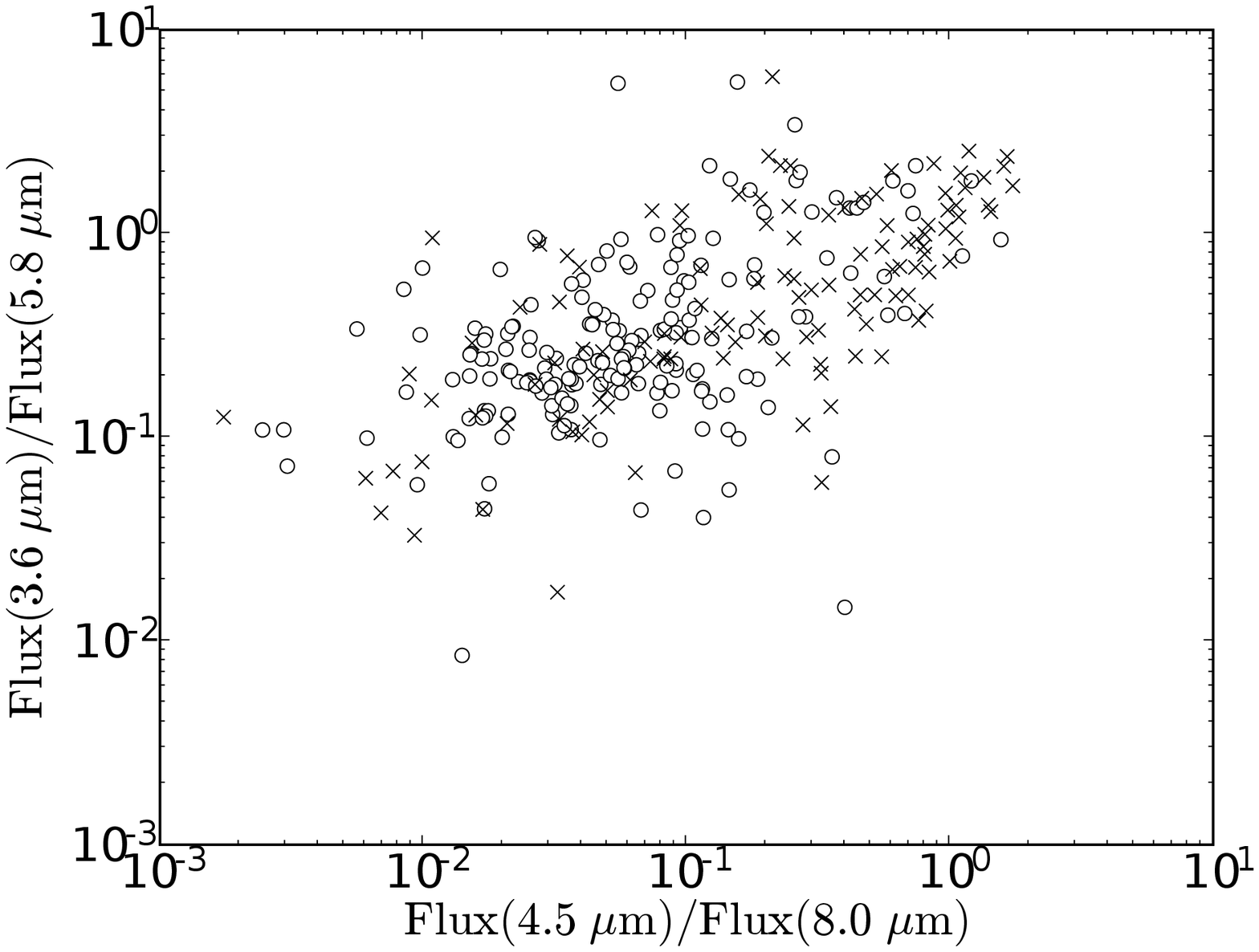}
\caption{Diagnostic diagram involving IRAC colours for the 24~\micron\ 
sources having a H$\alpha$ counterpart (open circles) and 
for the 24~\micron\ sources with no H$\alpha$ counterpart (crosses).}
\label{fig:ccDiagIRACMIPS_HaNoHa_fig1}
\end{figure}

We can  compare the IR colours 
of the latter sample with those of the complementary sample made up 
by the 121 24Ss with no H$\alpha$ counterpart, selecting sources detected 
in all IRAC and MIPS-24 bands. 
One such diagram is in 
Fig.~\ref{fig:ccDiagIRACMIPS_HaNoHa_fig1} and should be compared with the
equivalent diagnostic diagram of Sect.~\ref{sec:diagnostic}. The 24Ss
without H$\alpha$ actually seem to be less dusty sources than those 
emitting also in H$\alpha$, a trend confirming that 24Ss with no H$\alpha$ 
are not more embedded into dust.

\section{Summary and conclusions} \label{sec:disc}

This is the first in a series of papers aiming to investigate 
the star formation history in \mm\ by means of the IR data 
from \spitzer. We retrieved and reduced IRAC and MIPS 
data producing maps roughly the size of the entire star forming galaxy disk.
These were used to study the nature of the MIR emission, both 
globally (large scale structure of dust emission) and locally 
(by means of discrete sources). Our main results are the following:

\begin{itemize}
\item On large scales, the 3.6/4.5 and 5.8/8.0 
ratios are remarkably constant. The 3.6 and 4.5~\micron\ emission
mainly arises from point sources while the 5.8 
and 8.0~\micron\ bands present more diffuse emission following 
the morphological pattern of the galaxy (spiral arms). 
The 8.0/24 ratio is also rather constant suggesting a tight link 
between PAH and VSG properties. 
On the other hand, the 3.6/5.8 and 4.5/8.0 ratios reveal the flocculent
spiral structure in \mm, because of the increasing dominance
of the ISM toward longer wavelengths.
The distribution of cooler 
dust is investigated via the MIPS images: the 24 and 70~\micron\ 
emissions match closely, while the 160~\micron\ band reveals a more 
diffuse character in the cold-dust emission.  
The 24 and 70~\micron\ are also more tightly correlated
with the H$\alpha$ emission.

\item In the four IRAC and 24~\micron\ MIPS bands, 
we performed photometry of catalogued discrete sources of various 
type, including \hii\ regions, SNRs and PNe. 
We present the detection rates in each band for the
various types of object, together with their IR 
SEDs. For all types of objects, a drop at 4.5~\micron\ is 
observed, likely due to the contribution of emission features
or to the decreasing contribution of photospheric emission
from the central star. 

\item The IR colours of \hii\ regions, SNRs 
and PNe are used to define diagnostic diagrams. 
We observe a continuity among the various types: 
the \hii\ regions appear more dusty than the 
PNe, which behave more like stellar objects. 
We also compared these colours with colours predicted by models (such 
as Cloudy, \dusty, SB99, pure blackbody emission). 
The diagnostics based on IRAC alone cannot disentangle 
the different theoretical models, but usage of hybrid
IRAC+24~\micron\ colours allows the removal of
some of the degeneracies.
\hii\ regions are better modelled 
by pure Cloudy \hii\ regions (including PAHs) and by \dusty\ 
models with inner-shell temperatures around 600~K
and low extinction.

\item We have compared the colours of some of the known-type 
objects in \mm\ with similar objects in our Galaxy 
(\hii\ regions) or in the LMC (SNRs) and, 
although with some scatter, we found good agreement 
for the location of the sources in the colour diagrams.

\item From our 24~\micron\ image, we extracted a blind catalogue 
of 515 discrete sources, on which we performed photometry in 
the IRAC, and 24~\micron\ MIPS bands and in H$\alpha$.
Their colours match closely those of known-type sources, 
and in particular those of \hii\ regions. Consequently, 
the models which better describe the nature of the 
24~\micron\ sources are also pure Cloudy 
\hii\ regions and low-extinction warm (T$_{\rm in} = 600$~K) \dusty\ models.

\item The total 24~\micron\ flux of the 515 discrete sources 
amounts to $1.4 \times 10^{41}$~erg~s$^{-1}$ 
while the diffuse emission, that is the residual flux, amounts to twice this value.

\item We present the IR luminosity function of our 24~\micron\ 
sources, as a function of the 24~\micron\ flux 
and as well as a function of the total IR luminosity (estimated 
from the 8.0 and 24~\micron\ bands). The luminosity ranges from values
typical of the faintest ionising stars to those of very large star
formation complexes. The shape (double-sloped) is the same as that
observed for Galactic \hii\ regions and star complexes, though 
AGB stars could contribute significantly to
the faintest bins.

\item Within a radius of 5~kpc in \mm, 
we estimate a total IR luminosity 
of $1.0 \times 10^9$~L$_\odot$ and the FIR luminosity is about 
$5.8 \times 10^8$~L$_\odot$. This 
translates into an estimate of the SFR of 
0.2~M$_\odot$~yr$^{-1}$, in very good agreement with the SFR 
inferred from H$\alpha$ emission.

\item The relation between 24~\micron\ emission and 
\hii\ regions is also investigated by searching directly 
for H$\alpha$ counterparts of the 24~\micron\ 
sources. About half of the 24~\micron\ sources are 
very closely associated with H$\alpha$ while the remaining ones 
are more distant from an H$\alpha$ knot
than the allowance by positional uncertainties.
The SFR for the single sources in IR and H$\alpha$ are 
consistent but the scatter is
rather large and not imputable to extinction alone.

\item The IR colours of the 24~\micron\ 
sources with and without H$\alpha$ counterparts are quite similar.
If anything, upon comparison with the theoretical diagnostics,
those with H$\alpha$ appear more dusty than the others.

\end{itemize}

\begin{acknowledgements}
We would like to thank Rene Walterbos for providing us 
the H$\alpha$ image of \mm, Laura Magrini for the 
complete list of optical \hii\ regions, and Bruce Draine for 
the digital version of a diffuse ISM spectrum. 
The work of S.~V. is supported by a INAF--Osservatorio Astrofisico 
di Arcetri fellowship. The Spitzer Space Telescope 
is operated by the Jet Propulsion Laboratory, 
California Institute of Technology, under contract with 
the National Aeronautics and Space Administration. This research 
has made use of the NASA / IPAC Extragalactic Database, 
which is operated by JPL /Caltech, under contract with NASA.
\end{acknowledgements}

\bibliography{8179}

\end{document}